\documentclass[useAMS,usenatbib,usegraphicx]{mn2e}

\usepackage{amssymb}
\usepackage{hyperref}

\hypersetup{citecolor={black},linkcolor={black},urlcolor={black}}  

\hypersetup{
  colorlinks=true,             
  bookmarksnumbered=true,      
  bookmarksopen=false,         
  pdfpagemode=UseOutlines,     
  pdfdisplaydoctitle,          
  pdftitle={Wide-field IFS of NGC 628}
}

\newcommand{\tableline}{\hline}

\bibpunct{(}{)}{;}{a}{}{,} 
\newcommand{\funits}{10$^{-16}$ erg~s$^{-1}$~cm$^{-2}$~\AA$^{-1}$}
\newcommand{\Funits}{10$^{-16}$ erg~s$^{-1}$~cm$^{-2}$}

\newcommand{\nFUNITS}{10$^{-11}$ erg~s$^{-1}$~cm$^{-2}$}

\newcommand{\FunitsA}{10$^{-16}$ erg~s$^{-1}$~cm$^{-2}$~arcsec$^{-2}$}

\newcommand{\zh}{[Z/H]}

\newcommand{\hh}{H\,{\footnotesize II}~}
\newcommand{\nii}{[N\,{\footnotesize II}]~}
\newcommand{\sii}{[S\,{\footnotesize II}]~}
\newcommand{\siii}{[S\,{\footnotesize III}]~}
\newcommand{\oi}{[O\,{\footnotesize I}]~}
\newcommand{\oii}{[O\,{\footnotesize II}]~}
\newcommand{\oiii}{[O\,{\footnotesize III}]~}
\newcommand{\lam}{$\lambda$}

\newcommand{\ha}{H$\alpha$}

\newcommand{\ff}{\rm ff} 
\newcommand{\flux}{erg\,s$^{-1}$\,cm$^{-2}$}
\newcommand{\fluxA}{erg\,s$^{-1}$\,cm$^{-2}$\,\AA$^{-1}$}

\defcitealias{RosalesOrtega:2010p3553}{Paper~I}
\defcitealias{Fathi:2007p2409}{Fathi07}

\title[Wide-field IFS of NGC\,628]{PPAK Wide-field Integral Field Spectroscopy
  of NGC\,628: I. The largest spectroscopic mosaic on a single
  galaxy\thanks{Based on observations collected at the Centro Astron\'omico
  Hispano Alem\'an (CAHA) at Calar Alto, operated jointly by the Max-Planck
  Institut f\"ur Astronomie and the Instituto de Astrof\'isica de Andaluc\'ia
  (CSIC).}}

\author[S.~F.~S\'anchez et al.]{S.~F.~S\'anchez,$^{1}$\thanks{E-mail: sanchez@cefca.es}
F.~F.~Rosales-Ortega,$^{2}$ R.~C.~Kennicutt,$^{2}$  B.~D.~Johnson,$^{2}$ 
\newauthor A.~I.~Diaz,$^{3}$ A.~Pasquali,$^{4}$ C.~N.~Hao$^{2}$\\
$^{1}${Centro Astron\'omico Hispano Alem\'an, Calar Alto, (CSIC-MPG),
  C/Jes\'us Durb\'an Rem\'on 2-2, E-04004 Almeria, Spain }\\
$^{2}${Institute of Astronomy, Cambridge University, Madingley Road, Cambridge CB3 0HA, UK}\\
$^{3}${Departamento de F\'isica Te\'orica, C-XI, Universidad Aut\'onoma de Madrid, 28049 Madrid, Spain}\\
$^{4}${Max-Planck-Institut f\"ur Astronomie, K\"onigstuhl 17, D-69117 Heidelberg, Germany}
}

\begin{document}

\date{To be edited latter (June 2009)}

\pagerange{\pageref{firstpage}--\pageref{lastpage}} \pubyear{2010}

\maketitle

\label{firstpage}

\begin{abstract}

We present a wide-field IFS survey on the nearby face-on Sbc galaxy NGC\,628,
comprising 11094 individual spectra, covering a nearly circular field-of-view
of $\sim$\,6 arcmin in diameter, with a sampling of $\sim$\,2.7 arcsec per
spectrum in the optical wavelength range (3700--7000 \AA).  This galaxy is
part of the PPAK IFS Nearby Galaxies Survey, (PINGS, Rosales-Ortega et
al. 2009). To our knowledge, this is the widest spectroscopic survey ever made
in a single nearby galaxy. A detailed flux calibration was applied, granting a
spectrophotometric accuracy of $\sim$\,0.2 mag. The spectroscopic data was
analysed both as a single integrated spectrum that characterises the global
properties of the galaxy, and using each individual spectrum to determine the
spatial variation of the stellar and ionized gas components. The spatial
distribution of the luminosity-weighted ages and metallicities of the stellar
populations was analysed. Using typical strong emission line ratios we
derived the integrated and 2D spatial distribution of the ionized gas, the
dust content, SFR, and oxygen abundance.

The age of the stellar populations shows a negative gradient from the inner
(older) to the outer (younger) regions. We found an inversion of this gradient
in the central $\sim$\,1 kpc region, where a somewhat younger stellar
population is present within a ring at this radius. This structure is
associated with a circumnuclear star-forming region at $\sim$\,500 pc, also
found in similar spiral galaxies. From the study of the integrated and
spatially resolved ionized gas we found a moderate SFR of $\sim$\,2.4 M$\odot$
yr$^{-1}$. The oxygen abundance shows a a clear gradient of higher metallicity
values from the inner part to the outer part of the galaxy, with a mean value
of 12~+~log(O/H) $\sim$ 8.7. At some specific regions of the galaxy, the
spatially resolved distribution of the physical properties show some level of
structure, suggesting real point-to-point variations within an individual \hh
region. Our results are consistent with an inside-out growth scheme, with
stronger star formation at the outer regions, and with evolved stellar
populations in the inner ones.

\end{abstract}


\section{Introduction}
\label{sec:intro}

Galaxies in the local universe ($\sim$\,10 Mpc) are the fundamental anchor
points of any study of the evolution of these objects in cosmological time
scales. Therefore, it is important to understand their main properties,
including their morphology, ionized and neutral gas content, stellar
populations and metallicities. Due to their apparent scale-length they
represent the perfect laboratories to study the dependence of the stellar
population, the star formation history, and star formation rate on the
morphology and morphological substructures, the metallicity enrichment and the
mechanisms of metal transfer, as well as the nature of the gas ionization.

Powerful constraints on theories of galactic chemical evolution, on the
star formation history, and on the nucleosynthesis in galaxies can be derived
from the accurate determination of the nature of the ionization, the
star-formation rate and the chemical abundances at different locations within
a galaxy. Fortunately, these sort of studies can be addressed using ground-based
astronomy by observing galaxies with large apparent scale-lengths. 
Nearby galaxies may be more easily separated into a number of different
morphological components and several types of stellar populations. Given the
spatial variation in the star-formation histories (including violent episodes
in some cases), and the amount of dust attenuation within a galaxy, and their
relatively low surface brightness, it is difficult to study the stellar
populations of nearby galaxies using just their integrated light. More
information can be extracted by studying their resolved properties, although
it is a complex task to tackle.

Nearby galaxies have been observed for many decades using many different
techniques, such as multi-band optical and near-infrared broad-band imaging
(to derive the properties of their dominant stellar populations and population
gradients), narrow-band imaging and multi-object observations to derive their
gas content and gas kinematics
\citep[e.g.][]{Kennicutt:1980p3756,Belley:1992p3779,Scowen:1996p2663}, and
slit spectroscopy of the brightest \hh regions within the galaxy
\citep[e.g.][]{McCall:1985p1243,Zaritsky:1994p333,vanZee:1998p3468}. More
recent studies, using space-based observatories, have revealed new features in
these apparently well known objects, such as star-forming regions at very
large radii \citep[e.g. GALEX][]{GildePaz:2007p3757,GildePaz:2007p2488} and
obscured star-forming regions
\citep[e.g.][SPITZER]{Kennicutt:2003p1560,Prescott:2007p3758}.

Despite all these efforts, we still lack of a complete picture of the main
properties of these galaxies, especially those ones that can only be
revealed by spectroscopic studies (like the nature of the ionization
and/or the metal content of the gas). This is because previous
spectroscopic studies only sample a very few discrete regions in these complex
targets \citep[e.g.][]{Roy:1988p320,Kennicutt:1996p1603}, and in many cases
they were sampling very particular types of regions \citep[\hh
regions,][]{McCall:1985p1243,vanZee:1998p3468,Castellanos:2002p3372,Castellanos:2002p3374}.
Integrated spectra over large apertures are required to derive these
properties in a more complete way, but are difficult to obtain using classical
slit spectroscopy (although recent efforts have obtained integrated spectra of
some local galaxies by adopting a drift-scan procedure
\citep{Moustakas:2006p307}. Even in these cases, only a single integrated
spectrum is derived, and the spatial information is lost.

Recent studies have derived the integrated and spatially resolved properties of
certain portions of nearby galaxies, by using Integral Field Spectroscopy (IFS)
techniques \citep[e.g. SAURON,][]{Bacon:2001p2659,deZeeuw:2002p3161}. This
technique allows one to obtain spatially resolved, continuously sampled
spectroscopy over the field-of-view (FOV) of the instrument. These studies,
though leading to extremely important results, were focused on the study of
the central regions of galaxies, with a FOV of radius $\sim$\,20 arcsec,
corresponding on average to the inner $\sim$\,1 kpc. More recently,
\citet{Blanc:2009p3483} observed the central region of M\,51 ($\sim1.7$
arcmin$^2$) using the VIRUS-P instrument. However, to date there has been no
systematic study of a sample of local universe galaxies using IFS, covering a
substantial fraction of their optical sizes.

In order to fill this gap, we began an IFS survey of 17 nearby ($<$\,100 Mpc)
galaxies, called PINGS \citep[PPAK Integral-field-spectroscopy Nearby Galaxies
Survey,][hereafter Paper~I]{RosalesOrtega:2010p3553}. By their nature, most of
the objects in the survey cannot be covered in a single pointing with IFS
instruments, and a new observing-reduction technique had to be developed to
perform accurate mosaicking of the targets. In this article we present the
first scientific results based on one of our targets, NGC\,628 (or
M\,74). This is a local universe galaxy at $z\sim$\,0.00219 ($\sim$\,9 Mpc),
and the largest in projected angular size of the PINGS sample. 

In this first article of the series devoted to the IFS study of NGC 628, we
present a study of the small and intermediate scale variation in the line
emission and stellar continuum of NGC\,628. We derive these properties from
both the integrated spectrum of the galaxy and the spatially resolved spectra
(by means of pixel-resolved maps across the disk of the galaxy), and compare
the results.  Additionally, we include a description of the data acquisition
and reduction techniques, in particular on those details that are different
than the standard reduction of IFS data and that were not addressed in
\citetalias{RosalesOrtega:2010p3553}. The structure of the article is as
follows: In \autoref{sec:n628} we give an overview of the general properties
of NGC\,628. In \autoref{sec:obs} we present the observational details,
including the instrument, telescope and the observing technique. In
\autoref{sec:redu}, we explain the reduction technique, describing the
software packages used, and the sanity checks performed. In
\autoref{sec:analysis}, we describe the analysis performed on each spectrum
included in our dataset and consider also the integrated spectrum of the
galaxy. The spatial distributions of the different derived properties are also
presented here. In \autoref{sec:summary} we discuss the results and summarise
the main conclusions of this study. Finally, in
\hyperref[app]{Appendix~\ref{app}}, we describe in detail the technique used
to analyse the stellar populations in the galaxy, including simulations to
characterize its reliability.

\section{General properties of NGC\,628}
\label{sec:n628}

Morphologically classified as an Sbc \citep{Holmberg:1975p3759}, NGC\,628 it is
nearly face-on spiral, showing a typical grand-design structure. This galaxy
has been well observed at a variety of wavelengths and there is abundant
multiwavelength ancillary data from photographic plates and CCD imaging in the
optical \citep[e.g.][]{Holmberg:1975p3759,Boroson:1981p3760,Shostak:1984p3761,
Natali:1992p3762,Hoopes:2001p3417,Kennicutt:2008p3419}
to SPITZER in the NIR \citep[SINGS,][]{Kennicutt:2003p1560}, and GALEX in the UV
\citep{GildePaz:2007p3757}.
NGC\,628 is a good example of an isolated galaxy, previous studies have
established that NGC\,628 has not had an encounter with satellites or other
galaxies in the last 10$^9$ yr. \citep{Kamphuis:1992p3780}. However, it is
possible that the galaxy suffered some kind of interaction 1 Gyr ago, which
could explain the presence of a large-scale oval structure (weak bar)
discovered in the NIR \citep{James:1999p3781,Seigar:2002p3782},
and the disturbed morphology in the north part of the galaxy.

By means of observations in neutral hydrogen, an elliptical ring-like
structure was discovered well beyond the optical disc, at around 12 arcmin
from the nucleus of the galaxy \citep{Roberts:1962p3783,Briggs:1980p3784},
lying on a plane with $\sim15^\circ$ inclination with respect to the inner
disc. The presence of this warped velocity field is a puzzle, given the
apparent isolation of the galaxy which would rule out its origin by tidal
disruption. However, this feature is most likely the result of the interaction
with two large high-velocity clouds accreting onto the outer parts of the disc
\citep{Kamphuis:1992p3780,LopezCorredoira:2002p3785,Beckman:2003p3786}.

From the morphological and dynamical point of view, NGC\,628 displays one
prominent spiral arm to the south, and one or several disturbed spiral arms to
the north (although UV observations have shown spiral arms with a more
symmetrical appearance than in the optical), an inner rapidly rotating
disc-like structure \citep{Daigle:2006p3787}, a CO-discovered circumnuclear
ring of star formation at $\sim2$ kpc from the centre
\citep{Wakker:1995p3788,James:1999p3781} thought to be the result of a barred
potential \citep{Seigar:2002p3782}, and a nuclear (nested) bar on a $\sim100$
pc scale \citep{Laine:2002p3789}. All these ingredients seem to suggest that
the evolution of the structure in NGC\,628 has been driven by secular
evolution of the disc \citep{Kormendy:2004p3790}.

On the other hand, \citet{Cornett:1994p3791} found that the star formation
history in NGC\,628 varies with galactocentric distance (with a young stellar
population in the circumnuclear region), while \citet{Natali:1992p3762}
suggested a scenario with an inner and an outer disc with different stellar
populations, with a transition region located at 8-10 kpc from the centre, the
same radius at which \citet{Cepa:1990p3792} found that the star formation
efficiency is at its minimum, which they interpreted as the corotation radius
for the spiral arms.

More recently, \citealt{Fathi:2007p2409} (hereafter Fathi07) presented a
detailed kinematic analysis of the galaxy based on wide-field 2D Fabry-Perot
maps of H$\alpha$. They found that the velocity dispersion of the gas and for
individual \hh regions is practically constant at all galactocentric distances
covered by their study. This result, together with the fact that they were not
able to distinguish true diffuse gas emission given that the star formation is
widely distributed within the disc, let Fathi07 to suggest that the emission
in H$\alpha$ from the \hh regions dominates any emission from the diffuse
component. Furthermore, they confirmed the presence of a disc-like central
structure, which they argue could have been built up by inflow from large
galactocentric distances, suggesting that NGC\,628 is in the process of
forming a secular pseudo-bulge.

NGC\,628 has been also a classical target for optical spectroscopy, there is a
good spectroscopic coverage of many of its structures, from the core
\citep{Moustakas:2006p307} to many \hh regions located within and beyond the
optical disc of the galaxy
\citep[e.g.][]{McCall:1985p1243,vanZee:1998p3468,Ferguson:1998p224,Castellanos:2002p3372}. Integral
field spectroscopy of the central core ($\sim$\,33 arcsec$\times$41
arcsec) of NGC\,628 has been recently obtained using the SAURON Integral Field
Unit \citep{Ganda:2006p3135,Ganda:2007p3763}. These studies showed that the
stellar component rotates in the same direction as the gas, that the stellar
velocity dispersion drops in the central zones (suggesting a dynamical cold
inner disc), and that the H$\beta$ distribution is more extended than the
\oiii distribution, both suggesting a ring-like structure, confirming the
15-20 arcsec nuclear ring previously reported by \citet{Wakker:1995p3788}.

In terms of studies focusing on the gas-phase of the galaxy, and on particular
on the determination of the chemical abundance of the galaxy, previous long-slit
spectroscopic studies have derived the abundance gradient of NGC\,628 up to
relatively large galactocentric radii ($\sim2R_{25}$), using mainly empirical
metallicity indicators based on the ratios of strong emission lines
\citep[e.g.][]{McCall:1985p1243,Zaritsky:1994p333,vanZee:1998p3468,Ferguson:1998p224,Castellanos:2002p3372}.
These studies have shown a higher metallicity content in the inner part of the
galaxy, that the slope of the gradient is constant across the range of
galactocentric distances sampled by the different studies, and that the oxygen
abundance decrease is relatively small. In the dynamical scenario previously
described, this allows a moderate mixing of the disc material driven by the
large-scale oval distortion \citep{Zaritsky:1994p333}. However, these results
have been drawn from relatively few spectroscopically observed \hh regions.

An alternative method was developed by \citet{Belley:1992p3779}, who derived
reddenings, H$\beta$ equivalent widths, diagnostic line ratios and
metallicities for 130 \hh regions by the implementation of imaging
spectrophotometry, i.e. using narrow-band interference filters with the
bandpass centered on several key nebular lines. They found no trends in the
reddening nor for the H$\beta$ EW's with galactocentric distance. However they
found that the excitation, and some diagnostic line ratios are strongly
correlated with galactocentric radius. They also derived an oxygen abundance
gradient of NGC\,628 based on the [O\,{\footnotesize III}]/H$\beta$ ratio.

In summary, NGC\,628 represents an interesting object for a full 2D
spectroscopic study, considering that it is the prototype of a normal,
isolated, grand-design, face-on, nearby spiral, with an interesting
morphology and particular dynamical features. Furthermore, despite all the
previous studies towards a full understanding of NGC\,628, we still lack a
detailed knowledge of the gas chemistry, stellar populations and the global
properties across the surface of this galaxy, which can be only obtained by
spectroscopic means. The IFS data presented in this and future papers aim to
fill this gap, by deriving and comparing the global and spatially resolved
spectroscopic properties of NGC\,628 with previous results obtained by
multi-wavelength observations. The information provided by these 2D spectral
maps will allow us to test, confirm, and extend the previous body of results
from small-sample studies, and would provide a new path for the analysis of
the two-dimensional metallicity structure of disks and the intrinsic
dispersion in metallicity on this and other late-type normal spirals.

\section{Observations}
\label{sec:obs}

\begin{figure}
  %
  %
  \includegraphics[width=\hsize]{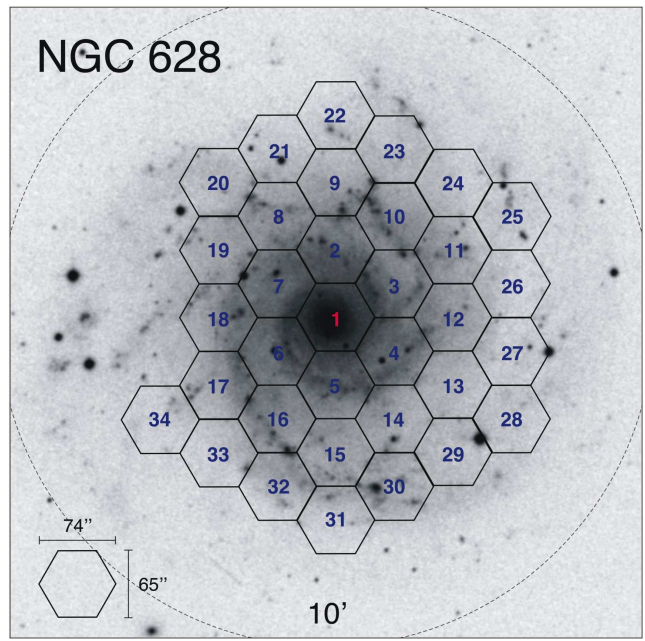} 
  \caption[PPAK mosaicking of NGC\,628]
  {
    $B$-band Digital Sky Survey image of NGC\,628. The mosaic of the PPAK
    pointings is shown as overlaid hexagons indicating the field-of-view of the
    central fibre-bundle. The identification number of each pointing is indicated,
    as listed in \autoref{tab:pointings}. The circle indicates the $D_{\rm
      25}$ radius (in the $B$-band). The image is 10'$\times$10' and it
    is displayed in top-north, left-east standard configuration.
    \label{fig:mosaic}
  }
\end{figure}

Observations were carried out at the 3.5m telescope of the Calar Alto
observatory with the Potsdam Multi Aperture Spectrograph, PMAS, \citep{Roth:2005p2463}
in the PPAK mode \citep{Verheijen:2004p2481,Kelz:2006p3341}.
The PPAK fibre bundle consists of 382 fibres of 2.7 arcsec diameter each
\citep[see Fig.~5 in][]{Kelz:2006p3341}. Of these 382 fibres, 331 (the science
fibres) are concentrated in a single hexagonal
bundle covering a field-of-view of 72''$\times$64'', with a filling factor of
$\sim$\,65\%. The sky background is sampled by 36 additional fibres,
distributed in 6 bundles of 6 fibres each, distributed along a circle
$\sim$\,90 arcsec from the center of the instrument FOV. The sky-fibres are
distributed among the science fibres within the pseudo-slit in order to have
a good characterization of the sky; the remaining 15 fibres are used for
calibration purposes. Cross-talk between adjacent fibres is
estimated to be less than 5\% when using a pure aperture extraction
\citep{Sanchez:2006p331}. Adjacent fibres in the pseudo-slit may come from
very different locations on the spatial plane \citep{Kelz:2006p3341}, minimizing
the effect of the cross-talk even more (although it does introduce an
incoherent contamination not important for the present study).

The V300 grating was used for all the observations, covering the wavelength
range $\sim$\,3700-7100 \AA, with a spectral resolution of $FWHM\sim$\,8
\AA. Due to the large size of NGC\,628 ($\sim$ 10.5\,$\times$\,9.5 arcsec)
compared to the FOV of the instrument a mosaicking scheme was adopted,
following the experiment by \citet{Sanchez:2007p1696} using the same
instrument. The initial pointing was centered on the centre of the
galaxy. Consecutive pointings followed an hexagonal pattern, adjusted to the
shape of the PPAK science bundle. Each pointing centre is 60 arcsec distant
from the previous pointing centre. Due to the shape of the PPAK bundle and by
construction of the mosaic 11 spectra of each pointing, corresponding to one
edge of the hexagon, overlap with the same number of spectra from the previous
pointing. This pattern was selected to maximise the covered area, minimise
large gaps, and allow enough overlap of fibres to calibrate exposures taken
under different atmospheric conditions. For the central pointing we adopted a
dithering scheme with three positions having offsets (0$\arcsec$,0$\arcsec$),
($+$0.78$\arcsec$,$+$1.68$\arcsec$), and ($+$0.78$\arcsec$,$-$1.68$\arcsec$)
following \citet{Sanchez:2007p3299}. This dither pattern allows us to cover the
gaps between fibres and to increase the spatial resolution in this region
(where more structure is expected). For each non-dithered pointing we obtained
3 exposures of 600s each, and for the central, dithered, pointings we obtained
2 exposures of 600s each.

\begin{table}
  \begin{center}
    \caption[Log of the observations]
    {
      Log of the observations.
      \label{tab:pointings}
    }
    \begin{tabular}{crr}\hline
      Date     && Pointings\\
      \tableline
      28/10/06   & & 1,2,3,9,10,22,23\\
      10/12/07   & & 1,4,5,6,7,8\\
      11/12/07   & & 14,15,16,17,18,19,30\\
      12/12/07   & & 11,12,13,25,29,31,32\\
      09/08/08   & & 11\\
      30/10/08   & & 24,26,27,28,29,33,34\\
      \tableline
    \end{tabular}
  \end{center}
\end{table}

The observations spanned six nights, distributed over three different
years. In \autoref{fig:mosaic} the field-of-view of each pointing, labelled with its
corresponding identification number, is overlaid on top of a $B$-band Digital
Sky Survey\footnote{The Digitized Sky Survey was produced at the Space
  Telescope Science Institute under U.S. Government grant NAG W-2166. The
  images of these surveys are based on photographic data obtained using the
  Oschin Schmidt Telescope on Palomar Mountain and the UK Schmidt
  Telescope. The plates were processed into the present compressed digital
  form with the permission of these institutions.} Note that a
substantial fraction of the galaxy is covered by the IFS
mosaic. \autoref{tab:pointings} gives a log of the observations, including the
date of observation and the pointings observed each night, following the
identification numbers shown in \autoref{fig:mosaic}. A concentric observing
sequence was adopted, starting from the inner to the outer regions. The
atmospheric conditions varied between the different observing runs, although
in general they were clear but non-photometric. The seeing varied between
$\sim$\,1 and $\sim$\,1.8 arcsec and the median of the seeing over all nights
was $\sim$\,1.4 arcsec. Different spectrophotometric stars were observed
during the observing runs, with at least two stars observed each night, in
order to perform flux calibration. In addition to the science pointings, sky
exposures of 300s were taken each night in order to perform a proper
subtraction of the sky contribution. Additional details on
the observing strategy can be found in \citetalias{RosalesOrtega:2010p3553}.

\section{Data reduction}
\label{sec:redu}

Data reduction was performed using {\sc R3D} \citep{Sanchez:2006p331}, in
combination with {\sc IRAF}\footnote{IRAF is distributed by the National
  Optical Astronomy Observatories, which are operated by the Association of
  Universities for Research in Astronomy, Inc., under cooperative agreement
  with the National Science Foundation.} packages and {\sc E3D} 
\citep{Sanchez:2004p2632}. The reduction
consists of the standard steps for fibre-based integral-field spectroscopy. A
master bias frame was created by averaging all the bias frames observed during
the night and subtracted from the science frames. The different exposures
taken at the same position on the sky were then combined, clipping the cosmic
rays, using {\sc IRAF} routines. Then the locations of the spectra on the CCD were
determined using a continuum illuminated exposure taken before the science
exposures. Each spectrum was then extracted from the science frames. In order
to reduce the effects of the cross-talk we did not perform a simple aperture
extraction, which would consist of co-adding the flux within a certain number
of pixels of location derived from the continuum illuminated
exposure. Instead, we adopted a modified version of the Gaussian-suppression
technique described in \citet{Sanchez:2006p331}.

The new technique assumes a Gaussian profile for the projection of each fibre
spectrum along the cross-dispersion axis. It basically performs a Gaussian
fitting to each of the fibres after subtracting the contribution of the
adjacent fibres in an iterative process. First, a simple aperture extraction
is performed with a 5 pixel aperture. This initial guess of the flux
corresponding to each spectrum is then used to model the profiles with a
Gaussian function, adopting as centroid the location of the peak intensity
determined from the continuum exposure. The width of the model Gaussian is
taken from the average width of all the fibre profiles ($\sigma\sim$\,2
pixels). In this first iteration the aperture extracted flux is used as the
integrated flux of the Gaussian function. This modelled profile is then used,
for each spectrum, to remove the contribution of the four adjacent fibre
spectra. The resulting {\it clean} profile is then fitted with a Gaussian
function, with the centroid and width parameters fixed, in order to derive a
better estimation of the integrated flux. This new flux is used as the input
for the next iteration of the process. It was found that with three iterations
the procedure converges, increasing the signal-to-noise ratio of the extracted
spectra and reducing the effect of cross-talk.

The extracted flux, for each pixel in the dispersion direction, is stored in a
row-stacked-spectrum file RSS \citep{Sanchez:2004p2632}. Wavelength calibration was
performed using HeHgCd lamp exposures obtained before and after each pointing,
yielding an accuracy of $rms\sim$\,0.3 \AA. Differences in the relative
fibre-to-fibre transmission throughput were corrected by comparing the
wavelength-calibrated RSS science frames with the corresponding frames derived
from sky exposures taken during the twilight.

\begin{figure*}  
  \centering
  %
  %
  \includegraphics[height=7cm]{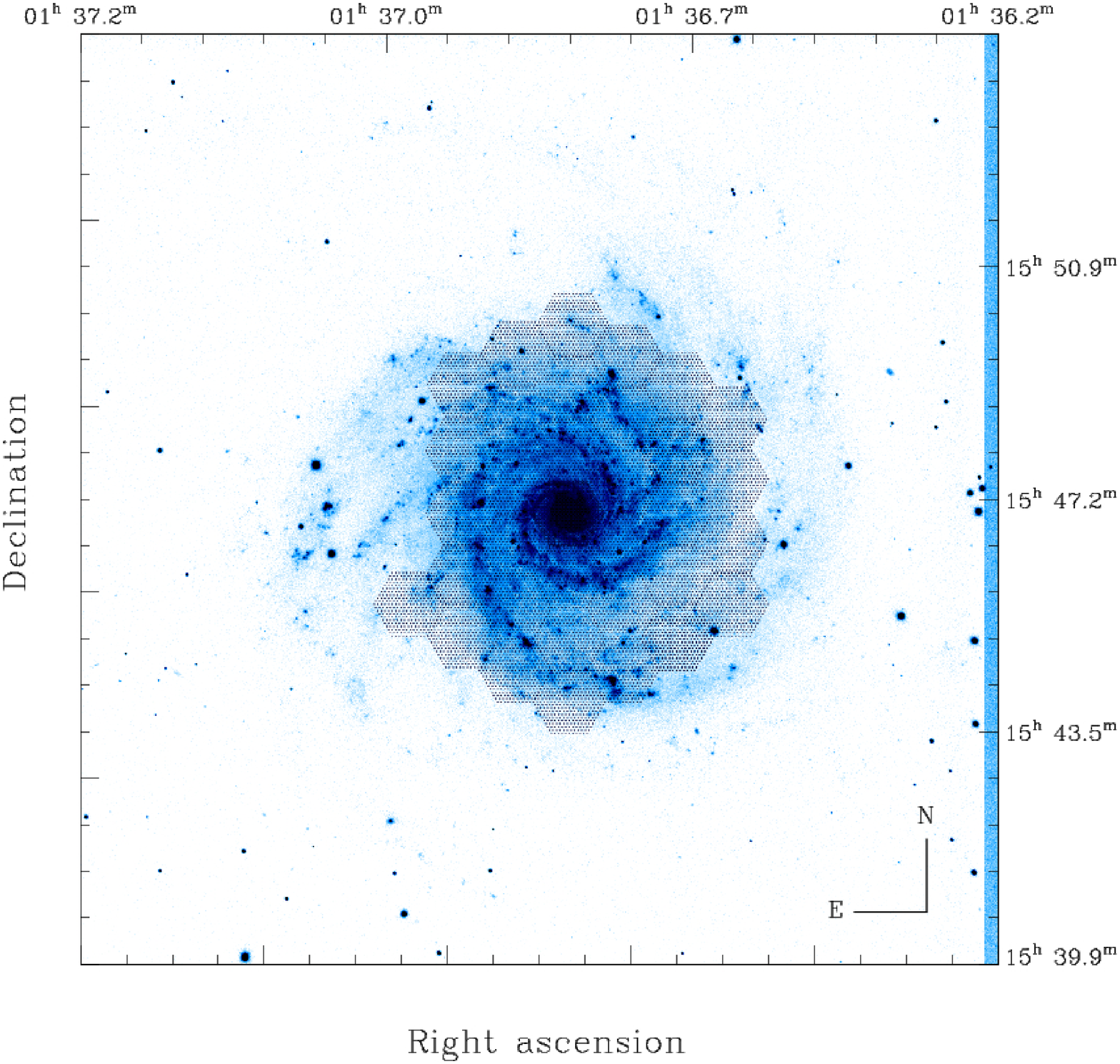}\hspace{0.2cm}  
  \includegraphics[height=7cm]{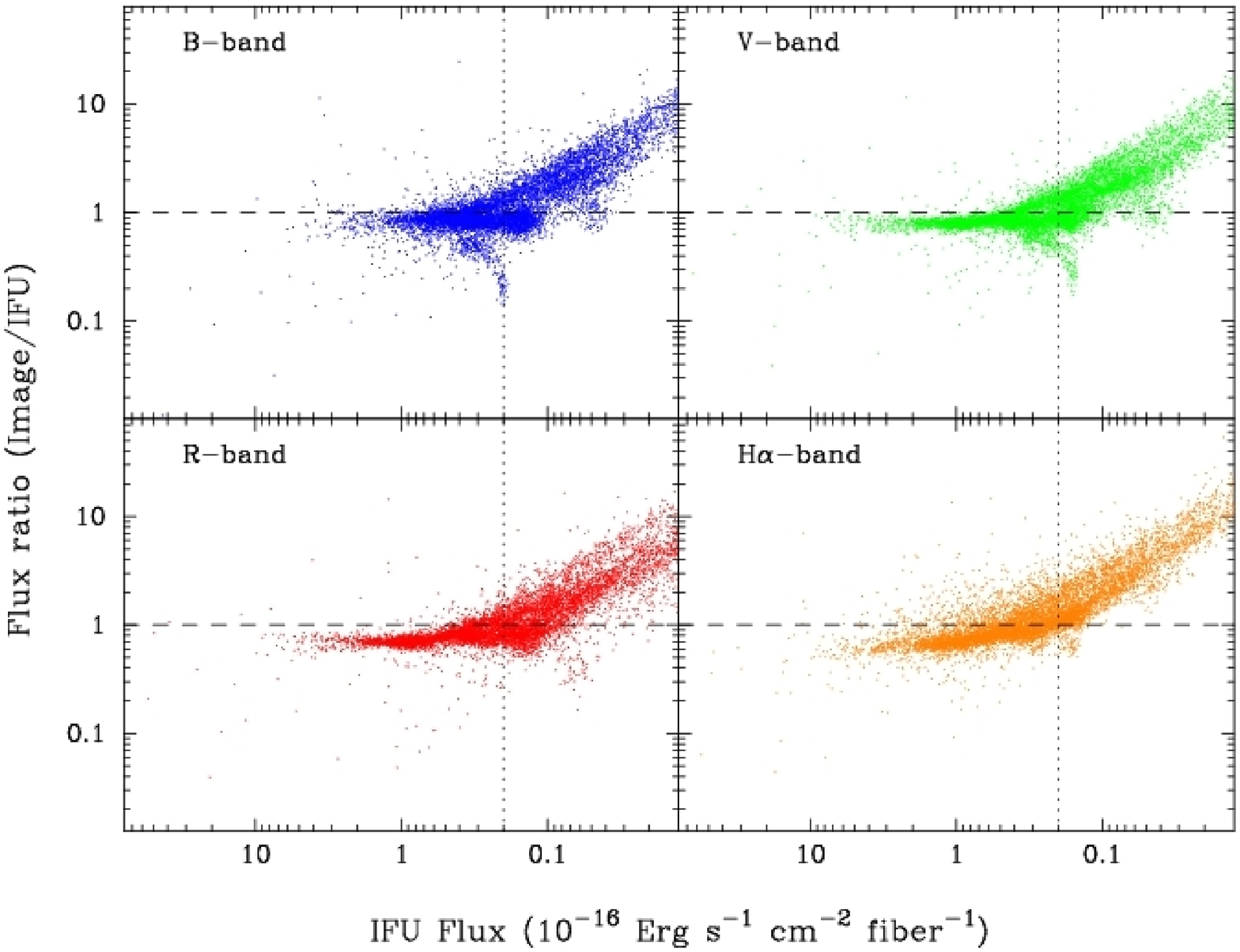} 
  \caption[IFU vs. broad-band imaging flux extraction]
  {
    Left: 
    SINGS $B$-band image of NGC\,628 used to perform the
    absolute flux re-calibration. The position of each individual fibre of the
    IFS mosaic is shown overlaid to the real scale.
    Right:
    Distribution of the ratio between the flux in the apertures corresponding
    to the PPAK fibres on each of the flux-calibrated broad-band images
    obtained from the SINGS ancillary data and the corresponding flux
    extracted from the IFU data. Each panel shows the results for the different
    imaging data: $B$-band (top-left panel), $V$-band (top-right panel), $R$-band
    (bottom-left) and H$\alpha$ (bottom-right panel). The horizontal
    dashed-line indicate the one-to-one ratio, which the vertical dotted-line
    indicate the adopted intensity cut for acceptable flux calibration.
    \label{fig:fluxcal}
  }
\end{figure*}

Although PPAK is equipped with fibres to sample the sky, in most of our
pointings the sky fibres are located within an area containing significant
signal from the galaxy. Accurate sky subtraction was therefore an important
issue in the data reduction. For the different nights and pointings we adopted
different sky subtraction schemes depending on the position of the
pointing. In some cases, a sufficient number of sky-fibres are located in
regions free from galaxy emission, and it was possible to perform an accurate
sky subtraction of the individual pointings using these sky spectra. Once a
certain frame is properly sky-subtracted, the sky spectrum of any adjacent
frame, observed in the same night and under similar weather conditions, can be
easily estimated. To do so the spectra of the 11 fibres in the sky subtracted
frame that overlap an adjacent frame (that has not been sky subtracted) are
combined. This combined spectrum is then subtracted from the corresponding
combined spectrum of the 11 fibres of the adjacent frame. This produces an
estimation of the sky spectrum in the adjacent frame that is then subtracted
from all the spectra in that frame. Prior to this subtraction it is necessary
to visually check that no residual of the galaxy is kept in the estimated sky
spectrum, which can occur if the atmospheric transmission changed
substantially between the observations of the two frames.

In other frames the sky observations taken during the night could be used to
subtract the sky in the science observations, especially when the sky and
target exposures are taken within a few minutes of each other. When the
exposures were more widely separated in time it was necessary to combine
different sky frames with different weights to derive good results. The
criterion adopted to decide when a subtraction was good or not was to minimise
the residuals in the typical emission features of the night spectrum. A
thorough analysis of the sky subtraction and the residuals found in the
reduction of this galaxy can be found in \citetalias{RosalesOrtega:2010p3553}.

After reducing each individual pointing we built a single RSS file for the
mosaic following an iterative procedure. The spectra of each pointing were
scaled to those of the previous pointing by the average ratio in the 11
overlapping spectra. Those overlapping spectra were then replaced by the
average between the previous pointing and the new rescaled spectra. The
resulting spectra were incorporated into the final RSS file, updating the
corresponding position table. By adopting this procedure the differences in
the spectrophotometric calibration night-to-night and frame-to-frame are
normalized to that of the first frame used in the process. For this reason the
mosaic was constructed starting from the central frame observed in the night
of the 10th of December 2007, under nearly photometric conditions according to
the Calar Alto extinction monitor. The final mosaic dataset comprises 11094
non-overlapping individual spectra, covering a field-of-view of
$\sim$\,6'$\times$7', i.e. $\sim$\,70\% of the optical size of the galaxy
(defined by $D_{25}$ mag arcsec$^{-2}$ radius in the $B$-band), and therefore represents the
largest spectroscopic survey of a single galaxy.

\subsection{Accuracy of the flux calibration}
\label{sec:test}

Although particular care has been taken to achieve the best spectrophotometric
calibration, there many effects that can strongly affect it. Among them, the
most obvious are the photon-noise from low surface brightness regions of the
galaxy, the sky-background noise or variations in the weather conditions
between the time when the spectrophotometric standards and the object were
observed. This latter effect was reduced by the adopted mosaicking procedure
in the data reduction, since the photometric calibration was renormalized to
that of a particularly good night. Less obvious is the effect of inaccuracies
in the sky subtraction. However, for low surface brightness regions this is
one of the most important effects.

Since it is our goal to provide accurate spectrophotometric data, we performed
a flux re-calibration on the data. To do so we used the flux calibrated
broad-band optical images from the SINGS legacy survey  (Kennicutt et
al. 2003). In particular, we compared our dataset with the $B$, $V$, $R$ and
H$\alpha$ images, since they are mostly covered by the wavelength range of our
spectra. The photometric calibration of those images is claimed to be
$\sim$\,5\% for the broad-band images and $\sim$\,10\% for the narrow-band
images. They reach a depth of $\sim$\,25 mag/arcsec$^2$ with a signal-to-noise
of $\sim$\,10$\sigma$. Therefore, for the structures included in the FOV of
our IFS data, the photometric errors of the imaging are dominated by the
accuracy of the calibration, and not by the photon-noise.

\begin{figure*}
  %
  %
  \centering
  \includegraphics[height=7cm]{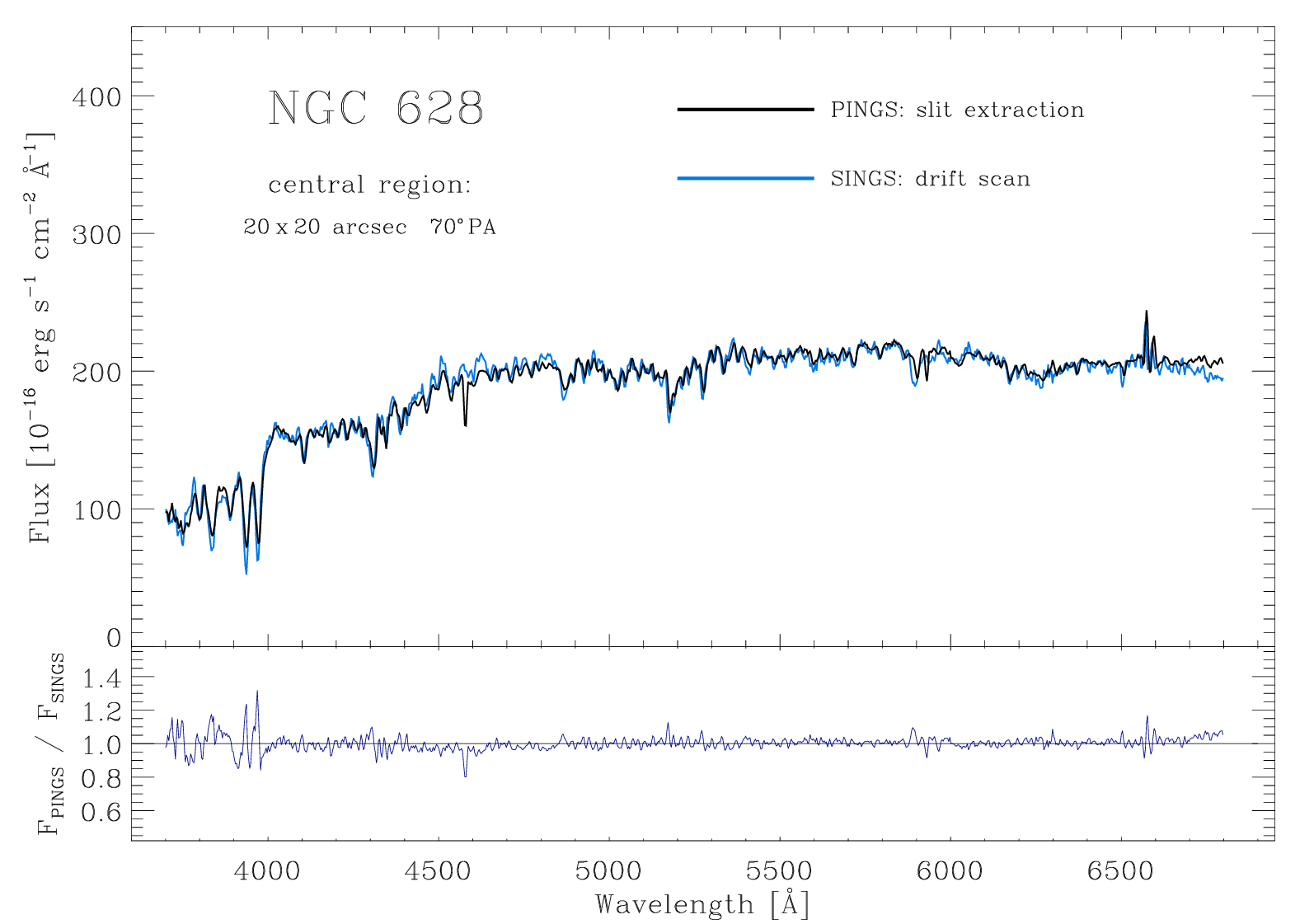}\vspace{0.5cm}
  \includegraphics[height=7.2cm]{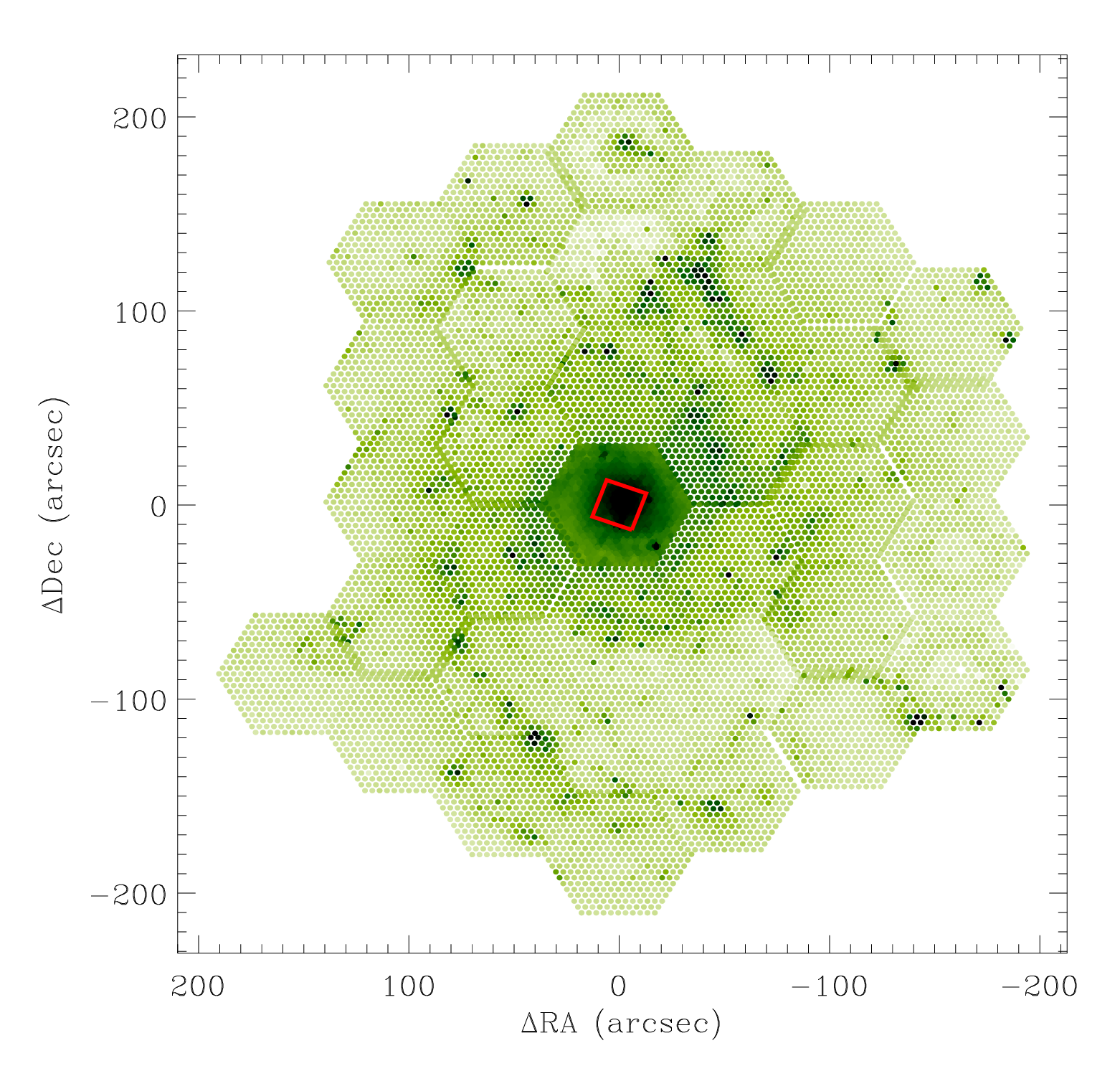}
  \caption[Comparison spectra of the central region of NGC\,628]
  { 
    Left: 
    Comparison between the spectrum extracted for the central
    region of NGC\,628 (shown in the right-panel) using the PPAK data (black-line)
    after co-adding the spectra within a simulated aperture with the same size,
    location and PA as the SINGS legacy survey drift-scan (blue-line, in the
    online version of the paper).
    The bottom-panel shows the ratio between the two spectra.
    Right:
    Narrow-band map of the PINGS mosaic of NGC\,628 centered at H$\alpha$, the red box
    in the centre corresponds to the simulated aperture from which the IFS
    spectrum was extracted.
    Despite the different techniques used to derive both spectra, there is a
    clear agreement between them in most of the wavelength range.
    \label{fig:speccomp}
  }
\end{figure*}

The IFS mosaic was registered to the SINGS optical data by matching the
coordinates of the galaxy bulge and foreground stars in the FOV (see the
left-panel of \autoref{fig:fluxcal}). After this
process the error in the astrometry of the IFS data was estimated to be
$\sim$\,0.3'', based on the $rms$ of the differences in the centroid of the
stars and galaxy bulge. Once registered, the mosaic position table was used to
extract aperture photometry from each broad-band image at the location of each
fibre and with an aperture similar to that of the fibres. The photometry was
transformed to flux (in cgs units) by using the counts-to-magnitude
prescription in the SINGS
documentation\footnote{http://tinyurl.com/SINGS-doc}, and the zero-points
included in \citet{Fukugita:1995p3366}. On the other hand, each spectrum in
the mosaic was convolved with the corresponding transmission curve of the
filters indicated before, in order to extract a similar flux fibre-to-fibre,
based on the IFS data. This procedure provides us with 11094 photometric
points per band to compare between the two datasets.

The right-panel of \autoref{fig:fluxcal} shows the ratio between the two
sets of photometric points versus the flux extracted from the IFU data, for
each of the considered filters. The figure shows the typical pattern obtained
when comparing the flux
ratio between two datasets with different depth, with the broad-band images
clearly deeper than the spectroscopic data, as expected. Down to 0.2 \Funits
($\sim$\,5772 fibres) the ratio is $\sim$\,0.9 for all the fibres, with a
standard deviation of $\sim$\,0.3 dex. Small differences in the transmission
curves of the filters used in this calculation and the ones used by
\citet{Fukugita:1995p3366} as well as the astrometric errors described before
introduce uncertainties in the derived fluxes which increase the standard
deviation. The uncertainties from the former source are difficult to
estimate. However the uncertainties introduced by astrometric errors are
estimated to be at least $\sim$\,10\%, by simulating different mosaic
patterns, moving the location of fibres randomly within 0.3 arcsec (the
uncertainty of our astrometry) and comparing the extracted photometry. Based
on all these results we estimate our spectrophotometric accuracy to be better
than $\sim$\,0.2 mag when we apply the re-calibration derived by this flux
ratio analysis.
The overall re-normalization factor applied to the IFS mosaic was 1.15,
derived as the average of the ratios found on each band. This difference in
the zero-point of the flux calibration lies within the expectations based on
the estimated accuracy of our spectrophotometric calibration.

\autoref{fig:speccomp} shows in the left-panel, the comparison between the
drift-scan spectrum of the central region of NGC\,628 (blue line), published
by Kennicutt et al. (2003)\footnote{http://irsa.ipac.caltech.edu/data/SPITZER/SINGS/galaxies/ngc0628.html}, and
the integrated spectrum extracted from the spectrophotometrically re-calibrated
PINGS mosaic (black line), after 
co-adding the spectra within a simulated aperture with the same size, location
and PA as the SINGS drift-scan. The SINGS drift-scan corresponds to a 20''
aperture and 70$^{\rm o}$ PA. The right-panel of \autoref{fig:speccomp}
shows a 100~\AA\ width narrow-band map of the mosaic of NGC\,628 centered at
H$\alpha$, the red box in the centre corresponds to the simulated aperture from
which the IFS spectrum was extracted. The coordinates, size and PA of the aperture
were obtained from the header of the SINGS data file. Some gaps in between the
edges of the pointings are seen in this map, they are due to the re-centering
of the individual pointings after comparing with the broad-band images as
discussed in above. 
As expected from the spectrophotometric re-calibration of the mosaic, there
is a very good agreement between both datasets; both the general shape of the
spectra and the strength of the spectral features match well. The left-panel of
\autoref{fig:speccomp} shows in the lower part the relative difference between
the two spectra, which is consistent with a null difference (in average) and
within the error of the absolute flux calibration for most of the spectral
range. As simulations with different position, apertures and mosaic versions
showed, the small deviations in the continuum are due to the presence of a
foreground star within the FOV of the simulated slit. The small disagreement
at wavelengths shorter than 4000~\AA\ is expected due to the degradation of the
CCD sensitivity in the blue, as explained in detailed in
\citetalias{RosalesOrtega:2010p3553}, where further comparisons with previous
spectrophotometric data can be found. A sligth missmatch of the wavelength
resolution at the edges of the spectra is also noticed, being in the range of
the expectations for such comparisons.

\begin{figure*}
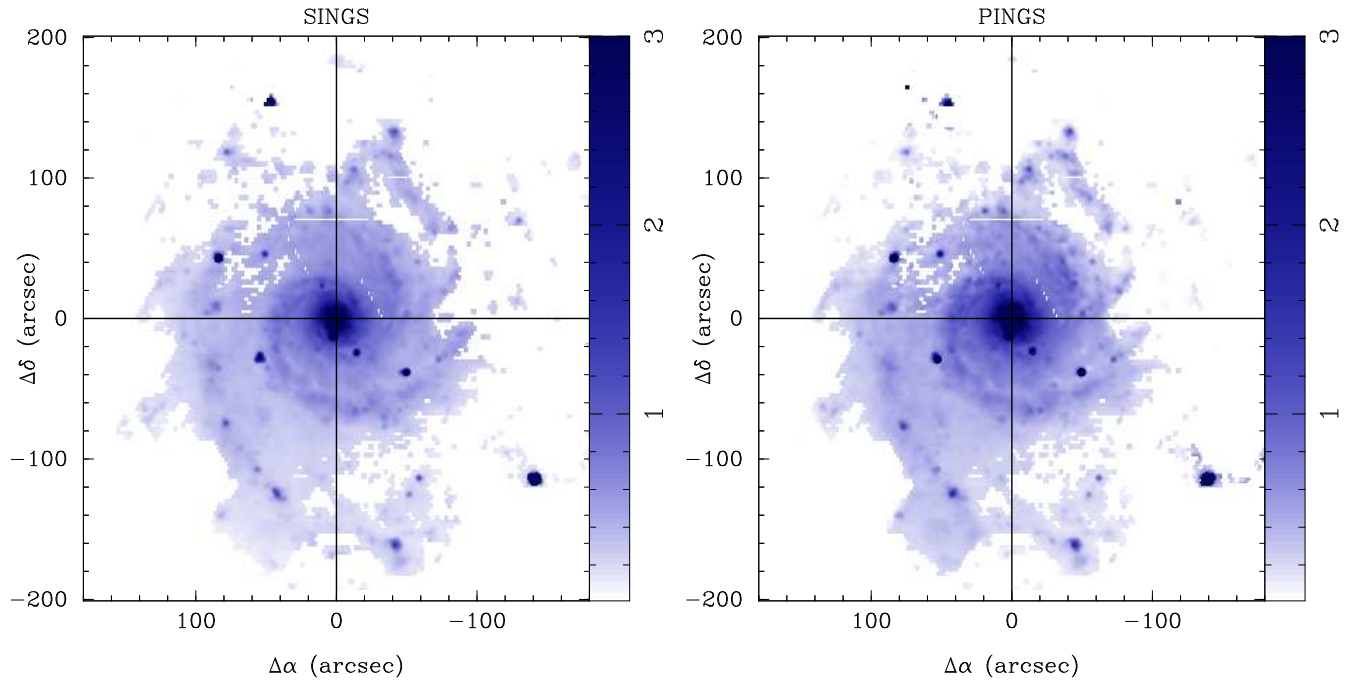

  %
  %
  %
  \includegraphics[angle=270,width=0.49\hsize]{V_band_SINGS}\hspace{0.2cm}
  \includegraphics[angle=270,width=0.49\hsize]{V_band_PPAK}
  \caption[Reconstructed IFS $V$-band image of NGC\,628]
  {
    Reconstructed $V$-band images of NGC\,628. Left: image created after
    interpolating the aperture photometry extraction of the SINGS broad-band
    image. Right: interpolated image derived after multiplying each
    spectrum of the IFS mosaic with the filter response curve. Both images were
    created with a regular grid of 1''/pixel. The areas not satisfying our
    criteria of accurate spectrophotometry were masked. The perpendicular
    lines are centered at the mosaic's reference point. Offsets are in arcsec.
    \label{fig:V_compar}
 }
\end{figure*}

\autoref{fig:V_compar} shows the comparison between two reconstructed $V$-band
images of NGC\,628. The image on the left was created after interpolating the
aperture photometry extraction of the SINGS broad-band image. The image on the
right was derived from the flux re-calibrated IFS dataset, once each spectrum
corresponding to a particular fibre was convolved with the filter response, as
described before. In order to create a regularly gridded image, the data were
interpolated using {\sc E3D} \citep{Sanchez:2004p2632}, adopting a natural-neighbor
non-linear interpolation scheme, and a final pixel scale of 1''/pixel. The
areas not fulfilling our criteria for accurate spectrophotometry
(F$_V<$\,0.2\Funits) were masked.

\section{Analysis and Results}
\label{sec:analysis}

In order to extract physical properties of the galaxy from the dataset it is
necessary to perform different kinds of analyses. In particular, for each
spectrum we need to identify the emission lines of ionized gas, and decouple
this emission from the underlying stellar population. Particular care has to
be taken in this decoupling technique, since some of the emission lines
(e.g. H$\beta$) may be strongly affected by underlying absorption features.

The decoupling of the stellar population from the emission lines was performed
adopting a scheme summarized here: (i) A set of detected emission lines was
identified in the integrated spectra of the stronger \hh regions in the outer
part of the galaxy. (ii) For each spectrum in the dataset, the underlying
stellar population was fitted by a linear combination of a grid of
single-stellar population (SSP) templates, after correcting for the
appropriate systemic velocity and velocity dispersion (including the
instrumental dispersion which dominates the total observed dispersion), and
taking into account the effects of dust attenuation. A spectral region of
30\,\AA\ width around each detected emission line was masked prior to the
linear fitting, including also the regions around the sky-lines
\citep{Sanchez:2007p1276}. (iii) Once we derived a first approximation of the
spectrum of the underlying stellar population, this was subtracted from the
original spectrum to obtain a pure emission-line spectrum. (iv) To derive the
intensity of each detected emission line, each of these {\it clean} spectra in the
dataset was fitted to a single Gaussian function per emission line plus a low
order polynomial function. Instead of fitting emission lines over the entire
wavelength range simultaneously, for each spectrum we extracted shorter
wavelength ranges that sampled one or a few of the analysed emission lines, in
order to characterise the residual continuum with a simple polynomial
function, and to simplify the fitting procedure. When more than one emission
line was fitted simultaneously, their systemic velocities and FWHMs were
forced to be equal (since the FWHM is dominated by the instrument resolution),
in order to decrease the number of free parameters and increase the accuracy
of the deblending process (when required). (v) Finally, for each spectrum in
the dataset a pure gas-emission spectrum was created, based on the results of
the last fitting procedure, using only the combination of Gaussian
functions. This model was then subtracted from the original data, spectrum by
spectrum, to produce a dataset of gas-free spectra. These spectra are then
fitted again by a combination of SSP, as described before (but without masking
the spectral range around the emission lines, in this case), deriving the {\it
  luminosity-weighted} age, metallicity and dust content of the composite stellar
population. The \hyperref[app]{Appendix~\ref{app}} gives a detailed description of the fitting
procedure, indicating the basic algorithms adopted, and including estimates of
the accuracy of the SSP-based modeling and the derived parameters based on
simulations. The 2D maps of the emission line intensities and physical
properties shown in the following sections were constructed based on the pure
gas-emission mosaics described above.

\subsection{Integrated spectrum}
\label{sec:integrated}

A particularly interesting use of IFS datasets is the combination of the
observed spectra to produce an integrated spectrum of the object, using the
IFU as large aperture spectrograph. This high signal-to-noise integrated
spectrum can be used to derive, for the first time, the real average
spectroscopic properties of the galaxy, as opposed to previous studies that
attempted to describe its average properties by the analysis of individual
spectra taken at different regions. The most similar approach would be the
spectrum derived by using a drift-scanning technique
(e.g. \citet{Moustakas:2006p307} and part of the ancillary data of the SINGS
survey), although in those studies (specially for the latter) the fraction of
galaxy covered by the spectra was much less than the for the spectrum
presented here. Another advantage of the use of an IFU with respect to the
drift-scan technique is that the former allows a comparison between the
integrated and the spatially resolved properties of the galaxy.

\begin{figure}
  %
  %
  \centering
  \includegraphics[width=\hsize]{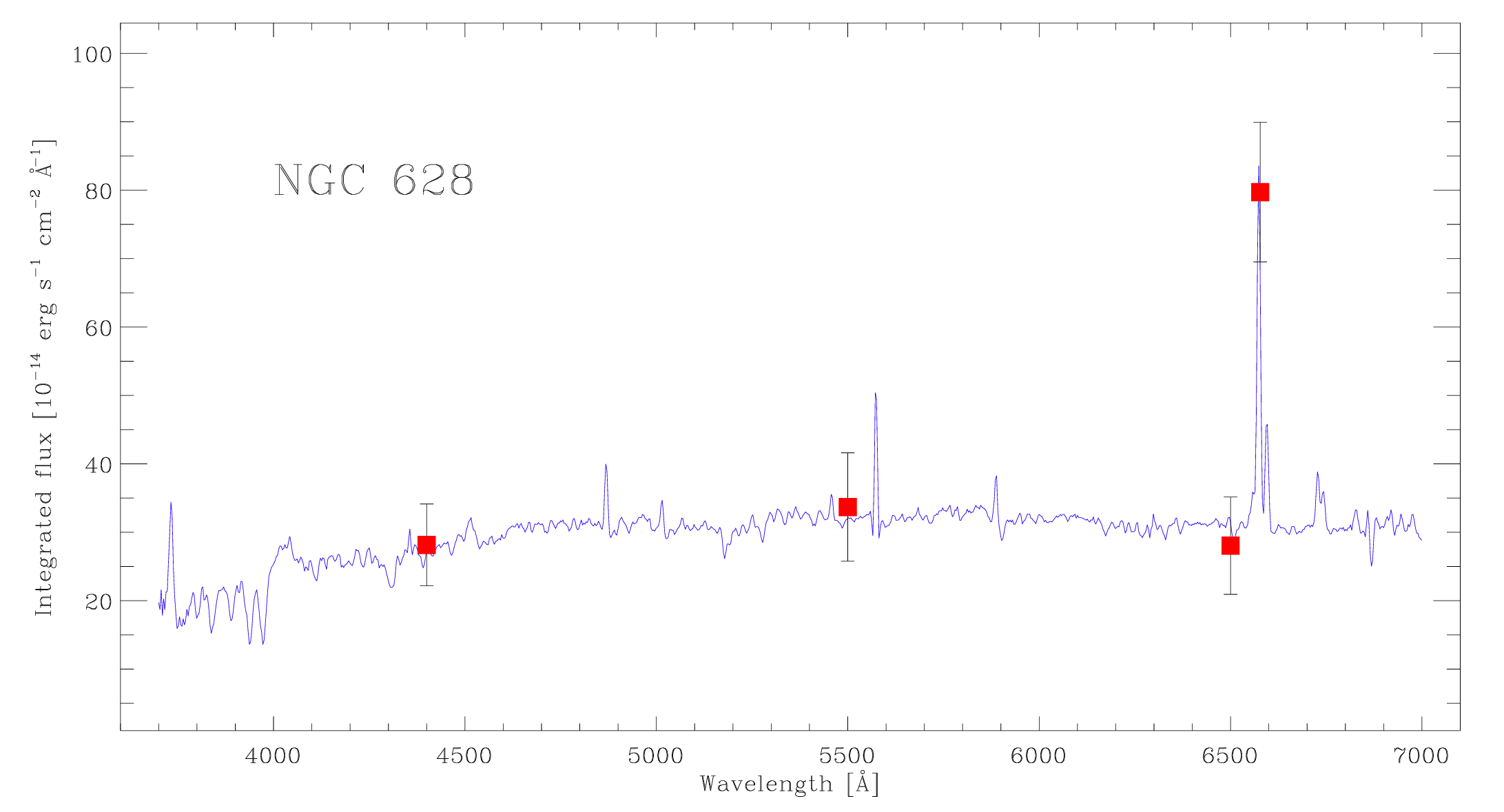}\vspace{0.2cm}
  \includegraphics[width=\hsize]{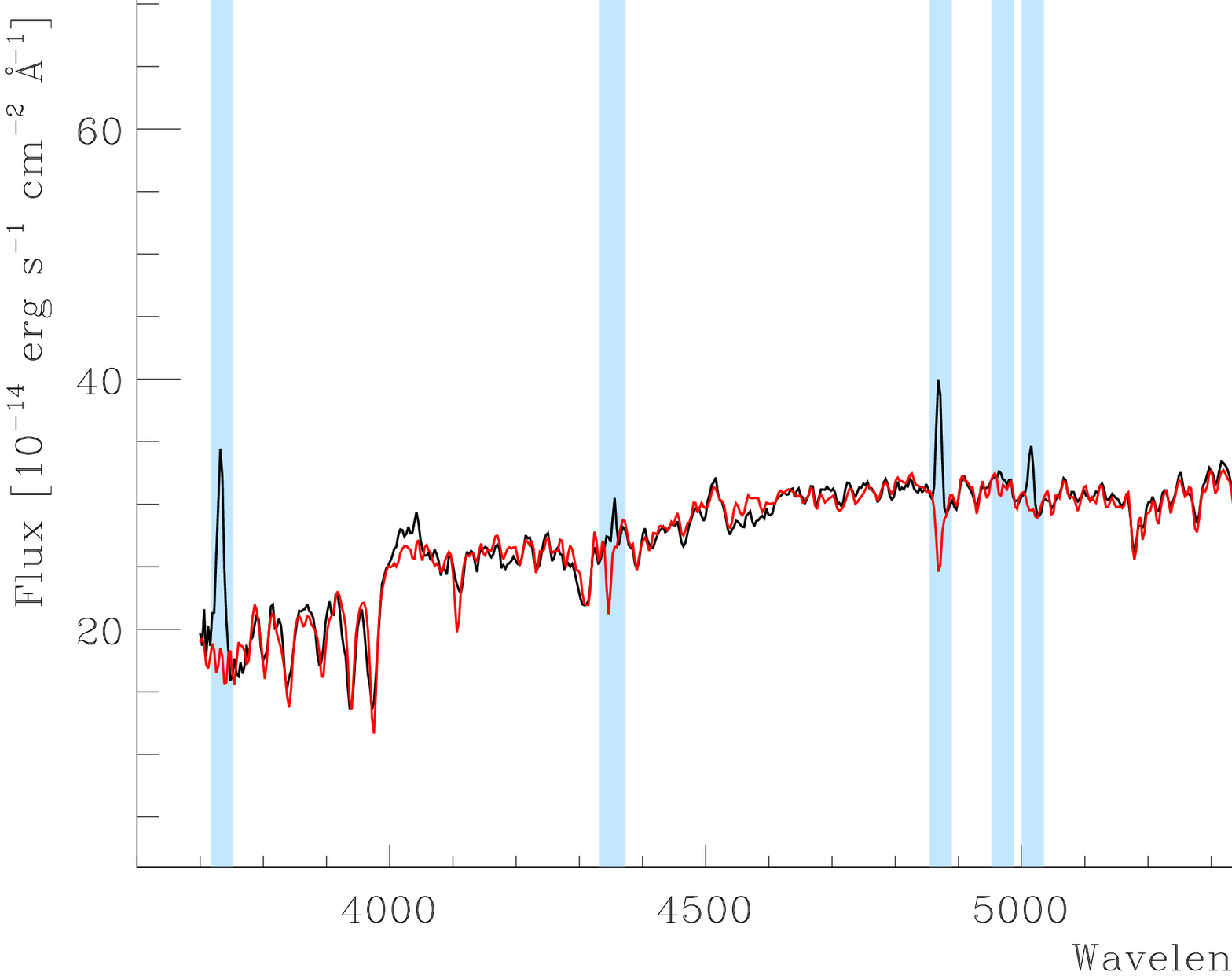}\vspace{0.2cm}
  \includegraphics[width=\hsize]{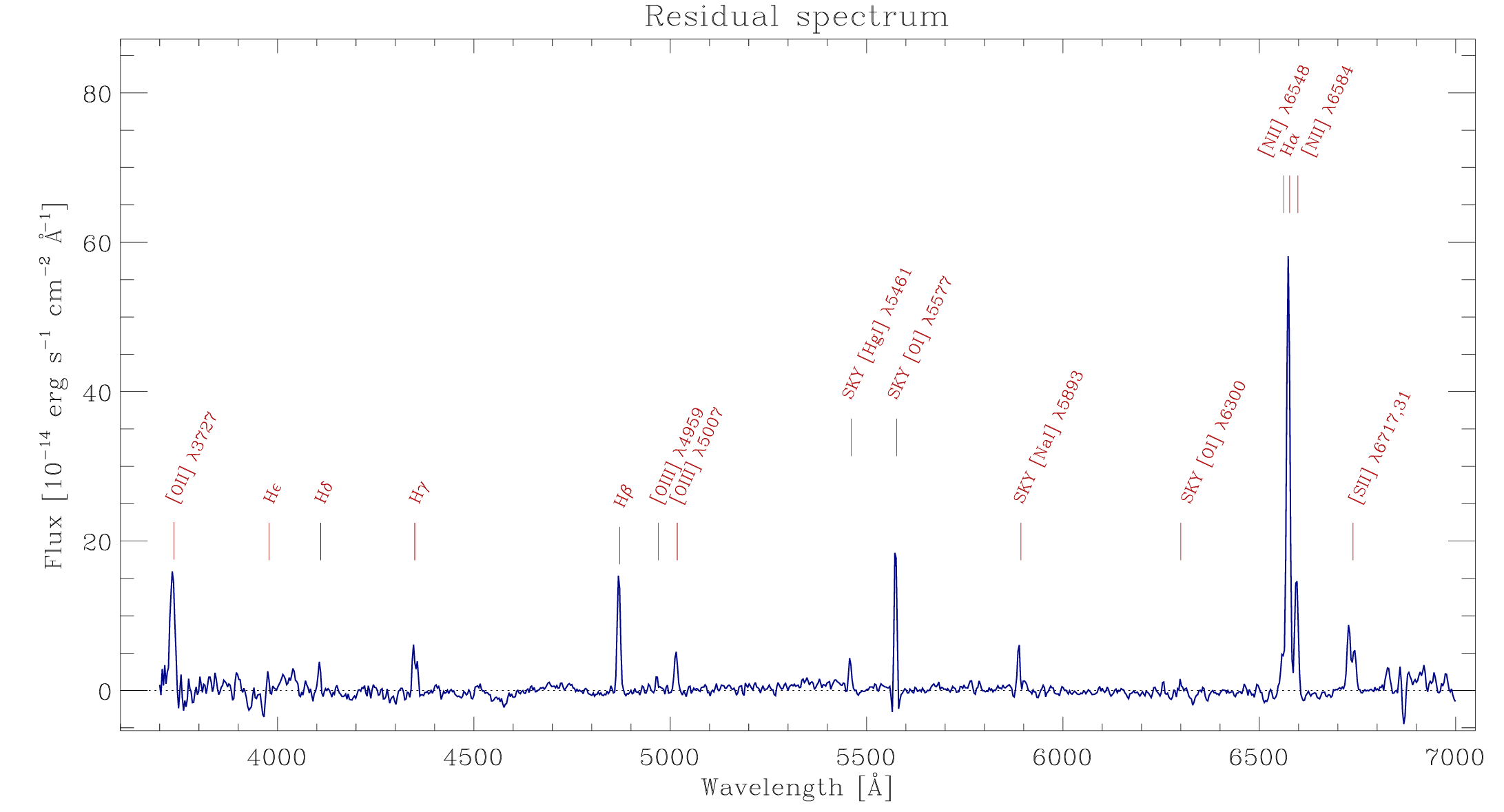}
  \caption[Integrated spectrum of NGC\,628]
  { 
    Integrated spectrum of NGC\,628. The red solid squares in the top-panel indicate
    the integrated flux derived from the $B$, $V$, $R$ and H$\alpha$
    images. The middle-panel shows the SSP model fitting (red-line, in the
    online version) to the
    spectrum, the light-blue bands correspond to the spectral regions masked
    during the fitting. The bottom panel shows the residual after subtracting
    the model from the original spectrum, the detected emission lines have
    been labeled.
    \label{fig:integ_plot}
  }
\end{figure}

A good amount of the fibres in the IFS mosaic of NGC\,628 do not contain enough
signal-to-noise or do not contain signal at all (i.e. spectra with a flat
continuum consistent with zero-flux), as the fibres were sampling regions
where the intrinsic flux from the galaxy is low or null (e.g. borders of the
mosaic, intra-arms regions, etc). In order to get rid of spectra where no
information could be derived, we obtained a {\em clean} version of the IFS
mosaic of NGC\,628 by applying a flux threshold cut choosing only those fibres
with an average flux along the whole spectral range greater than 10$^{-16}$
\fluxA. Furthermore, bad fibres (due to cosmic rays and CCD cosmetic defects)
and foreground stars (10 within the observed field-of-view of NGC\,628) where
removed from the mosaic. This procedure resulted in a refined mosaic version
which includes only those regions with high-quality spectrophotometric
calibration. The total number of spectra in the {\em clean} IFS mosaic version
of NGC\,628 is 6949. The top-panel of \autoref{fig:integ_plot} shows the
integrated spectrum of NGC\,628 derived by co-adding the spectra within the
{\em clean} IFS mosaic version of the galaxy.

The integrated spectrum of NGC\,628 shows a characteristic stellar continuum
with absorption features and emission lines
superimposed. H$\alpha$, H$\beta$, \oii \lam3727, \oiii \lam5007,
the \nii \lam\lam6548,84 and \lam\lam6717,31 doublets are
clearly identified. Less obvious are the H$\gamma$ and \oiii \lam4959
lines. Sky residuals are also present in the spectrum, especially the \oi
\lam5577 and [Na~{\footnotesize I}] \lam5893 lines. 
The red-solid squares (in the online version) correspond to the integrated
flux derived from the $B$, $V$, $R$ and H$\alpha$ images of the SINGS
ancillary data, obtained during the spectrophotometric re-calibration, using
the same apertures of the IFU data and co-adding them in a similar way as the
integrated spectrum. The position of these data-points with respect to the
continuum (in the case of the broad-band images) and the peak of the H$\alpha$
line in the spectrum, corroborates the accuracy of the absolute flux calibration.
The co-added region comprises $\sim$90\% of the total flux of the galaxy in
$V$-band, as estimated from corresponding SINGS image.
An analysis similar to that described above was performed on this
integrated spectrum to derive the main physical properties of both the ionized
gas and the stellar population.

\subsubsection{Integrated stellar populations}
\label{sec:integ_stellar}

The middle-panel of \autoref{fig:integ_plot} shows the best SSP model fit to
the integrated spectrum superimposed in red colour. The light-blue bands
correspond to the spectral regions masked during the fitting as explained
above. They coincide with the position of the strongest redshifted emission
lines and regions of bright sky residuals. Note the strength of the underlying
stellar absorptions in the Balmer lines. The multicomponent SSP model matches
accurately the continuum of the integrated spectrum, within an error of the
$\sim$3\%. Most of the discrepancies are in regions clearly dominated by
imperfections in the sky-subtraction, due to the strength of the night sky
lines (e.g. at $\sim$\,$\lambda$5577, see \citet{Sanchez:2007p1276} and
\citetalias{RosalesOrtega:2010p3553}).  In order to evaluate the possible
effects of these residuals in the derivation of the main properties of the
stellar population the analysis was repeated for the wavelength range between
$\lambda$4100 and $\lambda$5400 \AA, where there are no such strong
atmospheric features. For this wavelength range, the model matches the
continuum within a $\sim$2\%. In the UV-regime the errors are sometimes larger
than this value.


\autoref{tab:ssp_fit} lists the {\it luminosity-weighted} age, metallicity and
dust attenuation of the best fitting model for the two cases considered here:
one using the entire spectral range in the fitting process and the other using
the reduced range described before. There is a very good agreement between the
results derived in both cases. Based on the results from the simulations (see
Appendix), we expect an error of 10-20\% in the derivation of the Age, and of
a $\sim$5\% in the derivation of the metallicity. These errors do not take
into account the systematics, due to the applied algorithm, the current code
and the templates adopted, or the known degenerancies in the derivation of the
age and the metallicity, and therefore the expected discrepancies with
previous published results are much larger than the ones estimated on the
basis of the simulations. From these fits we may conclude that the average
stellar population of NGC\,628 is dominated by an old component of $\sim$\,9
Gyrs, but with a subsolar metallicity ($[Z/H]\sim -0.45$). Although synthesis
modeling is nowadays widely used, we have to take into account the
age-metallicity degeneracy problem that plagues most spectral fitting
techniques. The most frequently used technique to derive the age and
metallicity of the stellar population in galaxies is to measure certain line
strength indices, such as the Lick/IDS index system
\citep[e.g.][]{Burstein:1984p3764,Faber:1985p3766,Burstein:1986p3765,Gorgas:1993p3767,Worthey:1994p3434}.
For this purpose, one generally tries to use a combination of indices that are
most orthogonal in the parameter space (i.e. age and metallicity). In order to
cross-check the results based on the fitting procedure, we measured the
equivalent width of H$\beta$ (primarily sensitive to the age) and Mg$b$
(primarily sensitive to the metallicity), of the integrated spectrum.

%
%
%
%
%
%
%
%
\begin{table}
\begin{center}
\caption[Properties of the average stellar population]
{
  Properties of the average stellar population.
  \label{tab:ssp_fit}
}

\begin{tabular}{ @{\extracolsep{\fill}} lcccc }\hline

& Age &  \multicolumn{2}{c}{Metallicity} & A$_{V,\star}$\\ 
\cline{3-4}\vspace{-7pt}\\
\multicolumn{1}{c}{Analysis method} & {\scriptsize (Gyr)} &  Z  &
\multicolumn{1}{c}{[Z/H]}    &  {\scriptsize (mag)} \\

\hline
SSP fit, 3700-6800 \AA       & 8.95  & 0.007 & $-$0.44 &  0.4 \\
SSP fit, 4100-5400 \AA       & 8.40  & 0.007 & $-$0.44 &  0.8 \\
H$\beta$ vs. Mg$b$ indices   & 9.78  & 0.008 & $-$0.42 &  $-$  \\

\hline
\end{tabular}\vspace{-5pt}
\end{center}
$^{\star}$ Dust continuum attenuation.
\end{table}

The equivalent widths of the absorption lines were derived using the bandpass
definitions from the Lick index system revised by \citet{Trager:1998p3768},
shifted to the redshift of the object, as described in
\citet{Sanchez:2007p3299}.  To derive the age and metallicity from the
measured indices, we have adopted the model grid from
\citet{Thomas:2003p3769}, implemented in the {\sc rmodel}
code\footnote{N. Cardiel, http://tinyurl.com/rmodel}. The resulting estimates
of the age and metallicity, based on the absorption line index analysis, are
listed in \autoref{tab:ssp_fit}. Despite the strong conceptual differences
between these methods and the fitting technique described before, the results
are very consistent with the values obtained in both the full and reduced
wavelength range SSP fitting.

\subsubsection{Integrated properties of the ionized gas}
\label{sec:integ_gas}

\begin{figure}
  \includegraphics[width=6cm,angle=270]{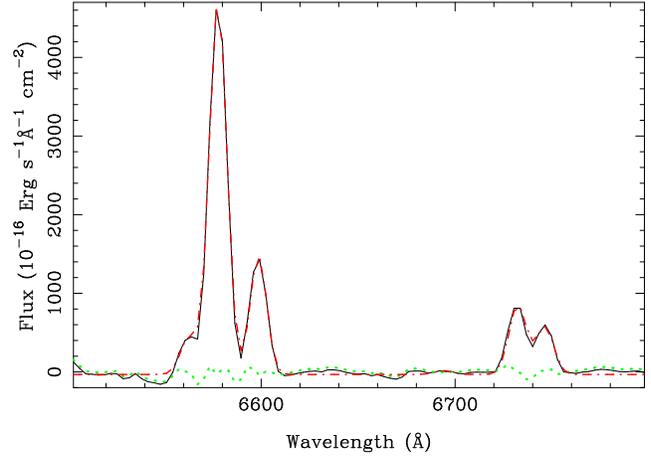}
  \caption[Example of the emission line fitting]
  { 
    Example of the fitting procedure applied to derive the intensity flux of the
    detected emission lines. The black solid-line corresponds to the pure emission line
    spectrum of NGC\,628, covering the wavelength
    range between 6500 and 6800 \AA. The red dashed-line shows the best fitting
    model describing the emission lines, which comprises a single Gaussian
    function for each of them. The green dotted-line shows the residual
    spectrum, once the previous model has been subtracted.
    \label{fig:eline_fit}
 }
\end{figure}

The bottom-panel of \autoref{fig:integ_plot}
shows the pure emission line spectrum of NGC\,628 obtained after
subtracting the underlying stellar population from the integrated spectrum. As
expected, the spectrum is dominated by a set of emission lines, plus a residual
continuum consistent with a zero average intensity. 
This procedure reveals additional emission lines, like H$\delta$ and H$\epsilon$.
All the detected lines have been labeled with their standard notation. Each of these
emission lines was fitted with a single Gaussian function, as described
before, in order to derive their strengths. \autoref{fig:eline_fit} shows an
example of the fits in the wavelength range between 6500\,\AA\ and
6800\,\AA. The black solid-line corresponds to the pure emission line spectrum
of NGC\,628, the red dashed-line shows the best fitted
model, consisting of six Gaussian functions, assuming a single Gaussian fit to
each of the emission lines detected in this wavelength range: the \nii
doublet, H$\alpha$, He~{\footnotesize I} $\lambda$6678 (very faint) and the
\sii doublet. In this particular case both the systemic velocity and velocity
dispersion were forced to be the same for all the emission lines, in order to
increase the accuracy of the derived parameters. The ratio between the two
\nii lines included in the spectral range were fixed to the theoretical value
\citep{Osterbrock:2006p2331}. A similar procedure was applied to the rest of
the emission lines. By adopting this procedure, it is possible to accurately
deblend the different emission lines.

\begin{table}
\begin{center}
\caption[NGC\,628 integrated line intensities]
{
  Integrated line intensities for NGC\,628. The first column corresponds to the
  emission line identification, with the rest-frame wavelength, the second
  one corresponds to the adopted reddening curve
  normalized to H$\beta$. The F(\lam)/H$\beta$ column
  corresponds to the observed flux, while the I(\lam)/H$\beta$ to the
  reddening corrected values; normalised to H$\beta$. The values in
  parenthesis correspond to the 1$\sigma$ errors calculated as explained
  in the text. The observed flux in H$\beta$ is in units of 10$^{-15}$
  \flux. The last row shows the number of fibres from which the integrated
  spectra was extracted.
}
\label{tab:n628_elines}

\begin{tabular}{@{\extracolsep{\fill}} lrrr }\hline

\multicolumn{1}{c}{Line} & \multicolumn{1}{c}{$f${\footnotesize ($\lambda$)}} &
 \multicolumn{1}{c}{F($\lambda$)/H$\beta$} &  \multicolumn{1}{c}{I($\lambda$)/H$\beta$} \\[2pt]\hline\\[-8pt]

                    [O{\footnotesize~II}]~~$\lambda$3727 &    0.32 &                1.494~~(0.046) &                2.138~~(0.320) \\[2pt]
  H{\footnotesize8}~+~He{\footnotesize~I}~~$\lambda$3889 &    0.29 &                0.158~~(0.032) &                0.217~~(0.054) \\[2pt]
                                H$\delta$~~$\lambda$4101 &    0.23 &                0.165~~(0.032) &                0.214~~(0.051) \\[2pt]
                                H$\gamma$~~$\lambda$4340 &    0.16 &                0.511~~(0.023) &                0.609~~(0.083) \\[2pt]
                                 H$\beta$~~$\lambda$4861 &    0.00 &                1.000~~(0.014) &                1.000~~(0.052) \\[2pt]
                   [O{\footnotesize~III}]~~$\lambda$4959 &   -0.03 &                0.111~~(0.010) &                0.108~~(0.017) \\[2pt]
                   [O{\footnotesize~III}]~~$\lambda$5007 &   -0.04 &                0.329~~(0.010) &                0.315~~(0.040) \\[2pt]
                      He{\footnotesize~I}~~$\lambda$5876 &   -0.20 &                0.299~~(0.012) &                0.239~~(0.031) \\[2pt]
                    [N{\footnotesize~II}]~~$\lambda$6548 &   -0.30 &                0.383~~(0.013) &                0.276~~(0.035) \\[2pt]
                                H$\alpha$~~$\lambda$6563 &   -0.30 &                3.998~~(0.041) &                2.870~~(0.352) \\[2pt]
                    [N{\footnotesize~II}]~~$\lambda$6584 &   -0.30 &                1.111~~(0.016) &                0.795~~(0.098) \\[2pt]
                    [S{\footnotesize~II}]~~$\lambda$6717 &   -0.32 &                0.619~~(0.011) &                0.435~~(0.054) \\[2pt]
                    [S{\footnotesize~II}]~~$\lambda$6731 &   -0.32 &                0.393~~(0.010) &                0.276~~(0.035) \\[4pt]
                    \hline\\[-8pt]
     [O{\footnotesize~III}]~~$\lambda$5007/$\lambda$4959 &         &                  2.96~~(0.28) &                  2.92~~(0.58) \\[2pt]
      [S{\footnotesize~II}]~~$\lambda$6717/$\lambda$6731 &         &                  1.57~~(0.14) &                  1.58~~(0.28) \\[5pt]

                              F(H$\beta$)~~$\lambda$4861 &         &                        1549.4 \\[2pt]     
                                             c(H$\beta$) &         &                  0.48~~(0.05) \\[2pt]
                                                   A$_V$ &         &                          1.04 \\[2pt]

Extraction fibres & & 6949\\[4pt]

\hline
\end{tabular}\vspace{-10pt}
\end{center}
\end{table}

\autoref{tab:n628_elines} lists the emission line intensities derived using
this procedure, including for each of the detected emission lines its standard
identification, the laboratory rest-frame wavelength, and the estimated
intensity normalised to the observed flux of H$\beta$ in units of
10$^{-15}$ \flux. They are shown in the columns labeled as F(\lam)/H$\beta$.
The observed \oiii ratio shown in \autoref{tab:n628_elines} is consistent with
the well-known theoretical value between these lines
\citep{Storey:2000p3365}. The associated 1$\sigma$ errors are solely due to
the statistical uncertainty $\sigma_{\rm stat}$ in the measurement of the flux
intensity. Thereafter, the observed line intensities were corrected for
reddening using the Balmer decrement according to the reddening function of
\citet{Cardelli:1989p136}, assuming $R \equiv A_V/E(B-V) = 3.1$.  Theoretical
values for the intrinsic Balmer line ratios were taken from
\citet{Osterbrock:2006p2331}, assuming case B recombination (optically thick
in all the Lyman lines), an electron density of $n_e = 100$ cm$^{-3}$ and an
electron temperature $T_e = 10^4$ K.  The last column of
\autoref{tab:n628_elines} shows the reddening-corrected
emission line fluxes for the integrated spectrum, designated by 
I(\lam)/H$\beta$. These flux ratios can be used to derive the average
properties of the ionized gas in the galaxy. The adopted reddening curve
normalized to H$\beta$, $f${\footnotesize (\lam)}, is shown in the second
column of the table. Although higher-order Balmer lines were detected in the
integrated spectrum, no Balmer lines beyond H$\gamma$ were used for the
determination of c(H$\beta$), as the associated error of the measurement of
these lines in the residual spectrum yielded high uncertainties in the
computed c(H$\beta$), due to their low signal-to-noise; therefore the
c(H$\beta$) value was derived using the H$\alpha$/H$\beta$ ratio solely. No
auroral lines were detected either in the integrated or residual spectrum. The
low strength of the \oiii lines suggests a relatively high metallicity for the
integrated abundance of NGC\,628, as it will be shown hereafter.

Although particular care has been taken in the flux
calibration of the spectra within the mosaic, the absolute flux intensity
listed in \autoref{tab:n628_elines} has to be treated with caution. On one
hand, not all of the galaxy surface has been covered by our IFS observations,
as can be seen in \autoref{fig:mosaic}. In
particular some of the brighter \hh regions, located to the east of NGC\,628
are not included in the FOV of our mosaic. This causes to underestimate the
intensity of all the lines. By a rough estimation, based on the $D_{25}$
optical radius ($B$-band), we consider that our IFS Mosaic covers
$\sim$\,70\% of the galaxy size. On the other hand, the central fibre-bundle
of PPAK has a filling factor of $\sim$\,65\%, as mentioned in \autoref{sec:obs},
which leads to a corresponding underestimation of the integrated flux. In the
particular case of this mosaic, a dithering scheme necessary to compensate for
this incomplete sampling was adopted only for the central pointing,
where the emission lines are in general weak. All together, we consider that
it is necessary to apply a correction of a factor $\sim$\,2.2 to take into
account the aperture and sampling effects described before. 
As mentioned before, NGC\,628 has been extensively studied before in several
publications. In particular, different authors have reported on the H$\alpha$
intensity flux, using different procedures, from photoelectric photometers to
narrow-band imaging. \autoref{tab:Ha_flux} lists a summary of these published
values, together with the value derived from our integrated spectrum and
considering the flux correction factor described above. Despite the different
biases introduced by the different methods, there is substantial agreement
between the previously published results and our reported value for the
integrated H$\alpha$ flux of the galaxy.

\begin{table}
  %
  %
  \begin{center}
    \caption[H$\alpha$ flux of NGC\,628]
    {
      Comparison between different H$\alpha$ fluxes reported for NGC\,628 in
      the literature. Fluxes in units of \nFUNITS.
      \label{tab:Ha_flux}
    }
    \begin{tabular}{rl} 
      \hline\hline        
      \multicolumn{1}{c}{Flux} & \multicolumn{1}{c}{Reference} \\
      \hline   
       1.07    & \citealt{Kennicutt:1983p3415}  \\
       0.87    & \citealt{Young:1996p3416}      \\
       1.51    & \citealt{Hoopes:2001p3417}     \\
       1.05    & \citealt{Marcum:2001p3418}     \\
       1.02    & \citealt{Kennicutt:2008p3419}  \\
       \hline \\[-4pt]
       1.14    & Current study\\
      \hline 
    \end{tabular}
  \end{center}
\end{table}


The derived dust extinction, A$_V$ = 1.04, larger than the one derived from
the analysis of the stellar populations (A$_V$ $\sim$ 0.4 mag). This result is
not surprising, since both methods sample different regions of the
galaxy. While the underlying continuum is dominated by the stellar components
of the central regions, clearly brighter, the ionized gas spectrum is dominated by the
star-forming regions in the spiral arms. These latter regions are known to be
more attenuated by dust, due to star forming process \citep[e.g.][]{Calzetti:2001p3421}.
Indeed, the extinction law derived by \citet{Calzetti:1997p3420} for star-forming
galaxies shows that the typical extinction in the emission lines of these
objects is approximately double that in their stellar continuum.

The integrated flux of H$\alpha$ and \oii \lam3727 can be used to
determine a rough value of the global star formation rate (SFR) in this
galaxy. The intensities of both lines were corrected by dust extinction,
adopting the $A_V$ value and the aperture correction mentioned
before. Absolute luminosities were derived by assuming a standard $\Lambda$CDM
cosmology with H$_0$=70.4, $\Omega_m$=0.268 and $\Omega_\Lambda$=0.73, and a
luminosity distance of 9.3 Mpc \citep{Hendry:2005p2408}.
The derived luminosities are L$_{\rm H\alpha}\sim$\,3.08 and L$_{\rm
  \oii}\sim$\,2.30, in units of 10$^{41}$ erg s$^{-1}$.
The values of the SFR were derived adopting the classical relations by
\citet{Kennicutt:1998p3370}, obtaining SFR $\sim$\,2.4 and 3.2 M$_\odot$
yr$^{-1}$, based on the H$\alpha$ and \oii luminosities respectively.

Different possible mechanisms can be responsible for the ionization in
emission line galaxies. The nature of the ionization can be derived from
ratios of the usual diagnostic lines \citep[BPT, ][]{Baldwin:1981p3310,
  Veilleux:1987p3423}. Based on the values listed in
\autoref{tab:n628_elines}, it was found that log$_{10}$(\nii
\lam6584/H$\alpha$) $\sim -0.56$ and log$_{10}$(\oiii \lam5007/H$\beta$) $\sim
-0.5$. These line ratios correspond to the expected values for star forming
galaxies and/or \hh regions
\citep[e.g.][]{Sanchez:2005p3138,Sanchez:2007p3299}, far away from any of the
boundary regions in the \oiii/H$\beta$ vs. \nii/H$\alpha$ diagnostic
diagram. Therefore, it is clear that the dominant ionization mechanism in the
integrated spectrum of NGC\,628 is due to hot OB stars (\hh regions), as
expected, since no previous study has reported any kind of nuclear activity in
this galaxy.

The (volume-averaged) ionization parameter, defined as the ratio of the
density of ionizing photons to the particle density: $ u =
\frac{Q_{H^0}}{{4\pi R_s^2nc}}$, where $Q_{H^0}$ is the flux of ionizing
photons produced by the exciting stars above the Lyman limit, $R_s$ is the
radius of the Strömgren sphere, $n$ the number density of hydrogen atoms, and
$c$ the speed of light.  This parameter determines the degree of ionization at
any particular location within the nebula. As discussed in the next section, many
of the empirical methods commonly used to derive the chemical abundance of a
star-forming region are sensitive to this parameter, and for some ranges of
metallicity, they are not useful unless the ionization parameter can be
constrained within a small range of possible values.

The ionization parameter is best determined using the ratios of emission lines
of different ionization stages of the same element. In general, the larger the
difference in ionization potentials of the two stages, the better the
constrain. A commonly used ionization parameter diagnostic is based on the
ratio [O~{\footnotesize II}]/[O~{\footnotesize III}] = \lam3727 / (\lam4959 +
\lam5007). However, this ratio is not only sensitive to the ionization
parameter, but is also strongly dependent on metallicity. 
Another possibility is to use the \siii \lam9069 and/or \siii \lam9532
together with the \sii  \lam6717, \lam6731 emission lines. The
[S~{\footnotesize II}]/[S~{\footnotesize III}] ratio provides a more reliable
useful ionization parameter diagnostic \citep{Diaz:1991p3390}.
Considering the available lines within the spectral range of the PINGS
observations \citep{Diaz:1994p3426}, a good approximation to the ionization
parameter can be determined from the [O~{\footnotesize II}]/H$\beta$ or
[S~{\footnotesize II}]/H$\alpha$ ratio , via:

\begin{equation}
  {\rm log}~u = -0.80~\log \left( {\left[ {\rm O~{\footnotesize II}} \right]/\left[  {\rm O~{\footnotesize III}} \right]} \right) - 3.02,
\label{eq:log_u}
\end{equation}

\noindent after \citet{Diaz:2000p3371}. Its uncertainty is estimated to be
$\pm$0.2 dex. The derived $\log u$ for integrated spectrum of NGC\,628 is
--3.58, which corresponds to a low value of the ionization parameter ($\sim$
10$^{-4}$).

\subsubsection{Integrated oxygen abundance}
\label{sec:integ_abun}


\begin{table*}
\label{tab:oxy}
\begin{minipage}{0.9\textwidth}
\centering
\caption[]
{Integrated oxygen abundances for NGC\,628 in units of 12+log(O/H). The
  columns designations correspond to the following abundance calibrators: M91,
  \citet{McGaugh:1991p314}; KK04, \citet{Kobulnicky:2004p1700}; N2,
      \citet{Denicolo:2002p361}; O3N2, \citet{Pettini:2004p315}; PT05,
      \citet{Pilyugin:2005p308}; (O/H)$_{\rm ff}$, ff--$T_e$ method (as explained in
      the text).
}

\begin{tabular}{@{\extracolsep{\fill}} ccccccc }\hline

log~R$_{23}$ &  M91 &  KK04 & N2 & O3N2 & PT05 & (O/H)$_{\rm ff}$ \\[2pt]\hline\\[-8pt]

0.41 &  8.90~$\pm$~0.14  &  9.06~$\pm$~0.11  &  8.62~$\pm$~0.16  &  8.71~$\pm$~0.11  & 8.32~$\pm$~0.15 & 8.55~$\pm$~0.09 \\[2pt]

\hline

\end{tabular}\vspace{-10pt}
\end{minipage}
\end{table*}


Galaxies in the local universe have been used as an anchor point to determine
the evolution of the metallicity (based on the gas phase oxygen abundance),
along different cosmological periods. 
The global metallicity of a given galaxy is represented by its oxygen
abundance. The rest of the elements vary within the same
fractions as those found in the Sun. However, while the determination of this
observable at high redshift normally describes the average value of the
galaxies as a whole (due to aperture effects), in the local universe most of
these determinations are based on studies of a number of discrete \hh
regions. The integrated spectrum of NGC\,628 can be used to perform an
integrated abundance analysis in a consistent way as the studies performed
over high redshift galaxies, with the advantage than the results of the
integrated study can be compared with the abundances of resolved regions
within the galaxy. In this section, we present a chemical abundance analysis
of the integrated spectrum of NGC\,628 based on a suite of different empirical
abundance diagnostic methods. A more detailed comparison between the
integrated and spatially resolved abundances will be presented in a subsequent
paper (Rosales-Ortega et al., in preparation).

Given the limitations imposed by the non-detection of auroral temperature
sensitive emission lines (such as \oiii \lam4363 or \nii \lam5755) 
in \hh regions of low-excitation and/or low surface brightness, empirical
methods based on the use of strong, easily observable optical lines have been
developed throughout the years. Although abundances derived in this way are
recognised to suffer considerable uncertainties, they are believed to be able
to trace large-scale trends in galaxies.
Following these ideas, several abundance calibrators have been proposed
involving different emission-line ratios and have been applied to determine
oxygen abundances in objects as different as individual \hh regions in spiral
galaxies, dwarf irregular galaxies, nuclear starburst, and the integrated
abundances of spiral galaxies \citep[e.g.][ and references
therein]{Kobulnicky:1999p1710,Pilyugin:2004p82}.

In order to derive the characteristic chemical abundance of NGC\,628, a set of
empirical calibrators were applied to the integrated spectrum. Different
abundance estimators were chosen in order explore the effect of a particular
calibration depending on the physical properties of the galaxy.
By far, the most commonly used such ratio is (\oii \lam3727 + \oiii \lam4959,
\lam5007)/H$\beta$, known as the $R_{23}$ method \citep{Pagel:1979p71}.
The logic for the use of this ratio is that it is not affected by differences
in relative elemental abundances, and remains essentially constant within a
given giant \hh region despite variations in excitation
\citep{Diaz:1987p1845}. However, there is an ambiguity inherent to this method
since there are two values of abundance corresponding to a given value of
$R_{23}$, i.e. the lower branch increases with increasing abundance, while the
upper branch shows an opposite behavior, i.e. $R_{23}$ decreases with
increasing abundance. Another drawback of using $R_{23}$ (and many of the
other emission-line abundance diagnostics) is that it depends also on the
ionization parameter.  From this category, two methods were applied: the
\citealt{McGaugh:1991p314} (hereafter M91) and the \citealt{Kobulnicky:2004p1700}
(hereafter KK04) calibrations, which are theoretical methods based on
photoionization models. Both take into account the ionization parameter to
produce an estimate of the metallicity. The estimated accuracy of these
methods is $\sim\pm$0.15 dex.

\begin{figure}
  \centering
  %
  %
  \includegraphics[width=\hsize]{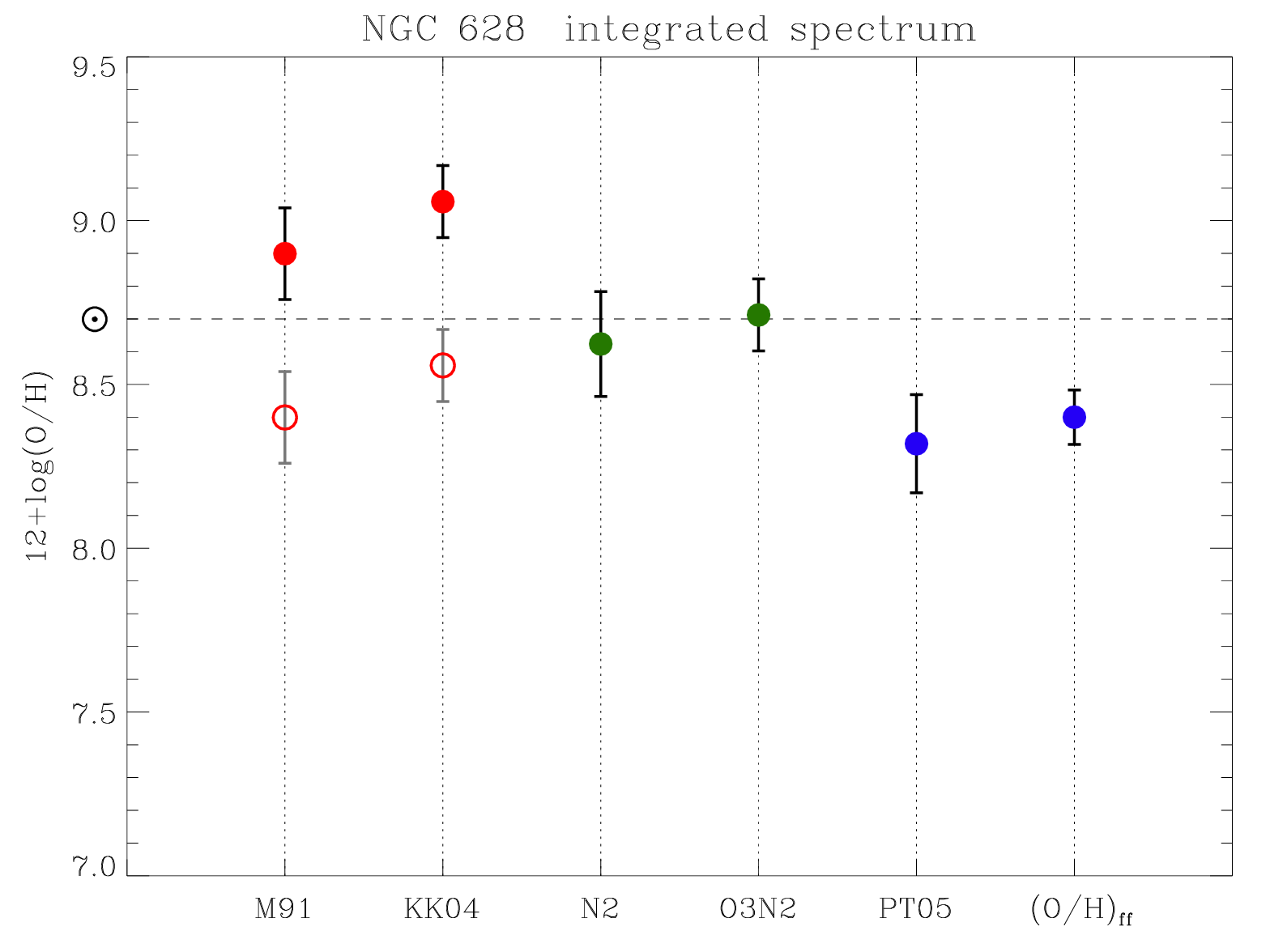}
  \caption[Azimuthally averaged radial profile of the oxygen abundance]
  { 
    Comparison of the integrated oxygen abundance of NGC\,628 derived for
    different estimators. Red-solid points correspond to $R_{23}$ calibrators,
    green-points to index-empirical methods, and blue-points to the different
    methods proposed by Pilyugin and collaborators. The open circles
    correspond to an arbitrary --0.5 dex offset of the $R_{23}$ based
    methods. The horizontal dashed-line correspond to the oxygen solar value.
    \label{fig:oxy}
  }
\end{figure}

In order to discriminate between the different $R_{23}$ branches, we used the
\nii~\lam6584/\oii~\lam3727 ratio following the
prescriptions by \citet{Kewley:2002p311}. The [N\,{\footnotesize
  II}]/[O\,{\footnotesize II}] ratio is not sensitive to the ionization
parameter to within $\pm$0.05 dex, and it is a strong function
of metallicity above log([N\,{\footnotesize II}]/[O\,{\footnotesize II}])
$\gtrsim$ --1.2, where the division between the upper and lower branches
occurs. The log([N\,{\footnotesize II}]/[O\,{\footnotesize II}]) value for the
integrated spectrum of NGC\,628 is $-0.43$, i.e. indicating that 
the $R_{23}$ value of the integrated spectrum for NGC\,628 ($\log R_{23} =
0.41$) corresponds to the upper branch of the $R_{23}$ relation.

Another subset of estimators was chosen from the category of empirical
strong-line methods, they correspond to the N2 calibration (first proposed by
\citealt{StorchiBergmann:1994p2885}, but using the definition after
\citealt{Denicolo:2002p361}), and the O3N2 calibration (first proposed by
\citet{Alloin:1979p2878}, but using the definition by
\citealt{Pettini:2004p315}). These two indices have the virtue of being
single-valued, however, they are affected by the low-excitation line \nii
\lam6584, which may arise not only in bona-fide \hh regions, but also in the
diffuse ionized medium, which is an issue for spectra integrated within
extended regions, such as the integrated spectrum of NGC\,628. The estimated
uncertainty of the derived metallicities is $\sim\pm$~0.2 dex.

The \citealt{Pilyugin:2005p308} (hereafter PT05) calibration is based on an
updated version of the \citet{Pilyugin:2001p1390} estimator, obtained by
empirical fits to the relationship between $R_{23}$ and $T_e$ metallicities
for a sample of \hh regions. This estimator was also considered to determine
the integrated abundance of NGC\,628.
This calibration includes and excitation parameter $P$ that takes
into account the effect of the ionization parameter. The PT05 calibration has
two parametrizations corresponding to the lower and upper branches of
$R_{23}$.  As in the case of the M91 and KK04 calibrators, the
[N\,{\footnotesize II}]/[O\,{\footnotesize II}] ratio was
used to discriminate between the two branches of the $R_{23}$ relation.

The last strong-line empirical method considered is a combination of the
flux-flux (or ff-relation) found by \citet{Pilyugin:2005p3401} and
parametrised by \citet{Pilyugin:2006p3376}, the $t_2-t_3$ relation between the
O$^+$ and O$^{++}$ zones electron temperatures for high-metallicity regions
proposed by \citet{Pilyugin:2007p3381}, and an updated version of the
$T_e$-based method for metallicity determination \citep{Izotov:2006p3429}.
According to these authors, the combination of these methods solves the
problem of the determination of the electron temperatures in high-metallicity \hh
regions, where faint auroral lines are not detected.
However, the abundances determined through this method rely on the validity of
the classic $T_e$ method, which has been questioned for the high-metallicity
regime in a number of studies by comparisons with \hh region photoionization
models \citep{Stasinska:2005p3430}. The abundances derived through this
method will be referred as (O/H)$_{\ff}$ or ff--$T_e$ abundances.

The advantages and drawbacks of all these different calibrators have been
thoroughly discussed in the literature (e.g. \citealt{Pagel:1980p1378},
\citealt{Kennicutt:1996p1603}, \citealt{Kewley:2002p311},
\citealt{PerezMontero:2005p306}, \citealt{Kewley:2008p1394}).
Regrettably, comparisons among the metallicities
estimated using these methods reveal large discrepancies. They are usually
manifested as systematic offsets in metallicity estimates, with high values
corresponding to theoretical calibrations and lower metallicities estimated by
electron temperature metallicities, with offsets as large as 0.5 dex in
log(O/H) units \citep{Liang:2006p2962}.

The calculated oxygen abundances (in units of 12~+~log(O/H)) derived for each of
the calibrators listed above are shown in \autoref{tab:oxy}.
The uncertainties correspond to the 1$\sigma$ error
found by propagating the errors through a Monte Carlo simulation by using
Gaussian distributions with a width equal to the errors of the emission line
intensities, modulated by recomputing the distribution 500 times.
These abundances are plotted in \autoref{fig:oxy}. The red-solid points (in
the online version) correspond to the $R_{23}$ methods, the green points to
the index-empirical methods, and the blue points to the different methods
proposed by Pilyugin and collaborators. The red-open circles correspond to an
arbitrary --0.5 dex offset of the $R_{23}$ methods (these offsets are included
given the well-known systematic offset of the $R_{23}$ theoretical-based
calibrations). The horizontal dotted-line corresponds to the oxygen solar
abundance 12~+~log(O/H)=8.7 \citep{Scott:2009p3377}.

The two $R_{23}$ methods derive super-solar metallicity values, while the PT05
oxygen value correspond to the lowest metallicity obtained. The index-empirical
methods, N2 and O3N2, stand in between the metallicities derived through the
$R_{23}$ and the PT05 + ff-$T_e$ methods. Interestingly, the $R_{23}$
calibrations metallicities which were shifted to a lower value (red-open
circles), are in close agreement (within the errors) to the pure-empirical and
the ff--$T_e$ based metallicities. The mean oxygen abundance derived from all
the methods is 12~+~log(O/H)=$8.69\pm0.31$.

\begin{figure*}
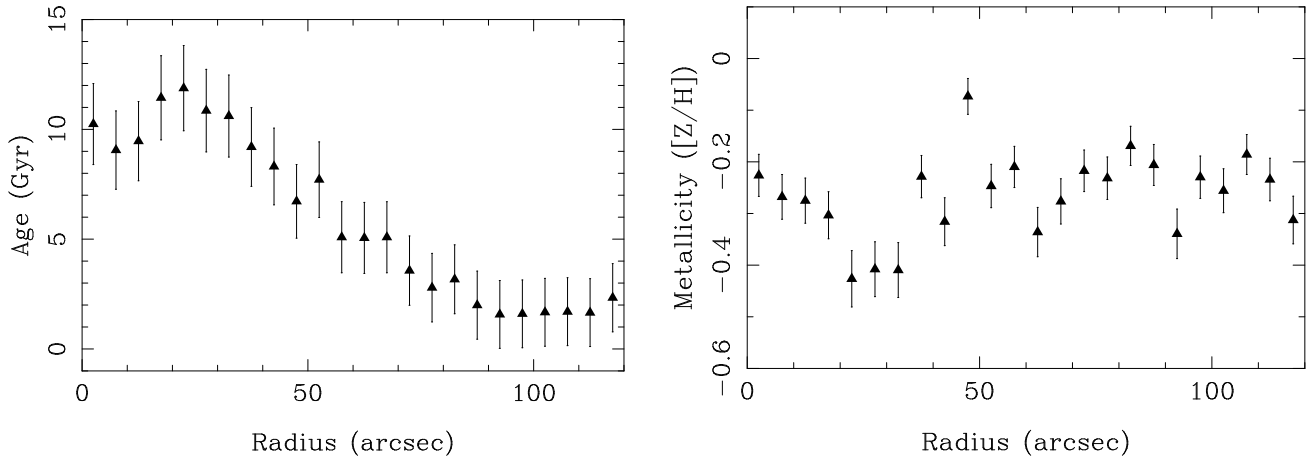

  %
  %
  \centering
  \includegraphics[width=6cm,angle=270]{Radial_Age.ps}\hspace{0.5cm}
  \includegraphics[width=6cm,angle=270]{Radial_ZH.ps}
  \caption[Radial distribution of the luminosity-weight age of the stellar
  populations]
  {
    {\it Left panel:} radial distribution of the luminosity-weight
    age (in Gyr), of the stellar population based on the fitting procedure
    using the combination of multiple SSPs. {\it Right panel:} similar distribution for the
    luminosity-weight metallicity, relative to the solar one (Z/H).
    \label{fig:rad_age_SSP}
 }
\end{figure*}

Previous studies have determined the oxygen abundance in this galaxy based on
spectroscopic imaging spectrophotometry observations of different \hh regions.
In particular, \citet{McCall:1985p1243} analysed a sample of \hh regions within a
radius of $\sim$\,200 arcsec, basically coincident with the area sampled by
our IFS survey. They reported a range of oxygen abundances between
12~+~log(O/H) $\sim 8.7-9.3$ (by employing a method based on the $R_{23}$ index),
with a considerable decline from the inner to the outer parts. 
\citet{Belley:1992p3779} obtained reddenings, H$\beta$ equivalent widths,
diagnostic line ratios and metallicities for 130 \hh regions by the
implementation of imaging spectrophotometry. They derived an abundance gradient
of NGC\,628 based on the [O\,{\footnotesize III}]/H$\beta$ empirical calibrator
\citep{Edmunds:1984p223}. Their values range between 12~+~log(O/H) $\sim
8.4-9.2$, covering a large baseline in galactocentric distances (up to $\rho \sim
2\rho_{\rm eff}$). Although this was not a strict spectroscopic study, given
the number of \hh regions analysed, the work of \citealt{Belley:1992p3779}
stood up to now as the most complete 2D description of the emission line
chemistry of NGC\,628. Subsequently,
\citet{vanZee:1998p3468} reported the oxygen abundances of the \hh regions in
an outer ring between $\sim150$ arcsec and $\sim300$ arcsec radius. They found
that the decline in the abundance continues, and the oxygen abundance ranges
between 12~+~log(O/H) $\sim$8.10 and 8.95 (using a modified version of the M91
calibrator). More recently, \citet{Castellanos:2002p3372} observed a reduced
set of \hh regions in the optical and near-infrared where they were able to measure
temperature-sensitive emission lines. They reported an average oxygen
abundance of 12~+~log(O/H) $\sim8.23$. However these \hh regions are located
beyond the FOV of our IFS mosaic, at galactocentric distances where, given the
well-known radial metallicity gradient of this galaxy
\citep{Ferguson:1998p224}, the oxygen abundance is expected to be lower than
the integrated metallicity derived at inner radii.
Therefore we cannot compare the integrated O/H abundance derived in this work with the
values of \citet{Castellanos:2002p3372} due to the non-coincident geometry. On
the other hand, the range of oxygen abundances reported by the previous
spectroscopic and imaging spectrophotometry studies agree perfectly with the
integrated abundances derived through the $R_{23}$ methods and with the mean
integrated abundance of NGC\,628.

The integrated properties of the ionized gas derived in this chapter need
to be compared with the resolved properties in order to analyse the validity
of the results obtained from the integrated analysis, taking into account
different effects, such as the extinction or the contribution of the diffuse
interstellar emission. These points will be addressed in the following
sections.

\subsection{Spatially resolved properties of the galaxy}
\label{sec:resolved}

In this section we present the results obtained by applying the spectra
fitting technique described in \autoref{sec:analysis} to each individual
spectrum of the dataset. As indicated in the
\hyperref[app]{Appendix~\ref{app}}, the technique allows the decoupling of the
SED of the underlying stellar population from that of the ionized gas. This
step is needed to derive the intensities of the different emission lines
detected in each spectrum accurately. In addition, when the signal-to-noise
ratio is high enough this fitting technique can be used to derive the physical
parameters that characterise the composite stellar-population: the {\it
  luminosity-weighted} age, metallicity and dust attenuation.

\subsubsection{Distribution of the stellar populations}

%
%

As shown in \hyperref[app]{Appendix~\ref{app}}, the results from simulations indicate
that an individual spectrum does not have, in general, the required
signal-to-noise ratio to derive accurate values of the physical parameters
that characterise the composite stellar population. However, the average
spectra in concentric annuli do possess the required signal-to-noise, up to
$\sim120$ arcsec ($\sim0.4\rho_{25}$ or $\sim5.4$~Kpc) in linear-projected
galactocentric radii.

In order to study the distribution of the properties of the stellar population
we obtained azimuthally averaged spectra by co-adding all the spectra in the
mosaic within successive rings of 5 arcsec, starting from the central emission
peak of the galaxy. These azimuthally averaged spectra were then analysed
using the fitting procedure described above. \autoref{fig:rad_age_SSP} shows
the radial distributions of the luminosity-weighted ages (left-panel) and
metallicities (right-panel), derived by this method. The adopted error bars
correspond to the typical errors estimated from the simulations
(\hyperref[app]{Appendix~\ref{app}}) for a signal-to-noise ratio of $\sim100$.

\begin{figure*}
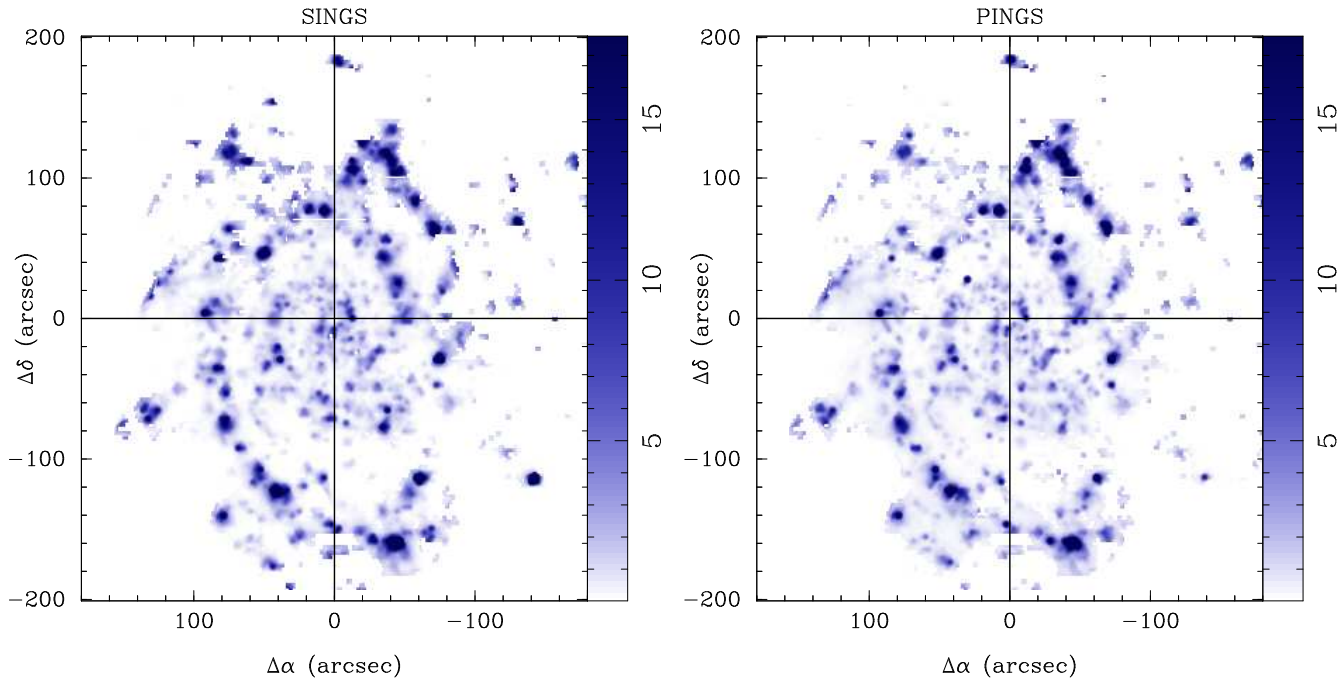

  %
  %
  \centering
  \includegraphics[angle=270,width=0.49\hsize]{Ha_SINGS_paper}\hspace{0.2cm}
  \includegraphics[angle=270,width=0.49\hsize]{Ha_PPAK_paper}
  \caption[H$\alpha$ line intensity maps]
  { 
    Left-panel: 
    Reconstructed, continuum subtracted H$\alpha$ narrow-band image of
    NGC\,628 obtained after interpolating the aperture photometry extraction
    of the SINGS narrow-band  image at the location of the mosaic fibres as in
    \autoref{fig:V_compar}. 
    Right-panel: 
    Pure H$\alpha$ emission lines map of NGC\,628 calculated from the IFS
    mosaic in units of 10$^{-16}$ \flux~arcsec$^{-2}$. No correction for dust
    extinction was applied to the map.
    All morphological structures of the galaxy are
    completely reproduced in the IFS image.
    \label{fig:Ha_maps}
 }
\end{figure*}

It is important to stress here that the accuracy of the absolute scale of the
ages is not as good as the relative one, since the former is much more model
dependent than the second. The age distribution shows, in general, a gradient,
with older stellar populations at the central regions of the galaxy and
younger ones at the outer regions. However, the oldest stellar populations are
not found in the centre of the galaxy, but in a circumnuclear ring at $\sim25$
arcsec radius, with an inversion of the gradient at smaller radii. This region
corresponds to the central kpc of the galaxy. At larger radii, the stellar
population becomes steadily younger, at a rate of $\sim0.20$~Gyr/kpc, up to a
radius of $\sim100$ arcsec ($\sim4.5$~kpc), and then becomes stable with an
age of $\sim1$~Gyr, on average. On the other hand, the radial distribution of
the metallicity of the stellar populations does not show such a clear trend,
remaining nearly constant at any radii with a larger dispersion in the
values. Both distributions show a similar shape than the one presented by 
\citep{MacArthur:2009p3412}, using a similar technique over classical
slit-spectroscopy data, altough their data show a larger dispersion.

The luminosity-weighted age and metallicity, derived from the plotted
values are $\sim8.2$~Gyrs and $Z\sim0.01$ ([Z/H]$\sim-0.27$), consistent with
the values derived from the integrated spectrum \autoref{tab:ssp_fit}. 
%
%

The gradient found in the stellar population of NGC\,628 (with the older
stellar populations in the center), is common in late-type
galaxies. \citet{Gadotti:2001p3772} found that $\sim60$\% of this kind of
galaxies show a negative color gradient in the $U-B$ and $B-V$ colors that it
is most likely due to a gradient in the age of the stellar populations, with
the younger populations in the outer regions. Differences between the colors
in the central bulges (redder) and the outer disks (bluer) are also reported
in other studies \citep[e.g.][]{Peletier:1996p3773}. NGC\,628 was included in
the sample studied by \citet{Gadotti:2001p3772}, and they reported a negative
color gradient in this galaxy up to a radius of $\sim$5 arcmin, qualitatively
in agreement with the trend found here based on the analysis of our IFS
dataset.

As mentioned before, \citet{Ganda:2007p3763} presented a study based on the IFS
observation of the central core of NGC\,628, using SAURON. The total field of
view of their observation was 33''$\times$41'', with a spatial sampling of
$\sim1$ arcsec. This FOV fits inside one of our single pointings, being
$\sim30$ times smaller in area than the FOV of the full dataset presented in
this article. Due to the dithering technique applied at the central pointings,
our final spatial resolution is just 2 times worse than their study. On the
other hand, their spectral resolution is 2 times better than ours, with the
penalty that their spectral coverage ($4800-5300$ \AA) is $\sim7$ times
shorter than ours. Despite these differences, their results can be compared
with the ones presented here for the inner regions.

Fig. 6 of \cite{Ganda:2007p3763} shows, among other things, the spatial
distribution of the age and metallicity of NGC\,628 derived from their
analysis. Despite the very small FOV of their observations and the more
reduced wavelength range sampled by their data, the basic structure described
above can be seen. In particular, the age distribution appears flat in the
inner 15 arcsec radius, with a ring of older stellar populations at $\sim20$
arcsec. On the other hand, the structure in the metallicity shown by
\cite{Ganda:2007p3763} is more clumpy. The azimuthally averaged radial
profile of both quantities, as presented in Fig A1a of that article, shows a
clearer picture, with a similar trend in both parameters as shown in
\autoref{fig:rad_age_SSP} for the same radii (although with an offset of the
absolute values). The decline in the age of the stellar population inferred
from our data at larger radii is out of the FOV of the SAURON observation.

Despite this qualitative agreement between both results, there are significant
differences in the quantitative parameters. The values of age and metallicity
derived by \cite{Ganda:2007p3763} for the central regions are younger
($\sim$2 Gyrs), and richer ($Z\sim$0.0-0.2), than the ones found by our
analysis. These differences are due to the different technique used in both
analysis. The analysis performed by \cite{Ganda:2007p3763} that we are
comparing with, is based on a single stellar population fitting
technique. Indeed, when we perform a similar analysis, adopting our fitting
techinique to compare with SSP, instead of allowing to mix multiple ones, we
derive an age of $\sim$1.4 Gyrs, in a better agreement with the reported value.

As noted by the authors, the SSP analysis does not reflect perfectly well the
complex starformation history of late-type galaxies. In particular, when a
galaxy has undergone two separate bursts of star formation, the age derived by
this analysis will be biased towards the one of the youngest stars, while the
metallicity will be biased towards the on of the older population, as studied
and described in detail by \cite{serr07}. Due to this effect, the authors
refined their analysis of the stellar populations, constraining the
metallicity by adopting the empirical relation between this value and the
velocity dispersion, derived by \cite{thom05}. By adopting this constrain,
they derive and age of 7.943 Gyrs and a metallicity of [Z/H]=-0.498 for the
stellar population of the NGC628 (Table 5 of \citeauthor{Ganda:2007p3763}
2007), in a complete agreement with our results, listed in Table 2.

\subsubsection{Distribution of the emission lines}

The ionized gas in any spiral galaxy exhibits a complex structure,
morphologically associated with the star-forming regions located along the
spiral arms. To study their properties, these regions have perviously been
targeted by narrow-band \citep[e.g.][]{Kennicutt:1980p3756} and Fabry-Perot
imaging (e.g., Fathi07). In many cases these narrow-band
images catch more than one single line (e.g. H$\alpha$, the \nii
\lam\lam6548,6584 doublet or the \sii \lam\lam6716,6731 doublet), reducing
their potential for the study the basic parameters of the ionized gas, or
requiring certain assumptions about the actual line ratios included within the
filter width \citep{Kennicutt:2008p3419}. IFS allows a much more refined
decoupling of the different lines, and the production of maps of each
individual emission line.

As in the case of the integrated spectra, in order to extract any physical
information from the data set, we need first to identify the detected emission
lines of the ionized gas and to decouple their emission from the stellar
continuum. We use for this purpose the {\em  clean} IFS mosaic version of
NGC\,628 (introduced previously in \autoref{sec:integrated}), which is free of
bad fibres, foreground objects, and includes only those regions with
high-quality spectrophotometric calibration, and fibres with an average flux
along the whole spectral range greater than 10$^{-16}$ \fluxA.

\begin{figure}
  %
  %
  \centering
  \includegraphics[angle=270,width=\hsize]{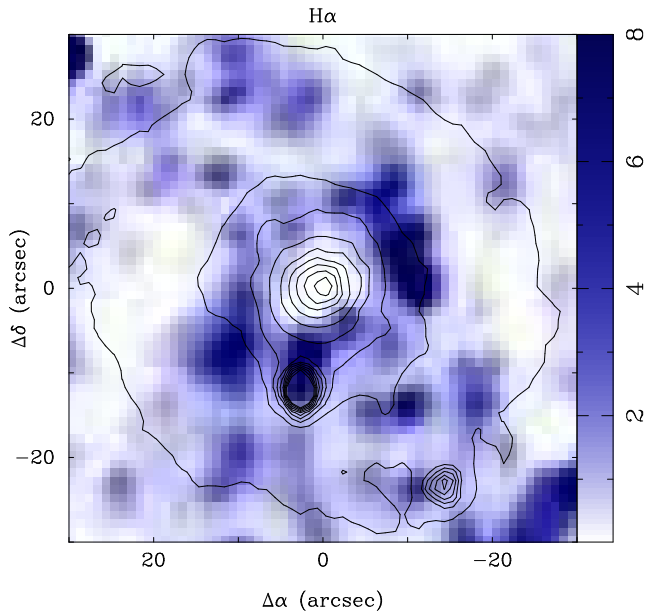}
  \caption[H$\alpha$ line intensity maps]
  { 
    Zoom of the central $\pm$30$\arcsec$ region of the
    pure H$\alpha$ emission lines map of NGC\,628 calculated from the IFS
    shown in previous figure. The countour plots shows the continuum emission
    in the same region. The ionized ring described by \cite{Wakker:1995p3788},
    and confirmed by \cite{Ganda:2006p3135} and Fathi07, is clearly seen.
    \label{fig:Ha_maps_cen}
 }
\end{figure}

\begin{table}
\begin{center}
\caption[Number of fibres in different mosaic versions]
{
  Number of fibres in different mosaic versions, the number in the
  {\em observed} mosaic refers to the total number of spectra considering
  all the pointings of NGC\,628. The {\em clean} mosaic number corresponds to
  the remaining fibres after the flux threshold cut as explained in the
  text. The percentage stands for number of {\em clean} fibres with respect
to the original mosaic.
}
\label{tab:num_fibres}

\begin{tabular}{@{\extracolsep{\fill}} ccc }\hline

 Observed mosaic & Clean mosaic & Percentage \\[2pt]\hline\\[-8pt]

11094 & 6949 & 63\%  \\[2pt]

\hline
\end{tabular}\vspace{-10pt}
\end{center}
\end{table}

After this (relatively conservative) flux threshold was applied, the number of
fibres remaining in the {\em  clean} mosaic was somewhat reduced, as shown in
\autoref{tab:num_fibres}. The final number of spectra in the {\em  clean}
mosaic accounts for $\sim$ 63\% of the total number of fibres in the original
mosaic. The SSP fitting method was then applied to each individual spectrum of
the clean mosaic, obtaining a residual spectrum per each fibre, in which the
intensities for each detected emission line were measured following the
techniques described in detail in \autoref{sec:analysis}. 
The statistical uncertainty in the measurement of the line flux was
calculated by propagating the error associated to the multi-component fitting
and considering the signal-to-noise of the spectral region.

As a result of the process described above, a set of measured
emission line intensities (plus associated errors) was obtained for each
observed spectrum of the final clean mosaic of NGC\,628.
From these sets of emission line intensities, emission line maps were created
by interpolating the intensities derived for each
individual line in each individual spectrum, based on the position tables of
the clean mosaics, and correcting for the dithering overlapping effects when
appropriated. The interpolation was performed using {\sc E3D}, adopting a
natural-neighbour, non-linear interpolation scheme, with a final scale
of 1''/pixel in the resulting maps. Regions in the borders of the mosaics
and/or large regions in between the mosaic without signal are prone to
artifacts created by the interpolation scheme, special masks were created in
order to deal with those regions. Further, as many of the derived maps are
based on a reliable measurement of H$\beta$, a flux threshold mask was created
for each mosaic for those regions with an integrated H$\beta$ flux per fibre
below 10$^{-16}$ \flux~ (i.e. 0.2 $\times$ 10$^{-16}$ \flux~ per pixel), which
corresponds to a detection limit of $\sim$ 5$\sigma$. This conservative limit
has been adopted to avoid contaminating our results with low signal-to-noise
data, and to grant the accuracy of the absolute spectrophotometry.

Practically any emission line map (or a combination of them) can be
constructed from the detected emission lines in the mosaic. In this section,
we present a number of maps corresponding to the most important emission lines
and derived quantities of interest \footnote{accesible at http://www.ast.cam.ac.uk/research/pings/html/public/}.
The right-panel of \autoref{fig:Ha_maps} shows the pure H$\alpha$
emission line intensity map in units of \FunitsA, obtained by the procedure
described previously. For a purely qualitative comparison purpose, the left-panel of the
same figure shows the continuum-subtracted H$\alpha$ narrow-band image of
NGC\,628 extracted from the SINGS ancillary data by interpolating the aperture
photometry extraction at the location of the mosaic fibres as described in
\autoref{sec:test}. A visual inspection of both maps shows that they match
remarkably well in terms of the morphology.
As the dominant gas ionization mechanism giving rise to the emission line in
this galaxy is due to hot (OB) stars (\autoref{sec:integ_gas}), the H$\alpha$
line intensity map shown in \autoref{fig:Ha_maps} traces the star-forming
regions in this galaxy. The map displays a distinctive spiral structure with
multiple \hh regions and \hh complexes of different sizes and morphology along
the spiral arms. The brightest sources are located at the outer regions of the
galaxy, with a particular giant region located at
($\Delta\alpha$,\,$\Delta\delta$) $\sim$ (--40,120) arcsec. From this map we
can also note the presence of diffuse emission along the spiral arms and in
the intra-arms regions.

\begin{figure}
  %
  %
  \includegraphics[width=\hsize]{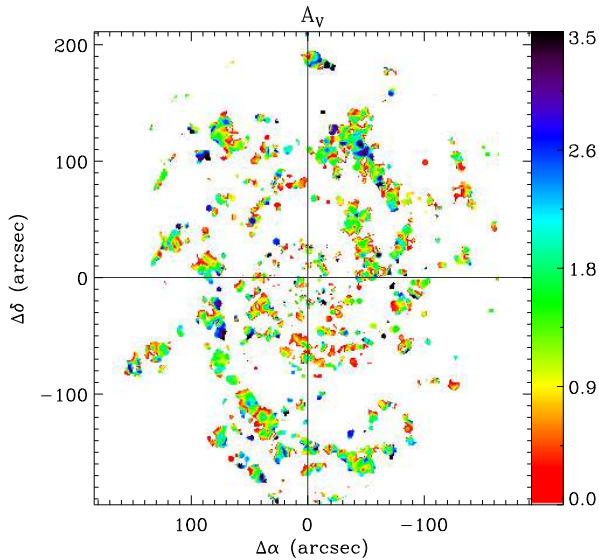}
  \caption[Dust attenuation map of NGC\,628]
  {
    Dust attenuation map of the ionized gas ($A_V$), derived by
    comparing the measured H$\alpha$/H$\beta$ line ratio map with the expected
    value for case-B recombination, assuming a homogeneous distribution of the
    electron density and temperature, and the attenuation law by
    \citet{Cardelli:1989p136}. The regions without accurate measurements of
    either emission lines were masked.
    \label{fig:av_map}
  }
\end{figure}

Fathi07 has recently provided with a similar H$\alpha$ map,
with twice the field-of-view of our IFS data, on the basis of Fabry-Perot and
complementary narrow-band images. The distribution shown by both maps are
remarkable similar. They identified 376 HII regions, many of them within the
field of view of our IFS data. A detail comparison of each of these individual
sources, and an analysis of their spectroscopic properties will be presented
elsewhere (Rosales-Ortega in prep.).

Their data confirm the presence of a central structure in NGC 628,
characterized by a ring of ionized gas at $\sim$15-20$\arcsec$, already
described by \cite{Wakker:1995p3788} and \cite{Ganda:2006p3135} in their
H$\beta$ and [OIII] emissin line distributions. Figure \ref{fig:Ha_maps_cen}
shows a zoom of the H$\alpha$ distribution already shown in Figure
\ref{fig:Ha_maps} (right panel), together with a counterplot of the continuum
emission in the V-band (fig. \ref{fig:V_compar}, right panel). The spatial
resolution for this region is better than the one in the overall Mosaic, due
to the dithering performed only in the central pointing. The slightly oval
ring structure in the ionized gas is clearly identified in the H$\alpha$ map,
showing a remarkable similarity with the H$\beta$ distribution shown by
\cite{Ganda:2006p3135}. It is interesting to note here that despite the
difference in the original sampling of the SAURON instrument (used by Ganda et
al. 2006, 2007), and PPAK, the apparent resolution of both maps are very
similar.  This ring is detected, at different levels, in all of the emission
line spices analyzed in the current article, following the same basic
morphological pattern.

\begin{figure*}
  %
  %
  \includegraphics[width=0.49\hsize]{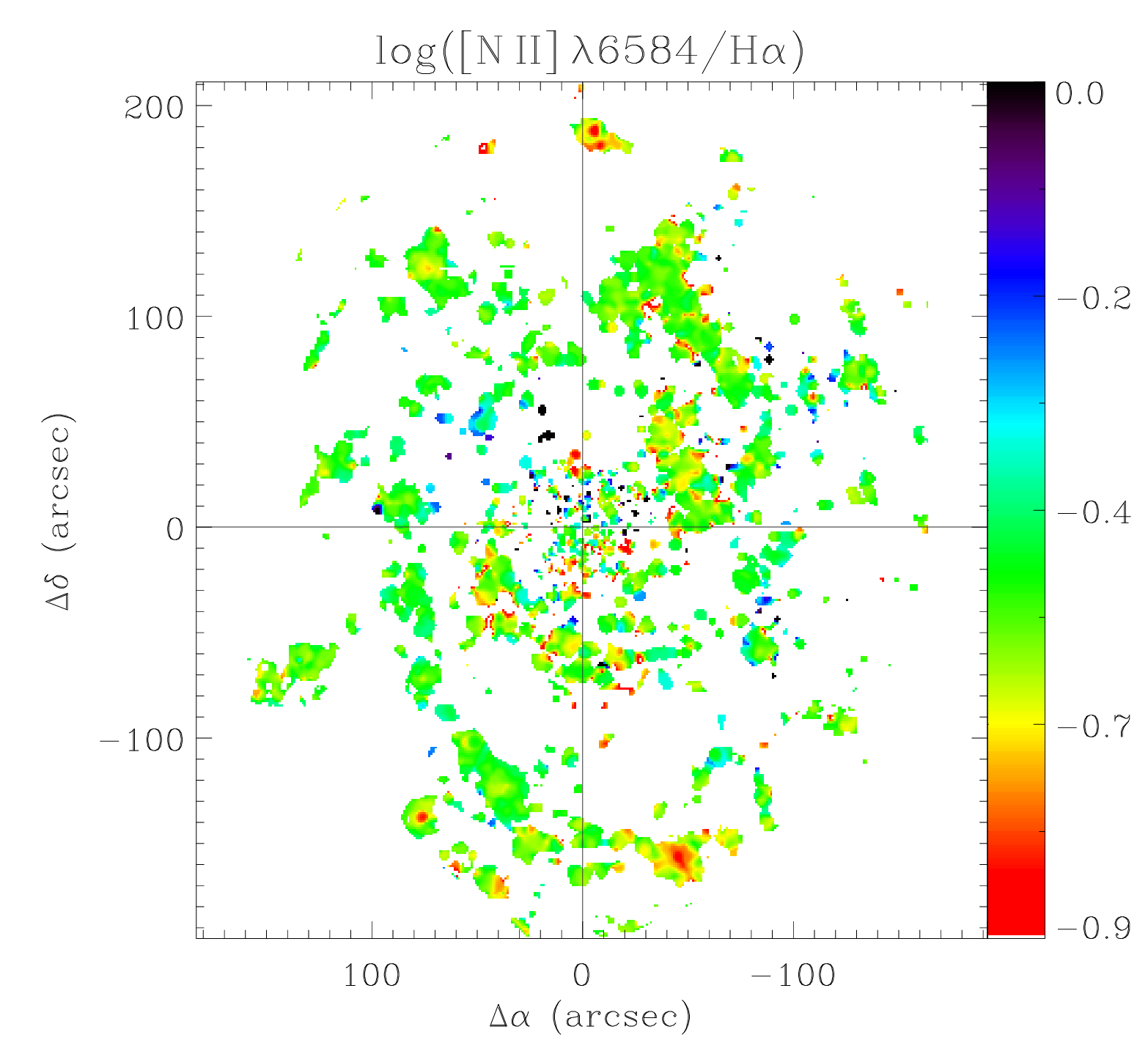}
  \includegraphics[width=0.49\hsize]{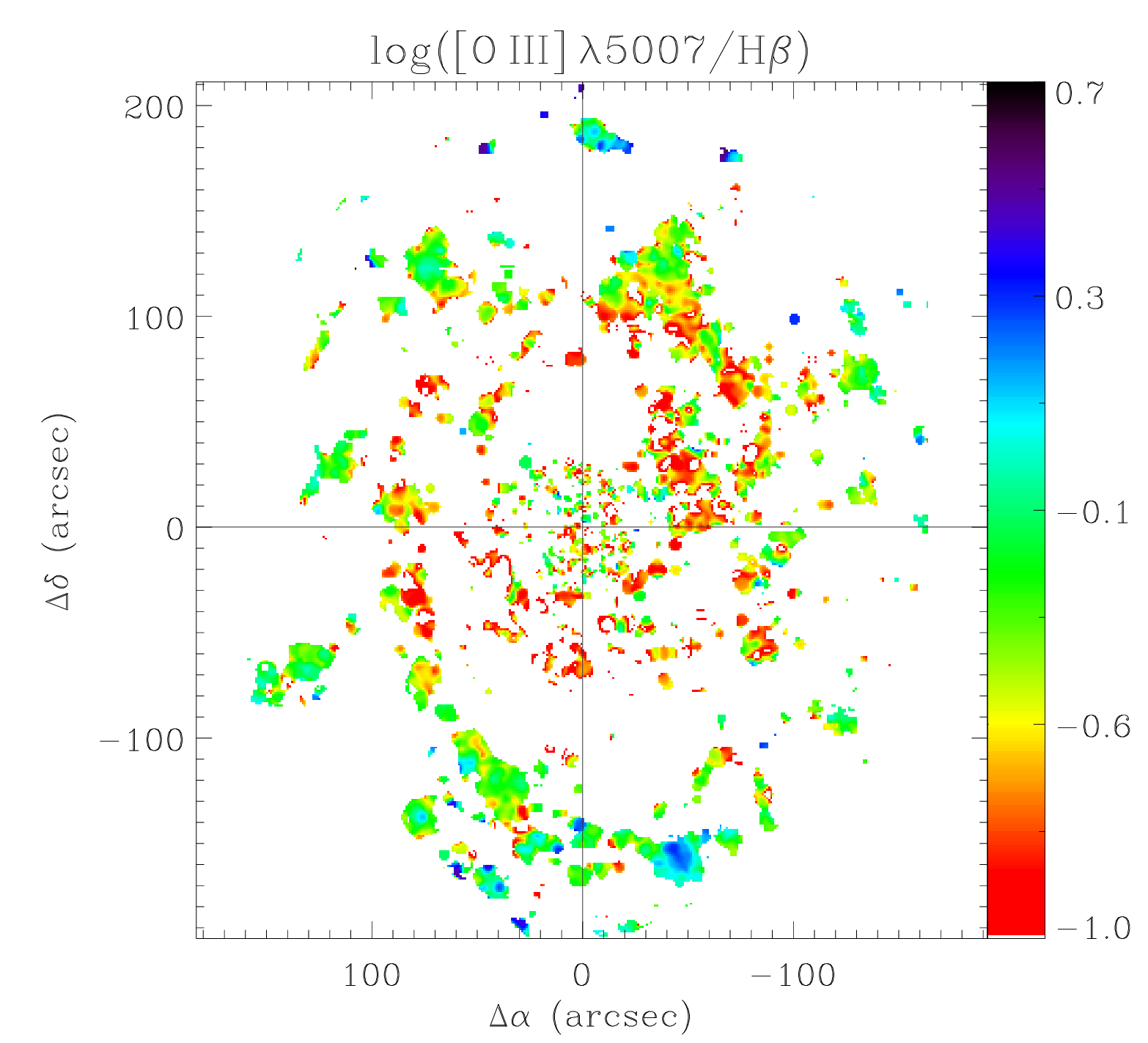}
  \caption[Classical diagnostic line ratio maps]
  {
    2D spatial properties of NGC\,628, classical diagnostic line ratio maps.
    Left-panel: \nii~\lam6584/H$\alpha$ line ratio map. Right panel:
    \oiii~\lam5007/H$\beta$ line ratio map; both in logarithmic scale. 
    \label{fig:bpt_maps}
  }
\end{figure*}

\subsubsection{Distribution of the dust attenuation}
\label{sec:dust}

Besides the intrinsic significance of the emission lines map presented
previously, the most interesting application of the 2D distribution
of the different line intensities resides in the calculation of well-known
line ratios that can be translated to physical properties of the emitting
gas. With the aid of the IFS data we can investigate for the first time, the
point-to-point variation of these physical properties over a considerable area
on the surface of a galaxy.

\autoref{fig:av_map} shows the distribution of the
dust extinction derived from the H$\alpha$/H$\beta$ line ratio, in terms of
the visual extinction $A_V$ (in magnitudes), assuming, as
in the case of the analysis of the integrated spectrum, case-B recombination
with an electron temperature of T$_e$ $\sim$ 10$^4$ K, and adopting the
\citet{Cardelli:1989p136} extinction law with R$_V$, the total to selective
extinction ratio, equal to 3.1. 
Although other Balmer lines were detected at different locations in the
galaxy, H$\alpha$ and H$\beta$ are obviously the lines with the highest
signal-to-noise and therefore are the most adequate for the determination of
$A_V$. 
The dust extinction map was derived for those locations
where the intensity of the H$\beta$ line was above the adopted flux threshold.
These regions clearly follow the spiral arm structure. The dust shows a clumpy
distribution, a rich structure and large variations even within the same \hh
region or complex. There is no apparent trend of the extinction along the
spiral arms or in any radial direction.
The average extinction derived from the values shown in the map is
$A_V\sim$ 1.24$\pm$\,0.76 mag, which is slightly larger but comparable to the
value derived using the integrated spectrum ($A_V\sim$ 1.04).
The reason for this discrepancy may reside in the fact that the
ionized gas component of the integrated spectrum is dominated by the spectra
of the outer regions, where the intensity of the emission lines is stronger
and the extinction is somewhat lower.
Extinctions greater than 2.5 are found in specific regions, e.g. the central
zone of the giant \hh complex at $\sim$ (--40,120), while others are found in
compact \hh regions along both spiral arms.
Previous studies have estimated the extinction at different locations
within this galaxy targeting individual \hh regions
\citep[e.g.][]{McCall:1985p1243,Belley:1992p3779,Petersen:1996p3462}, the
derived extinctions from these studies are consistent with the range of $A_V$
values found in this work.

\subsubsection{Ionization conditions}

As discussed previously in \autoref{sec:integ_gas}, the source and structure
of the ionization can be in principle investigated by exploring the line ratio
maps of typical diagnostic indices. \autoref{fig:bpt_maps} shows on the
left-panel the line ratio map of \nii \lam6584/H$\alpha$ ($N_2H_\alpha$) and
on the right-panel, the \oiii \lam5007/H$\beta$ ($O_3H_\beta$) ratio map, in
logarithmic scale. Note that, given the proximity of the emission lines, these
indices are almost reddening-independent.
The $N_2H_\alpha$ shows a very homogeneous behaviour, with small variations in
individual regions, some of them showing lower values in the central part of
the regions (e.g the knot at $\Delta\alpha$,\,$\Delta\delta$ $\sim$
40,--130). The average ratio derived from this map is --0.55\,$\pm$\,0.12 dex,
which is excellent agreement with the value derived from the integrated
spectrum ($N_2H_\alpha$ = --0.56).
However, some regions at the outer part of the galaxy show lower ratios
consistent with values $\sim$ $-0.75$, e.g
($\Delta\alpha$,\,$\Delta\delta$) $\sim$ (--50,--160), (0,200), which might
suggest a gradient of this index decreasing from the inner to
the outer regions of the galaxy.
On the other hand, the $O_3H_\beta$ shows a clear gradient along the spiral arms
with lower ratios towards the inner regions and greater values at the outer
part of the galaxy. The average value of this index derived from the map is
-0.50\,$\pm$\,0.25 dex, which again is in good agreement with the integrated
value of --0.48. It is interesting to note that the regions with the highest
values of $O_3H_\beta$ are coincident with the zones of the lowest $N_2H_\alpha$
ratios.
Given that the \nii emission originates in the singly ionized regions,
between the fully ionized and the partially ionized zones, the $N_2H_\alpha$
ratio traces the changes in the local ionization, while the \oiii originates
in the fully ionized zones, tracing the strength of the ionization. 
Therefore, the distribution found in these diagnostic maps may indicate that
the ionization is stronger in the outer parts of the spiral arms, than in the
central regions.
The values of both ratios at any location in the galaxy are consistent with
ionization produced by hot OB stars, as expected. In particular, there is no
evidence of ionization due to shocks and/or nuclear activity.

An additional way of studying the 2D distribution of the galaxy properties
consist in obtaining azimuthally-averaged radial spectra, from which radial
average properties can be derived. Taking again as a base the {\em clean} IFS mosaic
version of the galaxy, radial average spectra were obtained by co-adding all
the spectra within successive rings of 10 arcsec, starting from the central
reference point.
An average spectrum was obtained for each single annulus at a given
radius. The radial average spectra were then analysed using the same fitting
procedures described before. Although the derived spectra present more
signal-to-noise than the single-fibre case, the measured emission lines were
corrected by extinction using only the H$\alpha$/H$\beta$ ratio for
consistency with the previous method.

\begin{figure}
  \centering
  %
  %
  \includegraphics[width=\hsize]{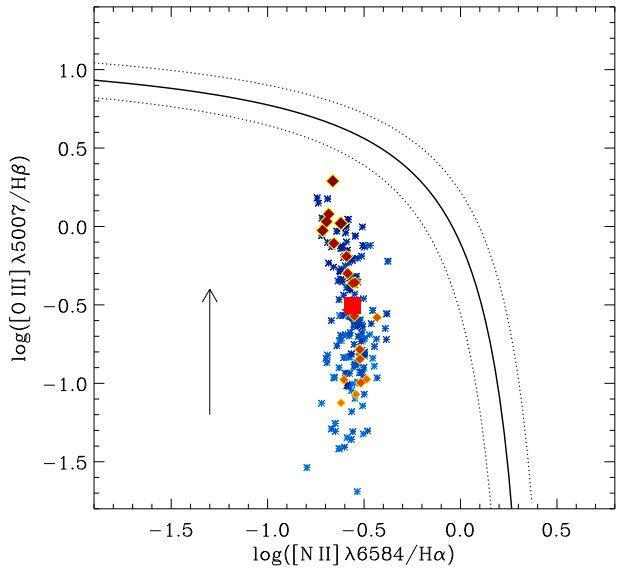}
  \caption[Diagnostic diagram maps]
  { 
    $N_2H_\alpha$ vs. $O_3H_\beta$ diagnostic diagram for NGC\,628. The bluish
    symbols correspond to the derived values of the emission line ratios for
    300 randomly selected spectra distributed uniformly over the surface
    covered by the emission line ratio maps shown in
    \autoref{fig:bpt_maps}. The reddish diamonds correspond to the
    azimuthally-averaged radial values. The red solid-square indicates the
    ratio derived from the integrated spectrum of the galaxy. 
    Lighter tones correspond to inner regions of the galaxy, darker
    colours to increasing galactocentric distance, outlined by the arrow.
    In the radial case, the size of the symbols increases with radius.
    \label{fig:bpt}
  }
\end{figure}

\autoref{fig:bpt} shows the BPT diagnostic diagram \citep{Baldwin:1981p3310}
for the $N_2H_\alpha$ and $O_3H_\beta$ line ratios. 
The bluish-asterisk symbols (in the online version) correspond to the derived
values of the emission line ratios for 300 randomly selected spectra
distributed uniformly over the surface
covered by the emission line ratio maps shown in \autoref{fig:bpt_maps}.
The reddish diamonds correspond to the radial average values introduced above,
while the red solid-square indicates the average line ratio derived from the
integrated spectrum of the galaxy.
The colour-coding of both the individual fibres and the radial-average samples
is related to the spatial position of a given
fibre/annulus. Lighter tones correspond to the inner regions of the galaxy,
while darker colours correspond to positions with increasing galactocentric
radius (arrow direction).
The dark-thick line corresponds to the theoretical boundaries
dividing the starburst region from other types of ionization using the
parametrization provided by \citet{Kewley:2001p3313}. The dashed lines
represent the $\pm$0.1\,dex variation.
As expected, the ionization is dominated by
hot stars (OB class) associated with star-forming regions, at any location in
the galaxy. A clear trend can be noticed in \autoref{fig:bpt} for both samples,
the spectra corresponding to the inner regions tend to have lower line ratios
for both indices; for regions at the outer part of the galaxy, the ratios
increase approaching the theoretical boundary. Part of the reason for this
behaviour is that inner parts of the galaxy lack emission in \oii and \oiii,
while towards the outer parts, the emission from these species is prominent,
increasing the two line ratios involved in this diagram

\begin{figure}
  %
  %
  \includegraphics[width=\hsize]{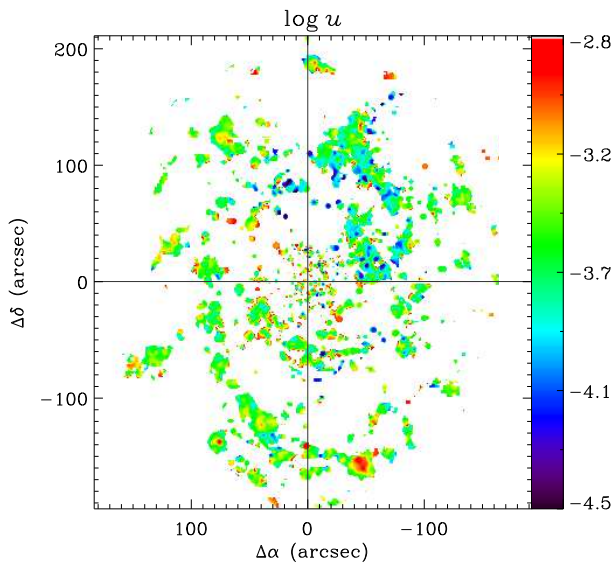}
  \caption[Distribution of the ionization parameter $u$]
  {
    Distribution of the ionization parameter $\log u$, in logarithm scale,
    derived from the dust corrected \oii~\lam3727/\oiii~\lam5007 line
    ratio.
    \label{fig:logu}
  }
\end{figure}

\begin{figure}
  %
  %
  \includegraphics[width=\hsize]{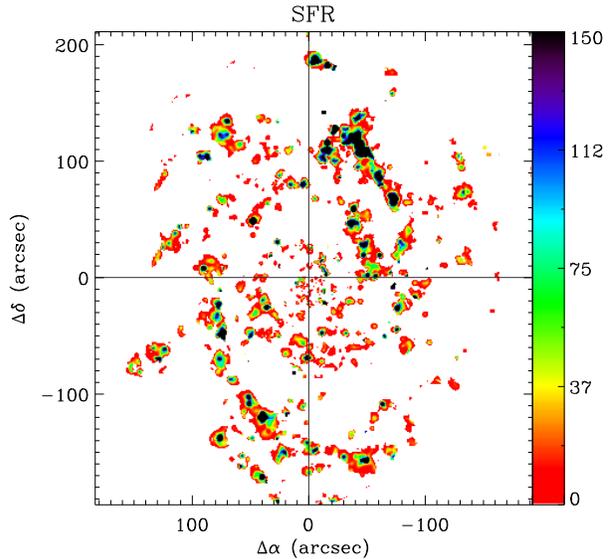}
  \caption[Surface SFR map of NGC\,628]
  {
    Surface star-formation rate (SFR) in units of 10$^{-5}$ M$_\odot$ yr$^{-1}$,
    derived from the H$\alpha$ line intensity map shown in Fig. 10, after
    correcting by dust attenuation. This map traces the distribution of the ionized
    gas, not the young stars, which form much more compact structures than the ones
    shown in the figure.
    \label{fig:sfr}
  }
\end{figure}

The differences in the line ratios at different locations in the
galaxy may be also driven by the strength of the ionization field, being
stronger in the regions with higher $N_2H_\alpha$ line ratio, and lower
$O_3H_\beta$ one.
In order to investigate this point, \autoref{fig:logu} shows the distribution
of the ionization parameter in logarithmic scale, $\log u$, calculated
accordingly to \autoref{eq:log_u} \citep{Diaz:2000p3442}, which is based on
the dust-corrected [O\,{\footnotesize II}]/[O\,{\footnotesize III}]
ratio. Both lines were corrected by extinction using the $A_V$ map discussed
previously. 
The $\log u$ map shows that the ionization is indeed stronger in the outer
parts of the spiral arms, as the previous maps suggested. However, we can also
distinguish a good degree of ionization structure in individual regions within
the spiral arms, with higher values of $\log u$ corresponding to the
geometrical centroids of the \hh regions, as one might expect in the scenario
of a central star cluster or association embedded within an \hh region.
Given that the dominant source of the ionization is radiation from OB stars,
this result indicates that the star-formation rate is stronger in the outer
regions of the galaxy.
A striking feature of this map is the presence of regions with low values of
the ionization parameter (blue colour) located mainly in an specific region,
corresponding to the first quadrant, north-west part of the galaxy. These low
$\log u$ regions are found mainly at the edges of giant \hh complexes, but they
are also found as individual regions. The nature of these low ionization
regions will be investigated thereinafter.
The mean value of $\log u$ derived from this map is --3.50\,$\pm$\,0.26 dex,
which is 0.08 dex higher than the value derived from the integrated spectrum
of NGC\,628.

\begin{figure*}
  %
  %
  \includegraphics[width=0.49\hsize]{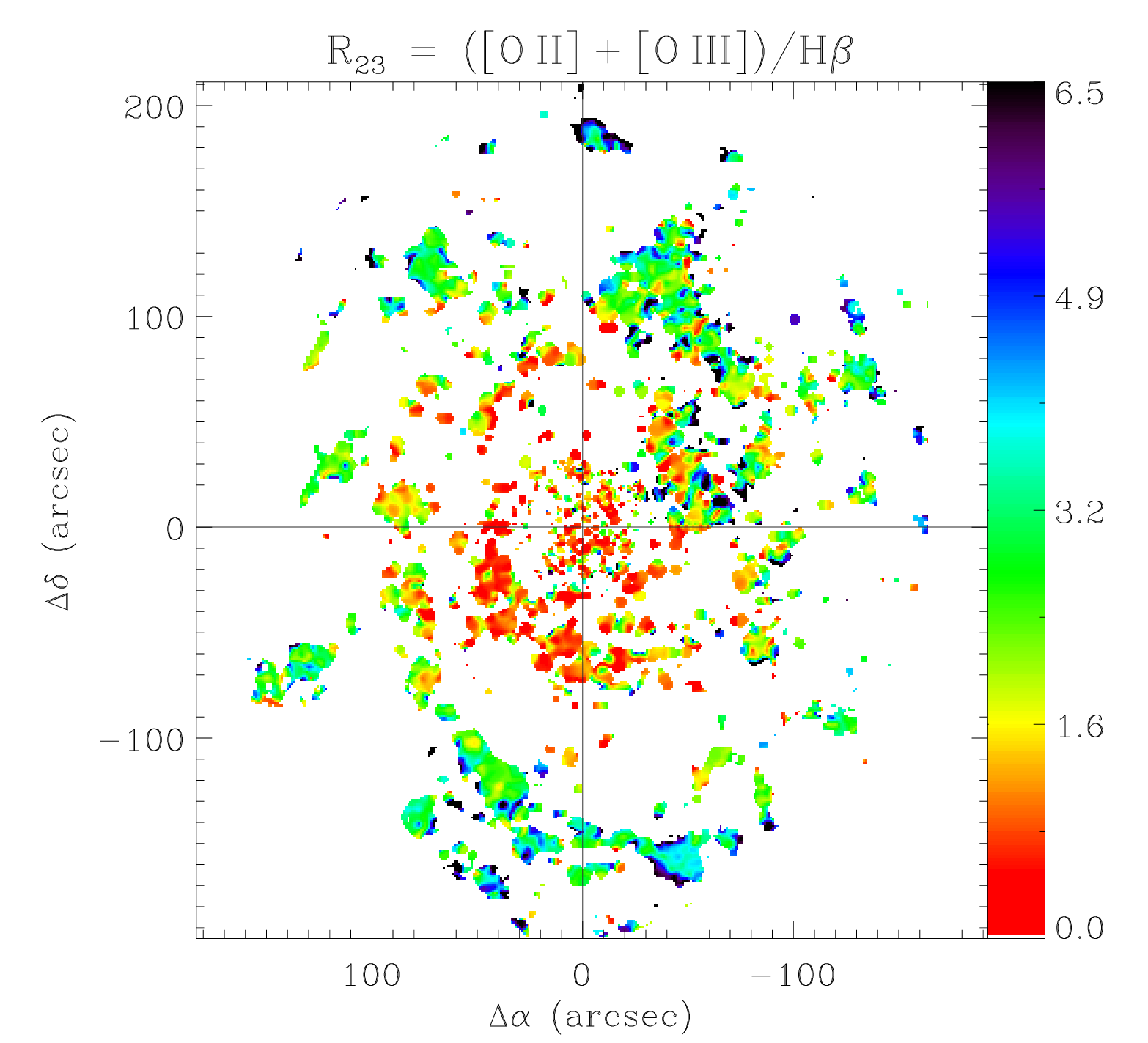}
  \includegraphics[width=0.49\hsize]{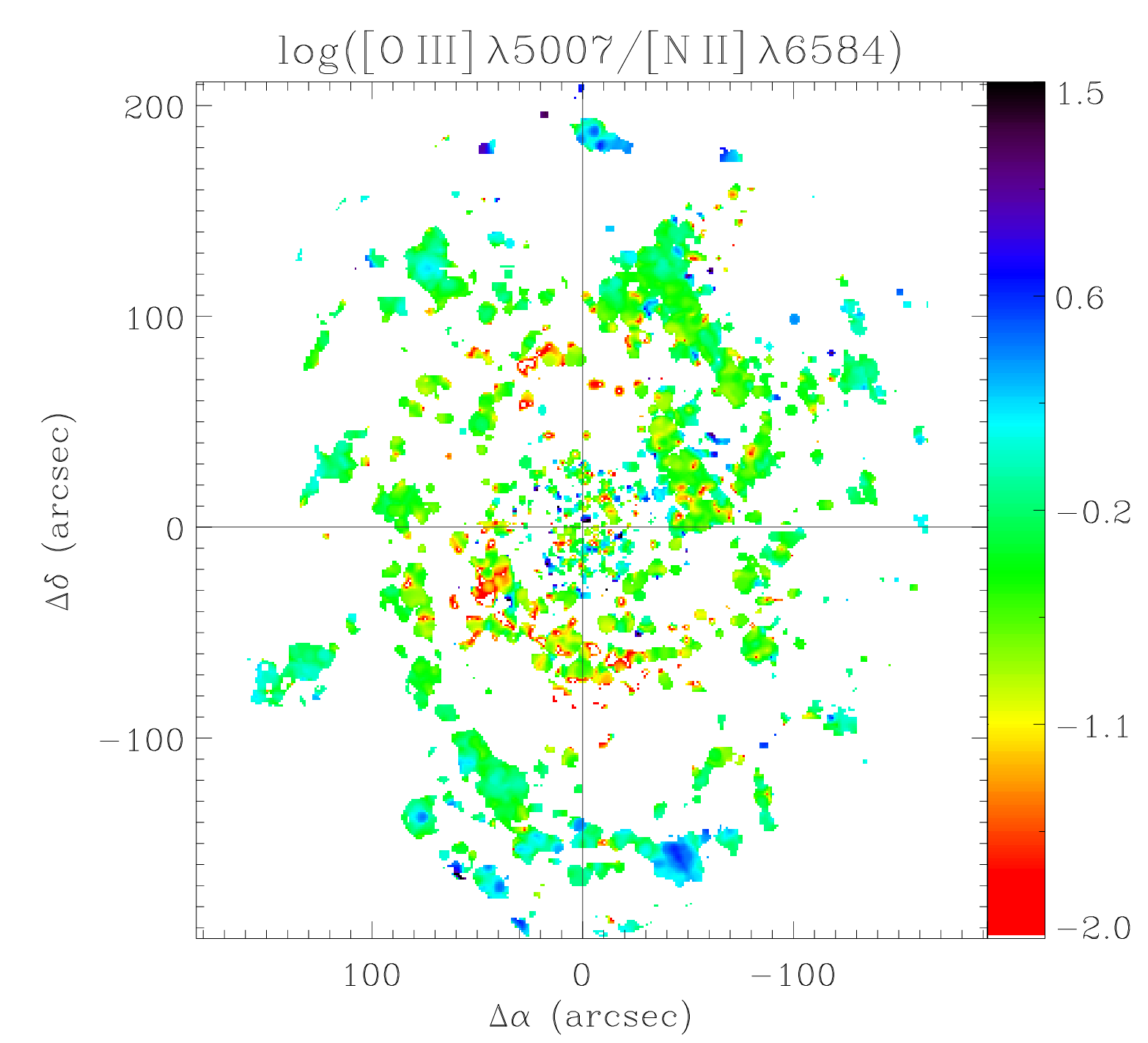}
  \caption[]
  {
    2D distribution of the $R_{23}$ and [O\,{\footnotesize
      III}]/[N\,{\footnotesize II}] metallicity indices for NGC\,628. Both
    maps show a gradient of lower values in the inner regions to higher values
    to the outer part of the galaxy.
    \label{fig:oxy_maps}
  }
\end{figure*}

\subsubsection{Surface star-formation rate}

The surface SFR can be derived using the emission line distribution of
H$\alpha$ or \oii~\lam3727. However, the latter line suffers from more
inaccuracies in both the spectrophotometric calibration and flux
determination, since it lies just at the blue edge of the spectral range
covered by our data set. Therefore, only the H$\alpha$ intensity map was used
in this analysis.
Following our previous prescription to derive the integrated SFR of the galaxy,
the H$\alpha$ intensity map was corrected for attenuation, adopting the dust
distribution described in \autoref{sec:dust}, and the same attenuation law
as in \autoref{sec:integrated}. Then, the dust corrected flux densities were
transformed to absolute luminosities, using the same distance and cosmology
described before. These luminosities were used to obtain the surface SFR, by
adopting the classical relations described by \citet{Kennicutt:1998p3370}.

\autoref{fig:sfr} shows the derived distribution for the surface SFR, in units
of 10$^{-5}$ M$_\odot$ yr$^{-1}$ arcsec$^{-2}$. As expected, the stronger
star-formation is distributed in clumps along the spiral arms, following a
similar morphology to the distribution of the H$\alpha$ emission (i.e. the \hh
regions). In addition, the direct transformation between extinction corrected
H$\alpha$ flux and SFR produce an apparent, low level, star-formation rate in
the inter-arms regions, two orders of magnitude lower than that of the \hh
regions. However, it is unclear how closely is associated this derived SFR
with the inter-arm H$\alpha$ emission. Ionizing photons leaking from the \hh
regions and stochastics in the conversion between both parameters may become
important at this intensity level. As already shown by the distribution of the
$N_2H_\alpha$ and $O_3H_\beta$ line ratios (\autoref{fig:bpt_maps}), and the
ionization parameter (\autoref{fig:logu}), there is an increase of the
star-formation rate from the inner to the outer regions of the spiral arms.

The global SFR can be derived by co-adding the surface SFR over the area shown
in the figure, obtaining a value of $\sim1.1$ M$_\odot$ yr$^{-1}$. By
applying the same aperture and sampling correction applied to the integrated
H$\alpha$ flux in \autoref{sec:integrated}, the estimated SFR is $\sim2.4$
M$_\odot$ yr$^{-1}$, which is practically identical to the estimate derived
based on the flux of H$\alpha$ in the integrated spectrum.

\subsubsection{Distribution of the gas-phase oxygen abundance}

One of the main purposes of the PINGS dataset is to study the metallicity
content of the galaxy in a 2D context. As discussed previously, the oxygen
abundance is normally derived using a variety of diagnostic methods based on
strong lines, from which the most common one is the $R_{23}$ indicator.
The dust-corrected emission line maps of \oii and \oiii provide the means to
explore the behavior of this indicator. The left panel of
\autoref{fig:oxy_maps} shows the $R_{23}$ map of NGC\,628. From a first visual
inspection, it is evident that the map presents a clear gradient of lower
values of $R_{23}$ in the inner regions of the galaxy, to higher values to the
outer parts, specially along the spiral arms. The average $R_{23}$ value
derived from the map is 2.41\,$\pm$\,1.37, which is in good agreement with the
value inferred from the integrated spectrum ($R_{23}$ = 2.57). The lower
values of $R_{23}$ in the central regions of the galaxy are expected, given
the low emission of \oii and (especially) \oiii, as inferred in figure
\autoref{fig:bpt}.
For well-defined \hh regions, the value of $R_{23}$ seems
to be constant in the majority of the cases; however, there are regions in
which the value of $R_{23}$ varies within the \hh region or complex, showing
some level of structure. Considering the O/H values
derived after different calibrators from the integrated spectrum of NGC\,628
in \autoref{sec:integ_abun} (12~+~log(O/H) $\sim$ 8.7), and previous
abundance determinations from individual \hh regions in this galaxy
\citep[e.g.][]{McCall:1985p1243,Ferguson:1998p224}, suggest that
the average oxygen abundance of NGC\,628 corresponds to the high metallicity
regime. In this scenario, the double-valued nature of the $R_{23}$ index can
be broken, and the gradient of higher-to-lower values of $R_{23}$ from the
inner to the outer parts of the galaxy shown in this map would correspond to
the decrement of the $R_{23}$ ratio in the upper-branch of a O/H vs. $R_{23}$
diagram, and therefore, to a true metallicity gradient of the galaxy, as
previous studies based on individual \hh regions have shown.

Another popular metallicity-sensitive index is given by the \oiii
\lam5007/\nii \lam6584 ratio. \citet{Pettini:2004p315} suggested the use of
this ratio (in a modified version) as abundance indicator suitable for the
analysis of high-redshift galaxies. However, the direct application of this
indicator may be too simplistic, the reasons being that the [O\,{\footnotesize
  III}]/[N\,{\footnotesize II}] ratio is strongly dependent on the excitation
of the nebula, and it is also sensitive to both the ionization parameter and
to the age of the cluster of exciting stars \citep{Dopita:2000p3448}.
As suggested by \citet{Dopita:2006p3449}, the \oiii \lam5007/\nii \lam6584
ratio can only be used as an abundance indicator only when a characteristic
age of the exciting clusters can be assumed, and its application in single
galaxies should be taken with caution. However, the combination of this ratio
with, for example, the \nii \lam6584/\oii \lam3727 ratio, can provide a good
diagnostic of both metallicity and age of the ionizing source
\citep{Dopita:2006p3449}.
The right panel of \autoref{fig:oxy_maps} shows the map of the \oiii
\lam5007/\nii \lam6584 ratio of NGC\,628 in logarithmic scale. It is worthy to
note that this ratio spans for more than two orders of magnitude. Similarly to
the $R_{23}$ map, this panel shows a gradient of lower values of the [O\,{\footnotesize
  III}]/[N\,{\footnotesize II}] ratio from the inner part of the galaxy, to
higher values in the outer parts. The average value derived from the map is
--0.35\,$\pm$\,0.36 dex, compared to --0.40 obtained from the integrated
spectrum. The map shows a smoother distribution within individual regions
compared to the $R_{23}$ map. With the corresponding cautions considering the
known dependences of this index with the functional parameters of the \hh
regions, the [O\,{\footnotesize III}]/[N\,{\footnotesize II}] gradient found
for NGC\,628 would correspond to a metallicity gradient of the galaxy,
with some level of inhomogeneity in individual regions as shown in this map.
In order to determine the existence of abundance variations within the same
region of the galaxy, it would be required to co-add the spectra corresponding
to regions with similar ionization conditions, and then perform an analysis
based on single regions. This detailed analysis, together with a complete
study of the radial abundance gradient of this galaxy will be presented in a
forthcoming paper (Rosales-Ortega et al. in preparation).

\begin{figure}
  \centering
  %
  %
  \includegraphics[width=\hsize]{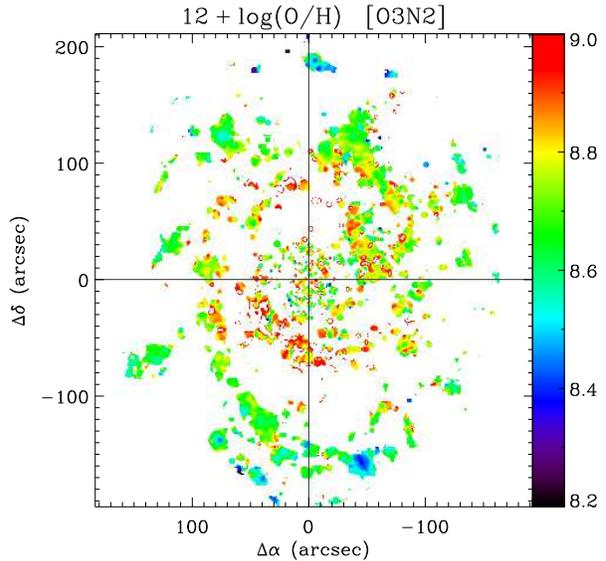}
  \caption[Oxygen abundance map of NGC\,628]
  {
    Oxygen abundance map of NGC\,628 derived by applying the O3N2 calibrator
    \citep{Pettini:2004p315} to the emission line maps of the galaxy. The
    figure shows a clear gradient in metallicity, with more abundant regions
    in the inner part or the galaxy.
    \label{fig:o3n2}
  }
\end{figure}

In \autoref{fig:o3n2} we present the oxygen abundance 2D map of
NGC\,628 based on the $O3N2$ calibrator in units of 12~+~log(O/H), obtained by
the manipulation of the emission line maps presented before. No other 2D
abundance maps were attempted since the parametrizations of the different
calibrators do not allow a simple image manipulation (large blank areas
produce convergence problems which limits the application of any interpolation
routine).
However, as discussed previously in \autoref{sec:integ_abun}, the $O3N2$ abundance
determination follows qualitatively the same pattern as the other more
elaborated calibrators based on $R_{23}$ and the ff--$T_e$ relation. The map
of \autoref{fig:o3n2} shows a clear gradient of higher oxygen
metallicity values from the inner part to the outer part of the galaxy,
and along the spiral arms. The most metal deficient regions are found at the
extremes of the spiral arms, a prominent region is found at
($\Delta\alpha$,$\Delta\delta$) $\sim$ (--40,--60). The average value of the
oxygen abundance derived from this map is 8.75\,$\pm$\,0.17, which is
equivalent to the oxygen abundance derived from the integrated spectrum using
the same calibrator, i.e. 8.71\,$\pm$\,0.11.

  %
  %

%

\begin{table}
\begin{center}
%
%
\caption[Summary of the main properties of the galaxy]
{
  Summary of the main properties of the galaxy.
  \label{tab:sum}
} 
\begin{tabular}{lrr}
\hline  
Parameter & \multicolumn{1}{c}{Integrated} & Average\\
\hline
Age (Gyr)                  &    8.95   &       \\
\zh                        &  --0.44   &       \\
A$_V$ stellar (mag)        &    0.4    &       \\
\hline
A$_V$  (mag)               &    1.04   &     1.24\,$\pm$\,0.76  \\
$\log($\nii \lam6584/H$\alpha)$    &  --0.56   &   --0.55\,$\pm$\,0.12  \\
$\log($\oiii \lam5007/H$\beta)$    &  --0.48   &   --0.50\,$\pm$\,0.25  \\
$\log u$                   &  --3.58   &   --3.50\,$\pm$\,0.26  \\
SFR (M$_\odot$ yr$^{-1}$)   &     2.4   &      2.4$^*$  \\
$\log R_{23}$              &     0.41  &     0.38\,$\pm$\,0.13 \\
$\log($\oiii \lam5007/\nii \lam6584$)$    &  --0.40   &   --0.35\,$\pm$\,0.36  \\
12~+~log(O/H)$_{\rm O3N2}$       &    8.71\,$\pm$\,0.11   & 8.75\,$\pm$\,0.17  \\
12~+~log(O/H)$\dagger$       &    8.69\,$\pm$\,0.31 \\

\hline             
\end{tabular}
\end{center}
$^*$ Co-added.\\
$\dagger$ Mean abundance among the various empirical calibrators.
\end{table}

\section{Discussion and conclusions}
\label{sec:summary}

In this article we presented wide-field IFS observations of a substantial
fraction of the nearby galaxy NGC\,628 ($\sim$\,70\% of its projected
optical size). The observations were performed by adopting a mosaicking procedure,
using the PPAK IFU (mounted at the 3.5m telescope at Calar Alto), which
provides a sampling of 2.7 arcsec and a FOV larger than 1 arcmin
(diameter). The total dataset comprises 11094 individual spectra, covering a
nearly circular FOV of $\sim$\,6 arcmin in diameter, centered on the bulge of
the galaxy. To our knowledge, this is the largest spectroscopic survey ever
made of a single nearby galaxy. A detailed flux calibration technique has been
applied, which granted a spectrophotometric accuracy of $\sim$\,0.2 mag, down
to a surface brightness of $\sim$\,22 mag arcsec$^{-2}$, comprising 6949
spectra from the original dataset.

The spectrophotometrically calibrated data have been analysed both as a single
integrated spectrum, which characterises the global properties of the galaxy,
and using each individual spectrum to determine the spatial variation of
galaxy properties. The stellar and ionized gas components have been decoupled
by a fitting technique which uses SSP templates and single Gaussian functions
respectively to characterise these components. The main properties extracted
from the analysis of the stellar continuum emission were the {\it luminosity-weighted}
ages, metallicities and dust attenuation of the composite stellar
population. On the other hand, for the gas content, we derived the morphology
of the ionized gas, the dust content, the spatially resolved SFR, the source
of the ionization, and the oxygen abundance.

\autoref{tab:sum} lists the main properties derived in this study for
NGC\,628. For each of the estimated parameters we list the values derived
based on the analysis of the integrated spectrum (labelled as Integrated), and
the average values derived from the 2D maps (labelled as Average). When more
than one technique was used to derive the values, we listed the values derived
using the same technique for both datasets. In all cases, the derived
values are compatible.

Regarding the spatial distribution of the different analysed properties, a
gradient was found in the {\it luminosity-weighted} age of the stellar populations from
the central regions, having older stellar populations, to the outer regions,
having younger components. On the other hand, the radial dependence of the
stellar metallicity is weaker, showing a slight or no gradient. If there is
any gradient, the richer stellar populations are located at the central
regions, while the poorer ones are at the outer regions. All together these
results are consistent with our current understanding of how the late-type
galaxies grow along cosmological epochs, that is inside-out
\citep{Barden:2005p3774}. The older and more metal-rich stellar populations,
formed before, on the basis of recycled gas from previous populations, are
located in the center, while the younger and less metal rich stars are in the
outer regions. It is interesting to note that the oxygen abundance in the
ionized gas follows nearly the same distribution (\autoref{fig:o3n2}), with
larger abundances in the inner regions and lower abundances in the outer
ones. This gradient was previously reported by \citet{McCall:1985p1243}, and
confirmed by \citet{Belley:1992p3779}, \citet{vanZee:1998p3468} and
\citet{Ferguson:1998p224}.

In the case of the stellar populations, there seems to be an inversion of the
age gradient at the very centre of the galaxy ($\sim$\,1 kpc), where
apparently there exist a ring of old stars at this distance, with a trend to
younger ones at the very center. These results were previously reported by
\citet{Ganda:2007p3763}, on the basis of their IFS observations at the very
centre of this galaxy. Similar results have been reported in other galaxies,
mostly Sa/S0, where the inner regions of their bulges present bluer colors,
consistent with younger stellar populations
\cite[e.g.][]{Deharveng:1997p3775}. According to this scenario, the
circumnuclear star-formation rings detected in Sa-Sbc galaxies might show
evidence of gas transfer due to radial motions in these galaxies
\citep[e.g.][]{Knapen:2006p3776}.

Spatially-resolved maps of the emission line intensities and physical
properties were derived for NGC\,628. Contrary to previous attempts to perform
a 2D wide-field analysis based on narrow-band (or Fabry-Perot) imaging, which
only allowed a basic analysis of the physical parameters and/or required
assumptions on the line ratios included within individual filters (e.g. \ha),
the emission line maps presented in this paper were constructed from
individual (deblended) emission lines at any discrete spatial location of the
galaxy, where enough signal-to-noise was found. This fact allowed to
investigate the point-to-point variation of the physical properties over a
considerable area on the galaxy. Extinction, ionization, and
metallicity-sensitive indicators maps were derived from reddening corrected
emission line maps. In general, they show that the ionized gas in these spiral
galaxies exhibits a complex structure, morphologically associated with the
star forming regions located along the spiral arms.

In general, it is found that the dominant ionization process is due to the OB
hot-stars associated with star-forming regions, based on the distribution of
the classical diagnostic lines. The ionization is stronger along the spiral
arms, associated with the \hh regions, and more intense in the outer than in
the inner ones. Indeed, the surface SFR is an order of magnitude stronger in
the outer \hh regions, at distance larger than $\sim$\,100 arcsec (4.5 kpc),
than in the inner ones. Considering that in these outer regions there is a
lower mass density, the growing rate of stellar mass is considerably larger
there than in the inner ones. Therefore, even taking into account the
circumnuclear star-formation rate, the growth of the galaxy is dominated by
the inside-out process.

There is growing evidence of a central structure in NGC 628, widely discussed
by Fathi07. This structure is characterized by a ring of ionized gas
\citep{Wakker:1995p3788,Ganda:2006p3135,Fathi:2007p2409}, and a drop in the
$B-V$ colors \citep{Natali:1992p3762,Fathi:2007p2409}. The kinematics analysis
of the stellar and gas component within this ring, in comparison with that of
the rest of the galaxy, indicates that there is a rapidly rotating component,
most probably a dynamically cold disc, which could be built by inflow from the
outer parts towards the center
\citep{Daigle:2006p3787,Ganda:2006p3135,Fathi:2007p2409}. Fathi07 found that
this circumnuclear ring is that the expected location of the inner Lindblad
resonance radius, where the gas is indeed expected to accumulate, due to
non-circular motions exerted by a bar or spiral arms. This gas concentration
may lead to a low luminosity starburst
\citep[e.g.,][]{Knapen:1995ApJ...454..623K}. Our results show, for the first
time, that (1) the gas in the circumnuclear ring is ionized by starformation,
and (2) the ring is spatially coincident with a decrease in the
luminosity-weighted ages of the stellar populations, which causes the
previously observed drop in the $B-V$ colors. All together, they support the
interpretation outlined by Fathi07.

The integrated spectrum of the galaxy was used to analyse the chemical
abundance of the galaxy by means of different empirical diagnostic
abundance calibrators. We confirmed the validity of the determination of the
integrated chemical composition of a galaxy through the analysis of the
global-emission line spectra, as previously found by other authors
\citep{Kobulnicky:1999p1710,Pilyugin:2004p82,Moustakas:2006p313}, under the
assumption that the observed emission lines arise via photoionization from
young, massive stars. From the set of calibrators used for this purpose, the
$R_{23}$ methods based on photoionization modelling provide higher values of
the oxygen abundance (i.e. M91 \& KK04), followed by those methods which
consider one of two emission-line indices (N2, O3N2) and the ff--$T_e$
method. The differences between different calibrators are as large as $\sim$
$\pm$0.7 dex, especially for the high values of O/H derived from the $R_{23}$
methods compared to the other ones. If an arbitrary negative offset is applied
(as suggested by the literature), the O/H values seem to be consistent for an
average integrated oxygen abundance of NGC\,628 of 12~+~log(O/H) $\sim$ 8.7,
in agreement with previous studies.
Note that the validity of the abundance determination depends mainly on the
chosen calibrator used to derive the chemical composition, and to a second
order, on the SSP fitting decoupling.

The spatially resolved distribution of the abundance shows a clear gradient of
higher oxygen metallicity values from the inner part to the outer part of the
galaxy, and along the spiral arms \autoref{fig:o3n2}. However, in some
instances, the value of the oxygen abundance (and other physical properties
like extinction and the ionization parameter) varies within what would be
considered a classical well-defined \hh region (or \hh complex), showing some
level of structure.
Indeed, the 2D character of the PINGS data allow us to study the small-scale
variation of the spectra within a given emitting area. As suggested in
\citetalias{RosalesOrtega:2010p3553}, the values of the emission
line ratios measured using different extraction apertures vary considerably as
a function of the aperture size, and that the scatter of the central value is
larger than the statistical error in the measurements, reflecting that this
might in fact be a physical effect.
If we take as a premise that for a sufficiently large \hh
region, the emission line measurements are aperture and spatial dependent,
i.e. that the light is emitted under different physical conditions, by gas in
different degrees of ionization, and modified by different amounts of
reddening (and therefore providing different elemental ionic abundances), 
the level of structure seen in the 2D maps of NGC\,628 may be due to the
intrinsic distribution of the ionizing sources, gas content, dust extinction
and ionization structure within a given region, i.e. we would be sampling real
point-to-point variations of the physical properties within a \hh region.

The emission line maps presented in this paper proved to be useful in
describing the general 2D properties of the selected galaxies. However, the
conclusions raised from them were based on general trends that depend, to a
certain level, on the interpolation scheme applied in order to derive the
pixel-resolved maps. More robust conclusions can only be drawn by analysing
specific individual regions within the given galaxy, or by co-adding spectra
of regions with the same physical properties and comparing the results in the
2D context. This approach will be followed in subsequent papers in order to
study the spatially-resolved spectroscopic properties of this and the rest of
the PINGS galaxies.


\section*{Acknowledgments}

SFS thanks the Spanish Plan Nacional de Astronom\'ia program
AYA2005-09413-C02-02, of the Spanish Ministry of Education and Science and
the Plan Andaluz de Investigaci\'on of Junta de Andaluc\'{\i}a as research
group FQM306, and the ARAID fundation for providing funds.

SFS thanks the director of CEFCA, Dr. M. Moles, for the inconditional support 
to this project.

FFRO would like to thank the Mexican National Council for Science and Technology (CONACyT)
and the Direcci{\'o}n General de Relaciones Internacionales (SEP-México).
We would like to thank the referee, R. Peletier, for the very valuable
comments and suggestions which improved the final content of this paper.
We acknowledge Dr. N.Cardiel for his help in the study of the stellar
populations.

\bibliographystyle{mnras}

\bibliography{mnras}

\appendix

\section{Multiple stellar population fitting}
\label{app}

A substantial fraction of the stellar continuum emission in early-type
galaxies is dominated by a single-stellar population, reflecting the fact that
their cosmological evolution is, in many cases, well described by a monolithic
collapse model. In this scenario most of the star formation took place in
these galaxies for a short time at early times, and therefore, their SFH can
be well described by a single star-formation process.  As a consequence, their
stellar continuum can be easily described by a single-stellar population.

It is well known that the spectral energy distribution (SED) of simple stellar
populations (chemically homogeneous and coeval stellar systems) depends on a
set of first principles (e.g. initial mass function, star formation rate,
stellar isochrones, metallicity, etc.), from which it is possible to generate
the spectra of synthetic stellar populations.  This technique, known as
evolutionary synthesis modeling \citep[e.g.][]{Tinsley:1980p3431}, has been widely used to
unveil the stellar population content of galaxies by reconciling the observed
spectral energy distributions with those predicted by the theoretical
framework. Unfortunately the variation of different physical quantities
governing the evolution of stellar populations produce similar effects in the
integrated light of those systems, leading to a situation in which the
observational data is affected by undesirable degeneracies, like the widely
mentioned one between age and metallicity 
\citep[e.g.][]{Oconnell:1976p3432,Aaronson:1978p3433,Worthey:1994p3434}.
However, the use of spectrophotometrically calibrated
spectra and the sampling of a wide wavelength range, as in the present data
set, helps to break the degeneracy and allows the derivation reliable physical
parameters by fitting the full spectral distribution with single-stellar
populations \citep{Cardiel:2003p3435}.

However, the simple assumption that a single-stellar population describes well
the SED of a galaxy is not valid for late-type galaxies. These objects present
complex star formation histories, with different episodes of activity, of
variable intensity and time scale. Therefore, a single-stellar population does
not reproduce well their stellar emission, in general terms. A different
technique, known as full-spectrum modeling, involving the linear combination
of multiple stellar populations and the non-linear effects of dust
attenuation, has been developed to reconstruct their stellar populations
\citep[e.g.][]{CidFernandes:2005p357,Ocvirk:2006p3413,Sarzi:2006p3133,
Koleva:2009p3414,MacArthur:2009p3412}.

In general, these reconstructions require a wide wavelength range to probe the
hot, young stars and the cool, old stars simultaneously. They also require the
best spectrophotometric calibration to disentangle the effects of age,
metallicity and dust attenuation. Although different implementations of this
technique have some differences, they are very similar in their basis, as
described above. The extracted information from the multi stellar-population
modelling differs in the different implementations. In some cases the
luminosity (or mass) weight ages and metallicities are derived, based on the
linear combination of different models \citep[e.g.][]{Sarzi:2006p3133}. In other
implementations the fraction of light (or mass) of different stellar
populations \citep[e.g.][]{Stoklasova:2009p3436}, or the fraction of light (or
mass) that corresponds to old or young stellar populations
\citep[e.g.][]{MacArthur:2009p3412} are derived.

For our particular analysis, it is necessary to model accurately the
underlying continuum not only to understand the nature of the stellar
population, but also to obtain a good representation of the continuum in order
to decouple it from the emission lines produced by the ionized gas. Therefore,
even in the case that the combination of SSPs is strongly degenerate, and the
created model has no physical meaning, it could be partially useful for our
purposes.

\begin{figure*}
  \label{templates}
  \caption[Spectra of the 72 SSP templates]{ 
    Spectra of the 72 SSP templates used in the population synthesis
    fits. Metallicity increases from the top-left to the bottom-right. Different
    ages are represented by different colours. All the spectra are normalized
    to the flux at 5000 \AA.
 }
\end{figure*}

The implementation of the multi stellar-population fitting technique used in
this article is part of the {\sc FIT3D} package
\citep[e.g.][]{Sanchez:2007p1696,Sanchez:2007p3299}, and it has been recently used
by \citep[e.g.][]{Stoklasova:2009p3436} in the analysis of the stellar population of
nearby galaxies based on IFS data.  This package enables linear fits of a
combination of SSPs, and non-linear ones of emission-lines plus an
underlying stellar population. It also includes routines to extract the 2D
distribution of the different derived parameters based on the analysis of the
stellar population (age, metallicity, dust) and/or the analysis of the
emission lines (flux intensity, systemic velocity and velocity dispersion).
The package will be delivered to the public in a forthcoming paper. The basic
steps of the fitting algorithm, spectrum by spectrum, are the following:

\begin{figure*}
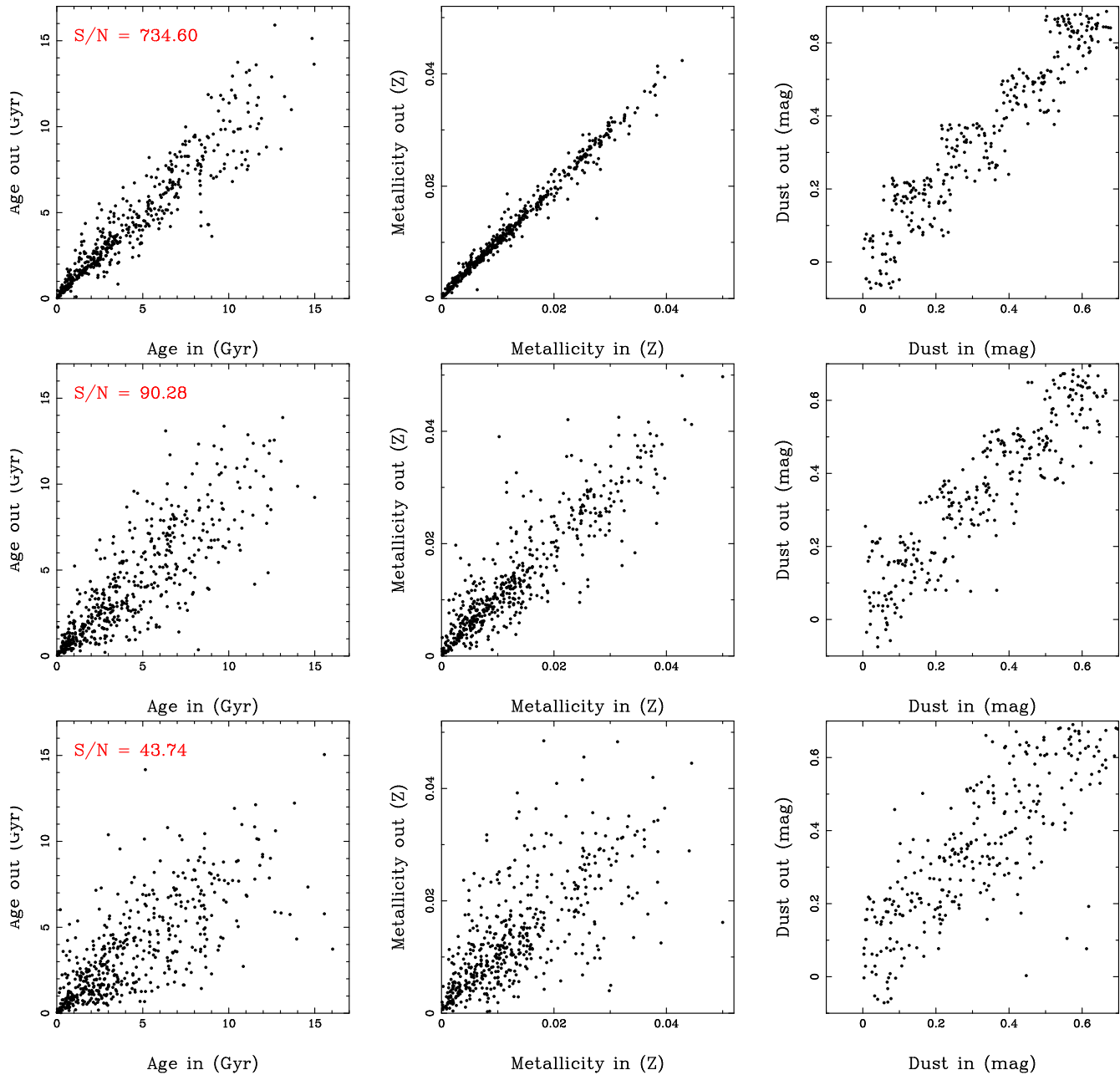

  \resizebox{\hsize}{!}{
    \includegraphics[width=\hsize,angle=270]{sim_age_met_0.ps}}
  \resizebox{\hsize}{!}{
    \includegraphics[width=\hsize,angle=270]{sim_age_met_1.ps}}
  \resizebox{\hsize}{!}{
    \includegraphics[width=\hsize,angle=270]{sim_age_met_2.ps}}
  \caption[Results from the simulations 1]
  {
    Results from the simulations. Each panel shows, from left to right, the
    comparison between the input luminosity-weighted age, metallicity and dust attenuation,
    with the recovered values by our fitting technique. The results are presented
    for different signal-to-noise ratios, decreasing from the upper to the lower
    panels. A random value of $\pm$0.08 mag has been added to output
    attenuation to show all the values, since they are originally quantized to
    steps of 0.1 mag.
  \label{sim1}

  }
\end{figure*}

\addtocounter{figure}{-1}

\begin{figure*}
  \resizebox{\hsize}{!}{
    \includegraphics[width=\hsize,angle=270]{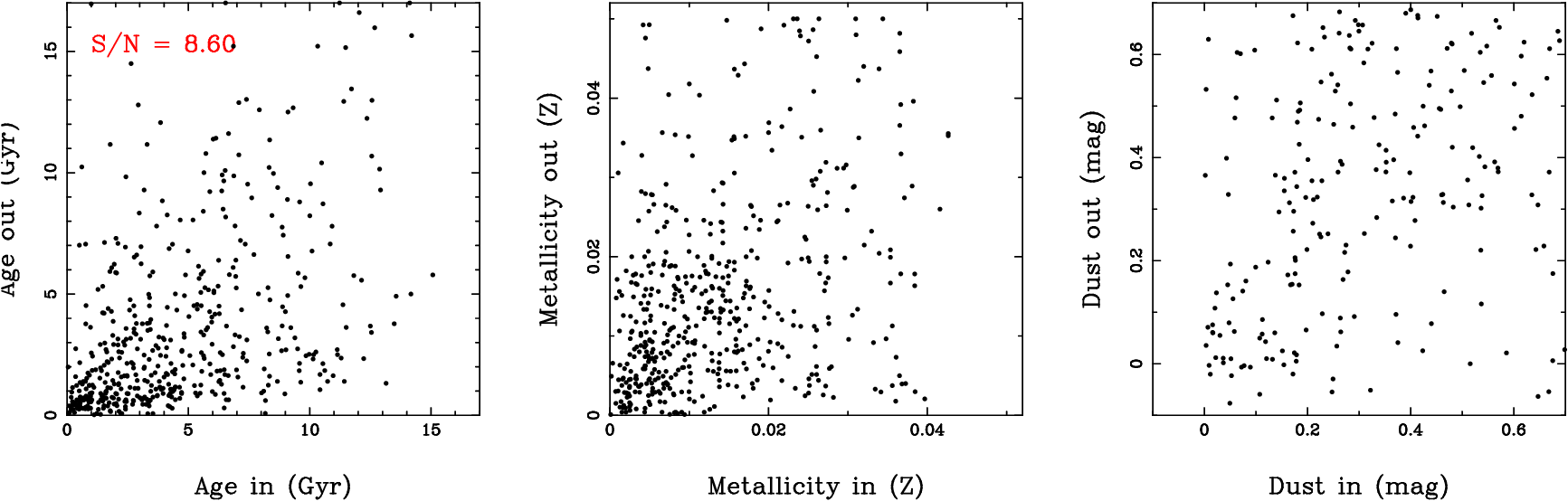}}
  \resizebox{\hsize}{!}{
    \includegraphics[width=\hsize,angle=270]{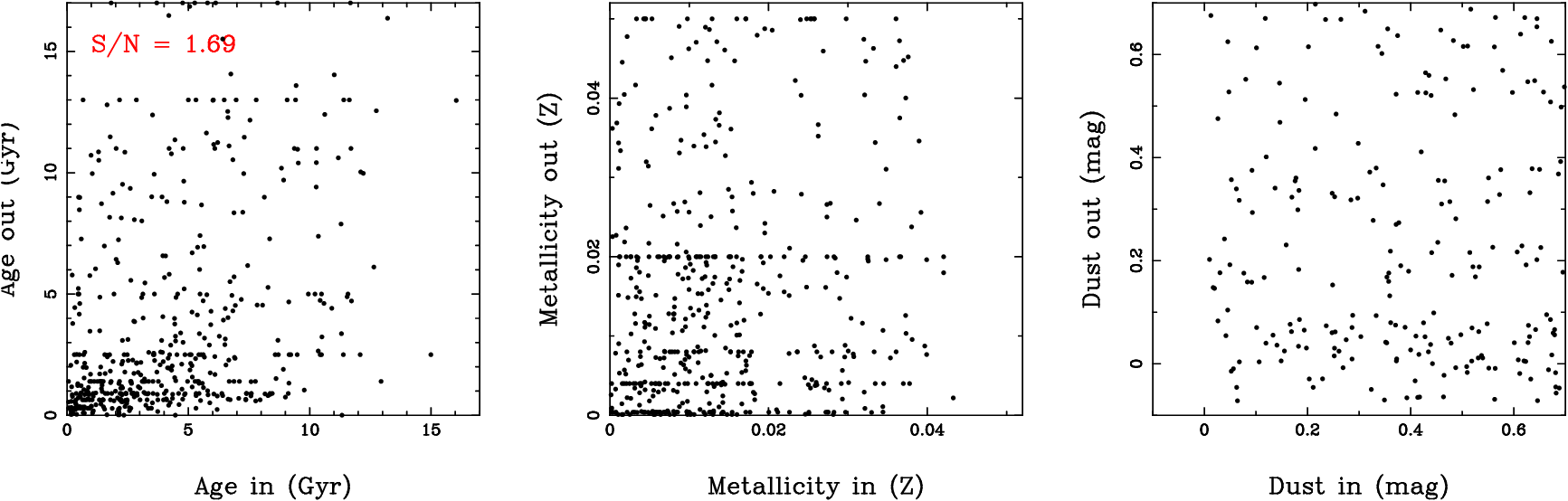}}
  \caption[Results from the simulations 1]
  {
    Continue
  \label{sim2} 
  }
\end{figure*}

\begin{enumerate}

\item Read the input spectrum, determine the areas to be masked, and construct
  the variance map. Lets define $G_i$ as the observed galaxy flux at
  wavelength $\lambda_i$, and $N$ the number of elements of the masked spectrum.

\item Read the set of SSP template spectra, shift them to the systemic
  velocity of the considered spectrum, convolve them to match its spectral
  resolution and velocity dispersion, and resample them to its wavelength
  solution. Lets define $F_{ji}$ as the flux of the $i^{th}$ wavelength of the
  $j^{th}$ template (once shifted, convolved and resampled), where $M_0$ is the
  total number of considered templates.

\item Apply a certain dust attenuation to the templates. The attenuation law of
  Cardelli, Clayton \& Mathis (1989) was adopted, with a ratio of total to
  selective attenuation of $R_V=$3.1 \citep{Jenkins:1987p3437}.  Lets define
  $F_{ji}^{A_V}$ as the flux of the $i^{th}$ wavelength of the  $j^{th}$
  template, after applying the dust attenuation corresponding to a certain
  attenuation of $A_V$ magnitudes.

\item Perform a linear least-square fitting of the input spectrum with the set
  of SSP templates, using a modified $\chi^2$ as a merit function to be
  minimised, with the form:

$$ \chi^2 = \frac{1}{N-M} \sum_{i=1}^{N} R^2_i, $$

\noindent where

$$ R_i = w_i \Big( G_i- \sum_{j=1}^{M} a_j F_{ji}^{A_V} \Big), $$

In the above expression, $w_i$ is the weight of the $i^{th}$ pixel, defined as

$$ w_i = \frac{1}{\sigma_i^2}, $$

\noindent and $a_j$ is the coefficient of the $j^{th}$ template in the final modelled
spectrum, and $M$ is the number of templates considered in the fitting
procedure.

\item Determine for which templates the fitting procedure produces negative
  coefficients in the linear combination (i.e. $a_j<0$). These templates will
  be excluded from the next iteration of the fitting procedure, that will
  be resumed in the step (iii). At each iteration, $M$ is decreased by the amount of
  excluded templates. This loop ends once all the coefficients are positive.

\item The modified $\chi^2$ corresponding to the considered dust attenuation is
  stored as $\chi^2_{A_V}$, and the fitting procedure is repeated again
  starting at the step (iii), modifying the considered amount of dust and
  starting the procedure with the full set of templates (i.e. $M=M_0$, again).
  This loop ends once the dust attenuation has covered a pre-defined range of
  possible values, that it is defined as an input parameter.

\item The best linear combination of templates and dust attenuations are
  selected based on the minimization of the stored $\chi^2$ parameters. The
  final modelled spectrum flux at $\lambda_i$ is given by

$$ S_i = \sum_{j=1}^{M} a_j F_{ji}^{A_V}, $$

\noindent where in this case, only the $M$ templates with positive coefficients $a_j$ are
considered, and $A_V$ corresponds to that value which minimises $\chi^2_{A_V}$.

\item The luminosity-weighted age ($H$) and metallicity ($Z$) of the underlying
  stellar population is then derived by the formulae:

$$ H = \sum_{j=1}^{M} a_j H_j, $$

\noindent and

$$ Z = \sum_{j=1}^{M} a_j Z_j, $$

\noindent where $H_j$ and $Z_j$ are the corresponding age and metallicity of the
$j^{th}$ SSP template. These luminosity-weighted ages and metallicities should
be considered as the {\it luminosity-weighted} ages and metallicities of the
modeled stellar population, since they would match with the corresponding values
if the population was composed by a single SSP.  

\end{enumerate}

This algorithm shares a basis with many other procedures described in the
literature (e.g. \citealt{CidFernandes:2005p357}, \citealt{Sarzi:2006p3133},
MacArthur et al. 2008). If required, low-frequencies in the spectrum can be
fitted by adding to the fitted templates a polynomial function, or by
multiplying them by that function. There are different origins for these
low-frequencies in the spectrum, mostly related with defects in the reduction
process. In the particular case of the data analyzed in the current article,
the inclusion of this additional step in the fitting procedure does not
improve considerably the results.

By construction, the fitting algorithm is useful for masking undesirable
regions in the spectrum. In general, it is needed to mask: (i) strong and
variable night sky-line regions, (ii) regions affected by defects in the CCD
(like dead columns) whose effect was not completely removed during the data
reduction process, (iii) regions affected by telluric absorption, not
completely corrected during the flux calibration process, and (iv) regions
containing emission lines from ionized gas.

As indicated above, one of the main goals of this fitting procedure is to
provide with and accurate modelling of the underlying stellar population, and
to derive a pure-emission line spectrum, given by

$$ C_i = G_i - S_i. $$

\subsection{Modelling the emission lines}

To derive the intensity of each detected emission line, the {\it clean}
spectrum ($C_i$) is fitted to a single Gaussian function per emission line
plus a low order polynomial function. This fit is performed using non-linear
fitting routines implemented in FIT3D, that were described in previous
articles. Instead of fitting all the wavelength range in a row, we extracted
for each spectrum shorter wavelength ranges that sampled one or a few of the
analysed emission lines, in order to characterise the residual continuum with
the most simple polynomial function, and to simplify the fitting
procedure. When more than one emission was fitted simultaneously, their
systemic velocities and FWHMs were forced to be equal (since the FWHM is
dominated by the spectral resolution), in order to decrease the number of free
parameters and increase the accuracy of the deblending process (when
required). These procedures are frequently used, and therefore, we will not
explain them in more detail.

A modelled gas-emission spectrum is created, based on the results
of the last fitting procedure, using only the combination of Gaussian
functions. This spectrum is given by

$$ E_i = \sum_{k=1}^{L} B_{k} * GAUSS_{ik} (\lambda_k,\sigma_k), $$

\noindent where $L$ is the number of emission lines considered in the model, $B_k$ is
the integrated flux of the $k^{th}$ emission line, and $GAUSS_{ik}$ is the
corresponding normalized Gaussian function (with central wavelength
$\lambda_k$, and dispersion $\sigma_k$), evaluated at the $i^{th}$ pixel.

Finally, the original spectrum ($G_i$) is decontaminated by the gas emission,
subtracting this modeled gas-emission spectrum, deriving a gas-free spectrum
given by

$$ GF_i = G_i - E_i. $$

\noindent This spectrum is then used to model the stellar population, by applying the
procedure described before, but without masking the emission lines spectral
regions. We adopted the results from this second iteration as a better
modeling of the stellar population.

\subsection{SSP template library}

As noted by \citet{MacArthur:2009p3412}, this kind of analysis is always limited
by the template library, which comprises a discrete sampling of the SSP ages
and metallicities. It would be desired for the stellar library to be as
complete as possible, and non-redundant. However, this would require an exact
match between the models and the data, which is not possible to achieve in
general terms, in particular if the stellar population comprises more than one
SSP. As a suitable solution for the analysis presented in this article, we
adopted a SSP template library that covers the widest possible range of ages
and metallicities. This template has already been demonstrated as useful for
modelling the stellar population of a varied set of galaxies
\citep[see][]{Sanchez:2007p3299}.

\begin{figure*}
\resizebox{\hsize}{!}{
\includegraphics[width=\hsize,angle=270]{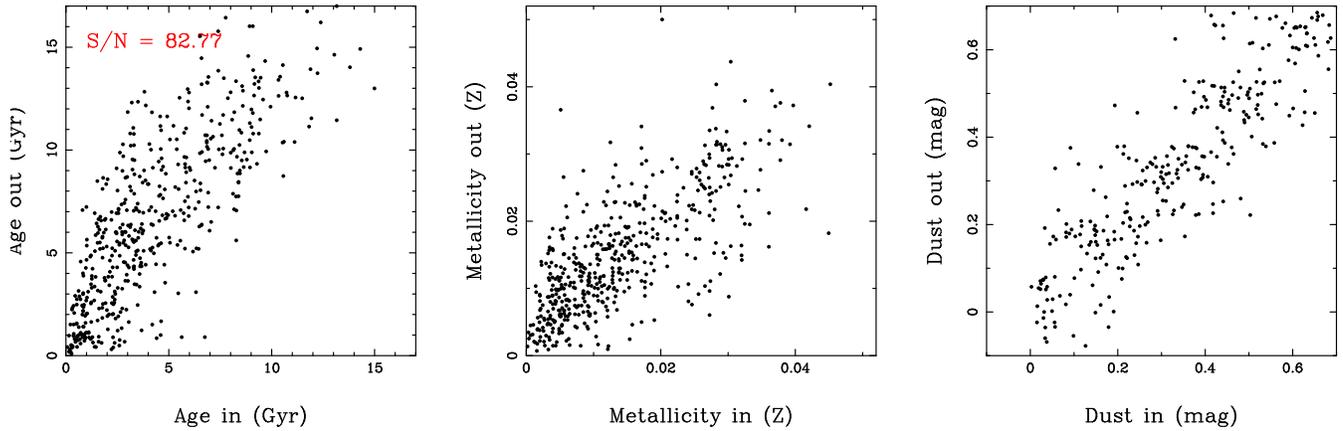}}
  \caption[Results from the simulations 2]{
    Results from the simulations. Each panel shows, from left to right, the
    comparison between the input luminosity-weighted age, metallicity and dust attenuation,
    with the recovered values by our fitting technique. The difference from
    Fig. \ref{sim1} is that in this particular case, a different
    library of SSPs was adopted to fit the spectra, compared to the one used
    to simulate them, as described in the text.
  \label{sim_bias} 

 }
\end{figure*}

The SSP models were created using the GISSEL code \citep{Bruzual:2003p3411}, assuming a
Salpeter IMF \citep{Salpeter:1955p3438} \citep[the use of][will not modify the
results]{Chabrier:2003p3777}, for different ages and metallicities. We create 72 models
covering a discrete grid of 12 ages (5 Myr, 25 Myr, 100 Myr, 290 Myr, 640 Myr,
0.9 Gyr, 1.4 Gyr, 2.5
Gyr, 5 Gyr, 11 Gyr, 13 Gyr, and 17 Gyr), and 6 metallicities ($Z=$0.0001,
0.0004, 0.004, 0.008, 0.02 and 0.05). Figure A1 shows the spectra
of all the templates, grouped by their metallicity, and normalized to their
flux at 5000 \AA. The adopted library is similar in many respects to the one
proposed by MacArthur et al. (2009). Both libraries have a similar number of
templates, although the latter library has a wider coverage of the ages and a
more reduced coverage of the metallicities of the stellar populations.

There is a number of caveats when applying model SSPs to the integrated light
of a star forming galaxy, which have been clearly identified by MacArthur et
al. (2009). The most important one is to assume that the parameter space
covered by the empirical library represents well that of the real
data. However in general, libraries are based on stars in the solar
neighborhood, and therefore it is not granted that they represent well the
stellar populations in other galaxies (or even in other regions of our
Galaxy). There are other potential problems related to the particular selected
templates, since it is well known that the Bruzual\& Charlot (2003) models
have problems when dealing with the non-solar abundance ratios. Most of these
problems are not particularly important in the context of our science case,
since (i) our primary goal is to model the stellar population to analyse the
ionized gas emission and (ii) we are interested in relative assessments about
the nature of the stellar populations in different regions of the studied
galaxy, and not in absolute values.

In addition, it is important to note that the treatment of the dust
attenuation may affect the resulting derived parameters (i.e. the luminosity-weighted
age and metallicity of the stellar population). In this particular
implementation of the analysis we adopted the \citet{Cardelli:1989p136}
attenuation law, which may not be the optimal solution to study the
dust attenuation in star-forming galaxies \citep[e.g.][]{Calzetti:2001p3421}.
MacArthur et al. (2009) adopted a completely different attenuation
law, based on the two-components dust model of \citet{Charlot:2000p3439}, which
is particularly developed to model the dust attenuation in star forming
galaxies. Despite the conceptual differences between the two attenuation laws,
their actual shapes are very similar in the wavelength range covered by our
data. The final range of dust attenuations explored by our fitting technique
comprises $A_V$ values up to 1 mag, with steps of 0.1 mag.

\subsection{Accuracy of the derived parameters}

As mentioned before, the basic parameters derived by the analysis of the
stellar component are the {\it luminosity-weighted} age, metallicity and dust
attenuation of the composite stellar population. In order to assess the
accuracy of these parameters we have performed a set of simulations. The
simulations were performed while trying to match as closely as possible the
original data, especially the noise pattern, which is composed of both white
noise corresponding to the photon-noise of the source and the background, and
electronic noise from the detector; and non-white noise corresponding to
defects/inaccuracies in the sky-subtraction, uncorrected defects in the CCD,
etc. These noise patterns are different spectrum-to-spectrum, and
wavelength-to-wavelength, and are clearly difficult to simulate on a simple
analytical basis.

\begin{table}
\begin{center}
\label{simtab}
\caption[Results of the simulations]{Results of the simulations.}
\begin{tabular}{rrrrr}\hline\hline
\multicolumn{1}{c}{S/N} & \multicolumn{1}{c}{rms} & \multicolumn{1}{c}{$\Delta$Age/Age} &
\multicolumn{1}{c}{$\Delta$Z} & \multicolumn{1}{c}{$\Delta$A$_V$} \\\hline
734.60 & 0.14 &  0.02\,$\pm$0.14 & 0.0001\,$\pm$0.0014 & 0.00\,$\pm$0.05\\
90.28  & 0.11 & -0.03\,$\pm$0.24 & 0.0006\,$\pm$0.0047 & 0.01\,$\pm$0.08\\
43.74  & 0.11 & -0.08\,$\pm$0.30 & 0.0004\,$\pm$0.0074 & 0.03\,$\pm$0.13\\
8.60   & 0.12 & -0.18\,$\pm$0.46 & 0.0014\,$\pm$0.0126 & 0.04\,$\pm$0.38\\
1.69   & 0.12 & -0.18\,$\pm$0.59 & 0.0006\,$\pm$0.0164 &-0.06\,$\pm$0.51\\\hline
 $^*$82.77   & 0.12 &  0.21\,$\pm$0.27 & 0.0010\,$\pm$0.0070&-0.01\,$\pm$0.10\\\hline
 $^{**}$81.97   & 0.13 &  0.12\,$\pm$0.47 & 0.0023\,$\pm$0.0075 &-0.01\,$\pm$0.23\\\hline
\end{tabular}
\end{center}
$^*$ Results when fitting the simulated spectra with a different library of
SSPs than the one used to create the simulation.

$^{**}$ Results when the input simulated spectra consists of a single SSP,
instead of a combination of them.
\end{table}

\begin{figure*}
\resizebox{\hsize}{!}{
\includegraphics[width=\hsize,angle=270]{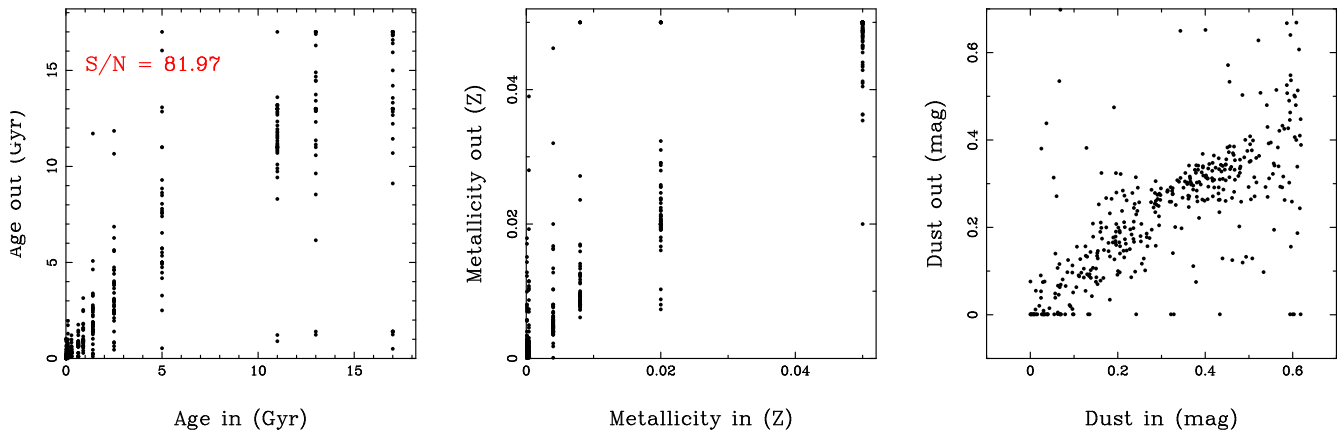}}
  \caption[Results from the simulations 3]{
    Results from the simulations. Each panel shows, from left to right, the
    comparison between the input luminosity-weighted age, metallicity and dust attenuation,
    with the recovered values by our fitting technique. The difference from
    Fig. \ref{sim1} is that in this particular case, the input model consists
    of a single SSPs, although it is fitted following the prescriptions shown
    in the text.
  \label{sim_ssp}
 }
\end{figure*}

Five different sets of simulations were performed. In each one, 500 simulated
spectra were created with the same normalized flux at 5000 \AA, corresponding
to 100, 10, 5, 1 and 0.2 \funits, respectively. For each simulation, a
composite stellar population was considered, comprising 4 different SSP
extracted randomly from the considered library template, with a different
relative contribution each one to the total flux: 53\%, 27\%, 13\% and 6\% at
5000 \AA, respectively. This procedure was adopted to resemble in a simple way
a possible star-forming history, with a dominant stellar population, and three
different bursts or merging events. Different experiments adopting a larger
number of intermix stellar populations, and relative contributions of each one
to the total were considered, with no significant variations on the derived
conclusions.

Once a particular spectrum was created, it was convolved with the instrumental
resolution of the observed data, and the corresponding noise pattern was
added. The noise pattern was created on the basis of the residuals of the
fitting procedure applied over the observational data. For each spectrum of
the original mosaic, the fitting procedure creates a residual spectrum, once
the derived composite stellar population and emission line models were
subtracted. This {\it noise} spectrum includes all the effects of the white
noise (for a particular flux level), and non-white noise (deficient sky
subtraction, CCD defects not corrected by the reduction, imperfections of the
modelling process, etc.). From this set of {\it noise} spectra, one was
randomly selected for each simulated spectrum, fulfilling the requirement that
the simulated and original fluxes at $\sim$\,5000 \AA\ match within a range of
$\sim$\,20\%. The selected {\it noise} spectrum was then added to the
simulated spectrum.

\begin{figure}
  \label{rad_flux}
  \resizebox{\hsize}{!}{
    \includegraphics[width=\hsize,angle=270]{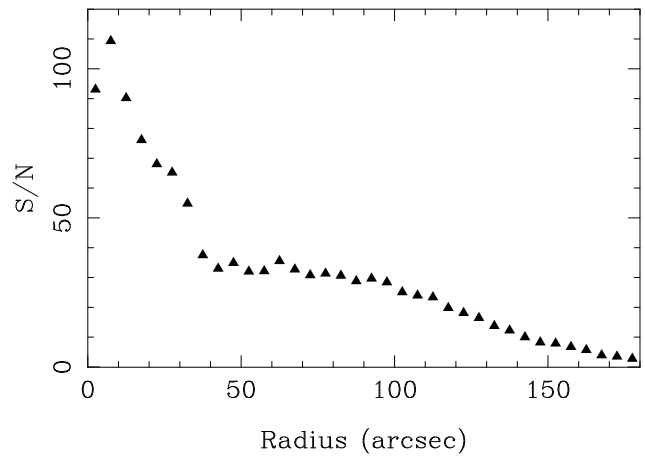}}
  \caption[Radial average S/N intensity of the IFS mosaic of NGC\,628]
  {
    Radial distribution of the average S/N at $\sim$\,5000 \AA\ of the spectra 
 within the flux calibrated mosaic of NGC\,628.
 }
  \end{figure}

The fitting procedure was then applied to the simulated spectra using the same
steps adopted to analyse the observational data, deriving, for each one, the
{\it luminosity-weighted} ages, metallicities and dust attenuation, and a new residual
spectrum. Table A1 lists, for each set of simulated spectra, their average
signal-to-noise ratio, the {\it rms} of the relative difference between the
input spectra (without noise) and the model derived by the fitting procedure
and the difference between the input and the recovered parameters that define
these models (luminosity-weighted age, metallicity and dust attenuation), and their
corresponding {\it rms}. As expected the parameters are better recovered for
the spectra with higher signal-to-noise ratios, requiring at least a S/N of
$\sim$\,50 per pixel to derive accurate results. Figure A2 illustrates
these results, showing for each set of simulations the comparison between the
input and recovered parameters for each individual spectrum in the dataset.

A possible source of error in the interpretation of these simulations is the
fact that the same set of templates was used to create the simulated data and
to model them. In principle, we assumed that the grid of templates is
representative of the real stellar population of the analysed spectra, but it
is very possible that our library is incomplete. This is not the case in our
simulations, by construction. To study the effects of this possible incomplete
representation of the observational spectra by the template library, we
created a new set of simulated data, corresponding to a flux level of 10
\funits, that was then fitted using a much reduced version of the template
library. This template comprises a grid of 12 SSPs, corresponding to 3 ages
(17 Gyrs, 900 Myrs and 25 Myrs), and 4 metallicities
($Z=$0.0001,0.004,0.02,0.05). The results from this simulation are listed in
Table \ref{simtab}, and shown in Figure A3. It shows that in the
case of an incomplete coverage of the spectroscopic parameters by the model
template, these parameters are still well reconstructed, albeit with a lower
accuracy (as expected).

Finally, we tested how well the code is able to reproduce a pure SSP, instead
of a mixture of them. With this test we try to overpass the intrinsic problems
in the interpretation of the luminosity-weighted ages/metallicities in
galaxies, and to test if we can recover the better understood properties of a
pure single-burst population. The results from this simulation are listed in
Table \ref{simtab}, and shown in Figure A4. Despite the discrete coverage of
the input parameters, it is clear that the parameters are recovered just
slightly worse than in the case of a multiple component stellar population.


On the other hand, the relative difference between the original spectra and
the recovered model is lower than $\sim$\,15\% for all the simulations, even
in those ones for which the accuracy of the recovered parameters is
deficient. These differences are of just 2-3\% for the spectra with the
highest signal-to-noise ($>$100). This basically means that the general shape
of the spectra is better recovered than the parameters that defines them. This
is easy to understand, since these parameters are characterised by particular
features in the spectra, rather than by their general shapes, which are more
degenerated.

Figure A5 shows the radial distribution of the average signal-to-noise ratio
at $\sim$\,5000 \AA\ of the spectra within the flux-calibrated mosaic of
NGC\,628, as described in Sec. 3. On the basis of this radial profile and the
results of the simulations discussed before it is possible to recover the
considered parameters of the composite stellar populations for individual
spectra only at radii less than $\sim$\,10 arcsec, for the central kpc of
the galaxy. The results are robust only for the integrated spectrum of the
galaxy (with a S/N\,$\sim$\,400), and for radii lower than $\sim$\,20 arcsec
(with S/N\,$\sim$\,100). However, for the azimuthally averaged spectra, where
the signal-to-noise increases by the root-mean square of the number of
individual averaged spectra, the simulations indicate that the recovery of the
considered parameters is robust even at radii as large as $\sim$\,120 arcsec,
above which both the S/N and the number of spectra with accurate
spectrophotometry also drops (see Figure 4). The combination of both effects
impose a limitation to the accuracy of the derived parameters at higher radii.

\label{lastpage}

\end{document}